\documentclass[journal]{IEEEtran}

\usepackage{graphicx}
\usepackage{float}
\usepackage{dirtytalk}
\usepackage{amsmath}
\usepackage{tikz}
\usepackage{tabularx}
\usepackage{blindtext}
\usepackage{hyperref}
\usepackage{cite}
\usepackage{stfloats} 

\begin{document}

\title{On the Intersection of Signal Processing and Machine Learning: A Use Case-Driven Analysis Approach}

\author{Sulaiman~Aburakhia,~\IEEEmembership{Member,~IEEE,}
       Abdallah~Shami,~\IEEEmembership{Senior~Member,~IEEE}
       and~George K. Karagiannidis,~\IEEEmembership{Fellow,~IEEE}

\thanks{Sulaiman Aburakhia and Abdallah Shami are with the Department of
Electrical and Computer Engineering, Western University, N6A 3K7, Canada (e-mail: saburakh@uwo.ca; abdallah.shami@uwo.ca).\newline George K. Karagiannidis is with Department of Electrical and Computer Engineering, Aristotle University of Thessaloniki, Greece and also with Artificial Intelligence and Cyber Systems Research Center, Lebanese American University (LAU), Lebanon  (e-mail: geokarag@auth.gr).}}

\markboth{}%
{Shell \MakeLowercase{\textit{et al.}}: Bare Demo of IEEEtran.cls for IEEE Journals}

\maketitle

\begin{abstract}
Recent advancements in sensing, measurement, and computing technologies have significantly expanded the potential for signal-based applications, leveraging the synergy between signal processing and Machine Learning (ML) to improve both performance and reliability. This fusion represents a critical point in the evolution of signal-based systems, highlighting the need to bridge the existing knowledge gap between these two interdisciplinary fields. Despite many attempts in the existing literature to bridge this gap, most are limited to specific applications and focus mainly on feature extraction, often assuming extensive prior knowledge in signal processing. This assumption creates a significant obstacle for a wide range of readers. To address these challenges, this paper takes an integrated article approach. It begins with a detailed tutorial on the fundamentals of signal processing, providing the reader with the necessary background knowledge. Following this, it explores the key stages of a standard signal processing-based ML pipeline, offering an in-depth review of feature extraction techniques, their inherent challenges, and solutions. Differing from existing literature, this work offers an application-independent review and introduces a novel classification taxonomy for feature extraction techniques. Furthermore, it aims at linking theoretical concepts with practical applications, and demonstrates this through two specific use cases: a spectral-based method for condition monitoring of rolling bearings and a wavelet energy analysis for epilepsy detection using EEG signals. In addition to theoretical contributions, this work promotes a collaborative research culture by providing a public repository of relevant Python and MATLAB signal processing codes. This effort is intended to support collaborative research efforts and ensure the reproducibility of the results presented.
\end{abstract}

\begin{IEEEkeywords}
Signal processing, signal preprocessing, signal denoising, feature extraction, signal transforms
\end{IEEEkeywords}

\IEEEpeerreviewmaketitle

\section{Introduction}

The rapid advancements in sensing and measurement represent a paradigm shift in how data is collected, processed, and interpreted. This opens the door for a wide range of signal-based applications, marking a transformative phase across various fields. Moreover, the development of computing technologies and the rise of the Internet of Things (IoT) have paved the way to leverage machine learning (ML) within signal-based applications, offering new insights and achieving unprecedented levels of accuracy and efficiency. This merge between signal processing and ML is expected to play a major role in the next generations of sensor-enabled systems across various areas \cite{rl22}. The integration of signal processing pipelines into ML models forms the fundamental core of these systems. Further, it represents a critical intersection in their advancement, motivating the research community to address the role of signal processing in ML. However, the diverse landscape of signal types and application requirements shapes the scope of the existing body of work to be application-centric, limiting their scopes to specific applications. For instance, The role of feature extraction in ML has been extensively reviewed within the context of vibration-based predictive maintenance (PdM). The work in \cite{aa21} presents a review that focuses on transforming traditional methods to ML techniques in applying vibration-based damage detection in civil structures. The article highlights traditional methods and presents a comprehensive review of the latest applications of ML algorithms used for this purpose. In \cite{al21}, a systematic review is conducted on adopting machine learning for failure prediction in industrial maintenance. The review covers the used datasets, preprocessing, and the training and evaluation of prediction models. In \cite{vz22}, a comprehensive review is presented on signal processing techniques for vibration-based feature extraction in structural health monitoring (SHM). The work in \cite{ar23} addresses the application of vibration-based condition monitoring techniques for the PdM of rotating Machinery. It provides a comprehensive review of vibration data acquisition and analysis, as well as the methods used for fault interpretation and diagnosis, including data acquisition, data transmission, signal processing, and fault detection. The work in \cite{sm24} introduces a tutorial on the same topic that describes relevant signal processing methods in this field. Furthermore, the tutorial provides Python and MATLAB code examples to demonstrate these methods alongside explanatory videos. \

In the biomedical field, the topic of signal processing in ML is an active area of research. For instance, the work in \cite{bs23} provides an end-to-end review of biomedical signal processing for health monitoring applications. It introduces a flow for developing biomedical signal processing systems. Further, it covers recent applications, types of low-cost, non-invasive biomedical sensors, signal processing techniques, and future perspectives for building reliable systems. The application of MI-based brain-computer interfaces (BCIs) in controlling external devices through EEG signal processing is addressed in \cite{ak23}. The article reviews recent ML models, identifies major challenges, and suggests potential solutions by focusing on feature extraction and classification methods. The work in \cite{aj21} reviews various feature extraction techniques for electrocardiogram (ECG) Signal analysis. In \cite{ab17}, the main steps in detecting and classifying EEG epileptic seizure activities are addressed, along with a review of related feature extraction techniques. The studies in \cite{cy23} and \cite{fj14} deal with the application of emotion recognition using EEG signals, where different existing feature extraction methods are analyzed and compared in terms of classification performance. A comprehensive review of methods and techniques that covers the entire process of EEG signal processing is presented in \cite{ec23}. The study analyzed numerous articles related to EEG signal processing, identified limitations, and analyzed future development trends. Besides biomedical and BdM fields,  the role of signal processing in ML has been addressed in various fields such as audio analysis and recognition \cite{rev1, rev2, rev3, rev4, rev5, rev6, rev7}, seismic signal analysis \cite{rev7, rev8, rev9}, and telecommunications\cite{rev10, rev11, rev12, rev13,  rev14, rev15, rev16, rev17}.\

Despite this diverse spectrum of articles addressing the role of signal processing within ML, there remains a noticeable gap in the literature, characterized by the following four main limitations in the existing studies:
\begin{itemize}
    \item Application-Specific Focus:  The analysis are often confined to specific applications, which limits the coverage  and depth in which the topic is addressed.
    \item Limited Audience Reach: Many articles assume a substantial background in signal processing, restricting accessibility to a limited audience.
    \item Task-Centric Approach: The exploration of signal processing in ML tends to focus on particular tasks, such as signal preprocessing or feature extraction. This neglects other critical tasks such as signal segmentation, smoothing, and denoising, which are essential components of a typical signal processing ML pipeline. 
    \item Theory-to-Practice Gap: There is a notable lack of practical demonstrations through use-case applications. Such demonstrations are critical to connecting theoretical concepts with real-world applications, and thus bridging the gap between theory and practice.
\end{itemize}\

This paper attempts to bridge this gap by implementing an integrated-article approach that addresses the aforementioned shortcomings through the following contributions:
\begin{itemize}
    \item Comprehensive Tutorial for Diverse Readership: The paper starts with a comprehensive tutorial on the fundamentals of signal processing that is aimed at readers from different domains, allowing the interested reader to develop a proper background before delving into the review.
    \item Application-Independent Approach: The paper adopts a broad, application-independent review, providing a comprehensive overview of signal processing in ML that is not shaped to a specific use case.
    \item End-to-End Overview of Signal Processing Workflow: The article thoroughly discusses the key tasks in a typical signal processing pipeline, grouping them under three main categories: preprocessing, processing, and application.
    \item Exhaustive Review with Novel Taxonomy: The paper conducts a detailed review of feature extraction techniques, presented and categorized through a new taxonomy that presents new insights and enriches the reader's understanding. 
    \item Addressing Major Challenges and Potential Solutions: The paper identifies the primary challenges faced in implementing signal processing-based ML applications and discusses both existing and potential solutions.
    \item Bridging Theory with Practice: The paper addresses the practical application of signal processing in ML through two use cases. In the first use case, a spectral-based method is introduced for vibration-based condition monitoring of rolling bearings. In the second case, wavelet-energy analysis is utilized for epilepsy detection using EEG signals.
    \item Public Repository of Signal Processing: The paper introduces a public repository of the Python and MATLAB codes used throughout the article, along with additional codes relevant to signal processing, thereby fostering a collaborative research environment and ensuring reproducibility of work.
\end{itemize}
\begin{figure*}[!h]
\centerline{\includegraphics[width=0.75\textwidth]{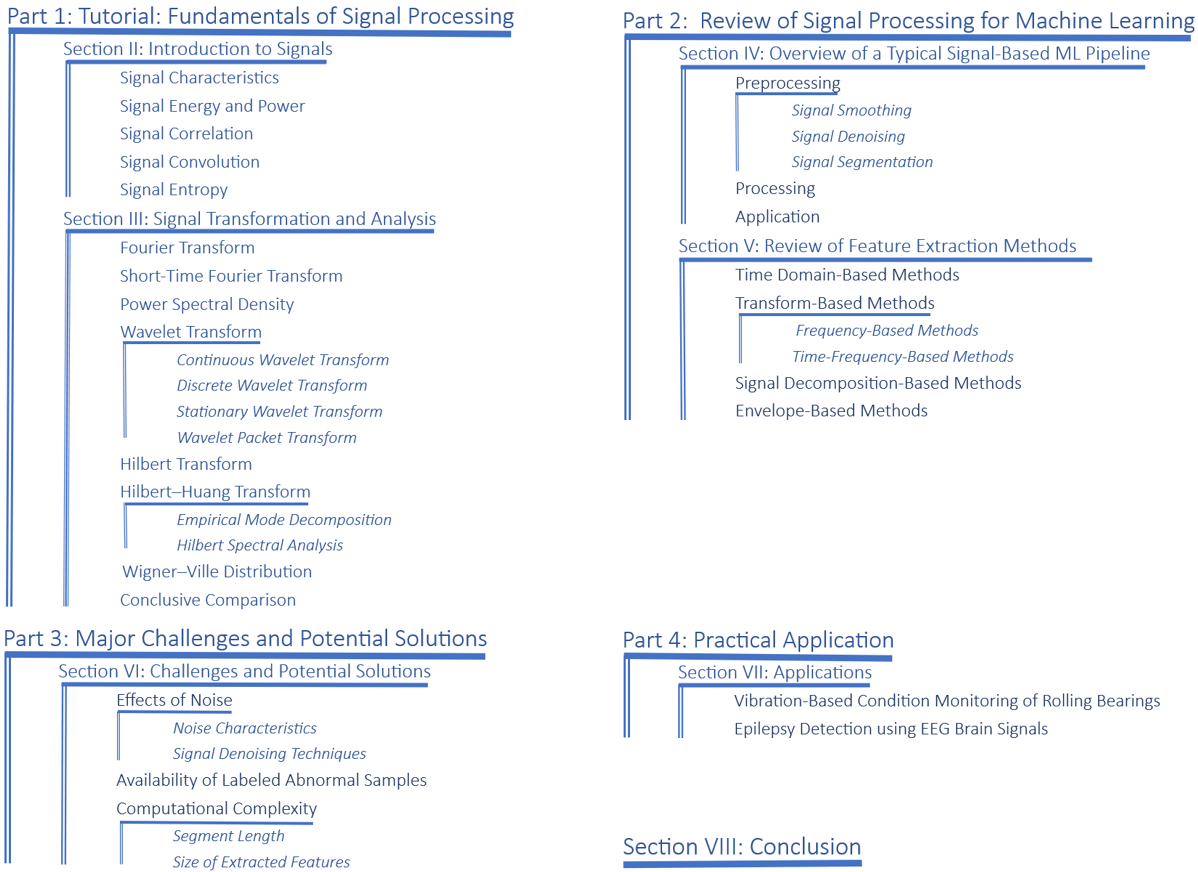}}
\caption{Visual representation of the paper's contents.}
\label{content}
\end{figure*}
With these contributions, the paper aims to become a foundational reference on the topic of signal processing in ML, serving a wide range of readers and offering new perspectives on the intersection of signal processing and ML. Each of these contributions has its own section in the paper. The sections are logically grouped under four parts.  For reader convenience, a visual representation that shows the various sections grouped under each part is depicted in Fig. \ref{content}. Although these parts are organized to build knowledge progressively, they are written in a self-contained manner, allowing for selective reading based on interest or need.\

The article is structured as follows:  The tutorial is carried out in Section II and Section III. A typical signal-processing pipeline for signal-based ML applications is presented in Section IV. The review of feature extraction techniques is introduced in Section V. Section VI discusses the main challenges faced in implementing signal processing-based ML applications and addresses potential solutions. The practical application use cases are introduced in Section VII. The paper is finally concluded in Section VIII.

\section{Introduction to Signals}
This tutorial provides an in-depth- introduction to signal processing, highlighting concepts, mathematical formulation, applications, advantages, and limitations of common signal processing tools. Additionally, the tutorial offers insights into implementation considerations and highlights programming libraries that offer functionalities for implementing these tools. Throughout the tutorial, illustrative examples are generated using Python and MATLAB codes. These codes are publicly available on the Github site of the Optimized Computing and Communications (OC2) Laboratory\footnote[1]{\url{https://github.com/Western-OC2-Lab/Signal-Processing-for-Machine-Learning}}

\subsection{Signal Characteristics}
A signal is a function of time; it represents the value of a physical entity or phenomenon as it evolves over time, such as voltage, current, acceleration…etc. Depending on the field, there are more definite definitions of the term "signal". For instance, within a signal processing context, a signal can be defined as a function that conveys information about the behavior of a system or attributes of some phenomenon \cite{ip90}. In manufacturing, the term “signal” refers to a physical quantity that carries a certain type of information and serves as a means for communication \cite{sg11}. Examples of such signals include a vibration signal generated by an accelerometer attached to the rolling-bearing element of a rotating machinery and a torque signal generated by a torque Sensor in a computer numerical control (CNC) milling machine. The changes in these signals are directly related to the operation of the machine and can therefore be used to communicate the operating status of the machine to the machine operator \cite{sg11}.\

\begin{figure}[!htbp]
\centerline{\includegraphics[width=0.5\textwidth]{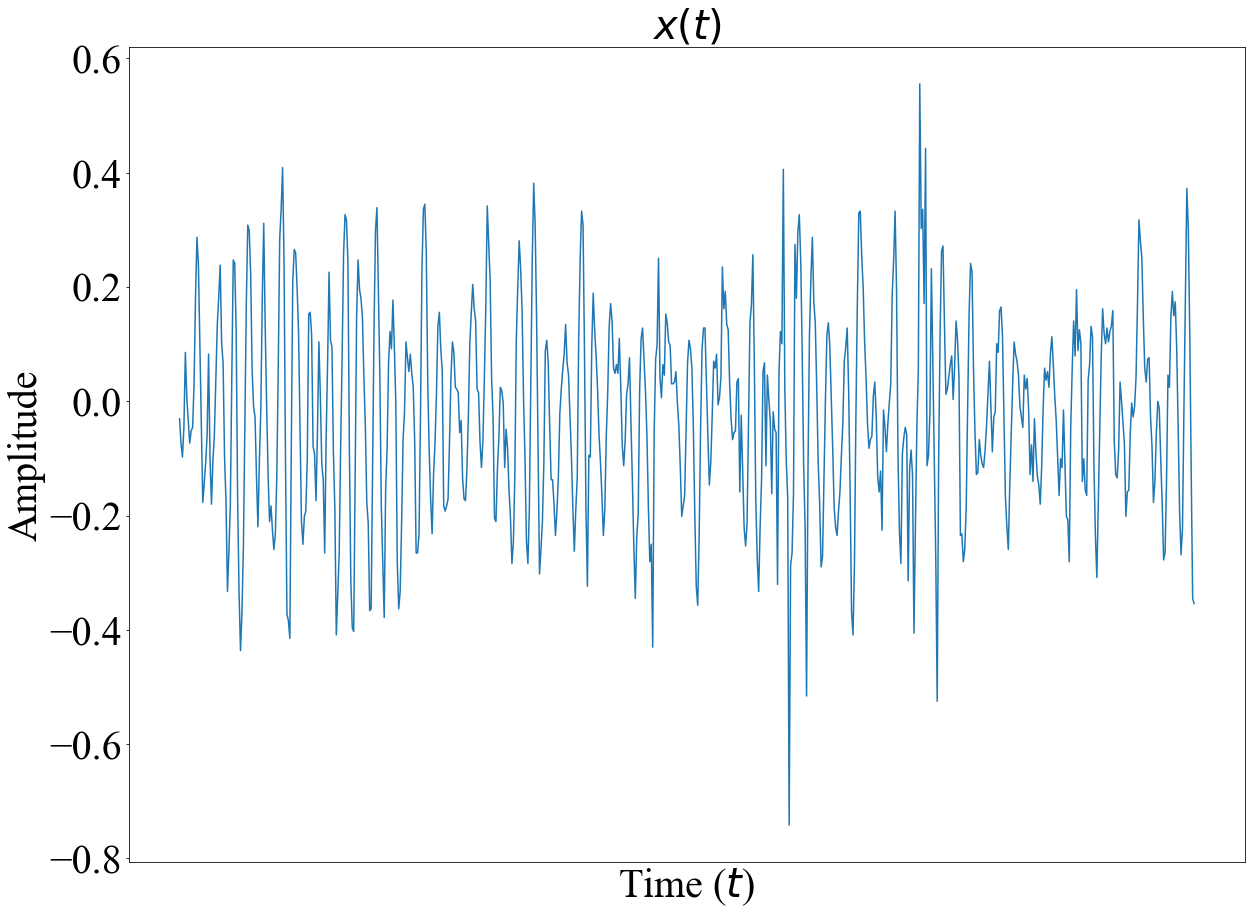}}
\caption{Waveform of a signal, $x(t)$, as it evolves over time $t$.}
\label{wavefrom}
\end{figure}

Signals are typically represented by their time waveforms; Fig. \ref{wavefrom}. displays the waveform of a given $x(t)$ as it evolves over time $t$. 
Amplitude, frequency, and phase are the main characteristics of a signal. Amplitude measures the amount and direction of change in the signal with respect to a reference value, as shown in Fig. \ref{wavefrom}. Frequency is the number of cycles or oscillations that occur in a given unit of time. It represents the rate at which the signal oscillates. Frequency is measured in \textit{Hertz (Hz)} where $1$ \textit{Hz} is equal to $1$ cycle per second. The phase of a signal refers to the position of a point (time instant) on the signal's waveform cycle. The phase of a wave refers to its position within its cycle and is measured in either degrees ($0-360$) or radians ($0-2\pi$). It helps describe the relative position and timing of two signals at a given moment. The concept of frequency of a signal is closely related to the rate at which its phase changes. Specifically, the frequency at a particular moment in time indicates the speed at which the phase is changing at that moment. A higher frequency indicates a faster rate of phase change. Therefore, frequency and phase are mathematically linked through the concept of the derivative. By knowing the phase information of a signal, the frequency can be calculated as the first derivative of the phase with respect to time. The period of a waveform refers to the time it takes to complete one cycle. A periodic signal is a signal that repeats its pattern or the sequence of values exactly after a fixed duration time, known as the period $T$. This can be expressed mathematically as:
\begin{equation}
    x(t)= x(t+T)
\end{equation}
Since the period is the time duration a signal takes to complete one cycle, it is related to the signal's frequency, which represents the number of cycles per second, through the following relation:
\begin{equation}
    T = \frac{1}{f} \; \; (seconds)
\end{equation}
While purely periodic signals do not exist in practice \cite{fc19}, they represent an essential theoretical concept in signals theory \cite{sb01}. A sinusoidal wave, $s(t)$, represents the basic form of a periodic signal; it is expressed mathematically as:

\begin{equation}
  s(t) = A \sin(2\pi ft + \phi) 
\end{equation}

\begin{figure}[!htbp]
\centerline{\includegraphics[width=0.5\textwidth]{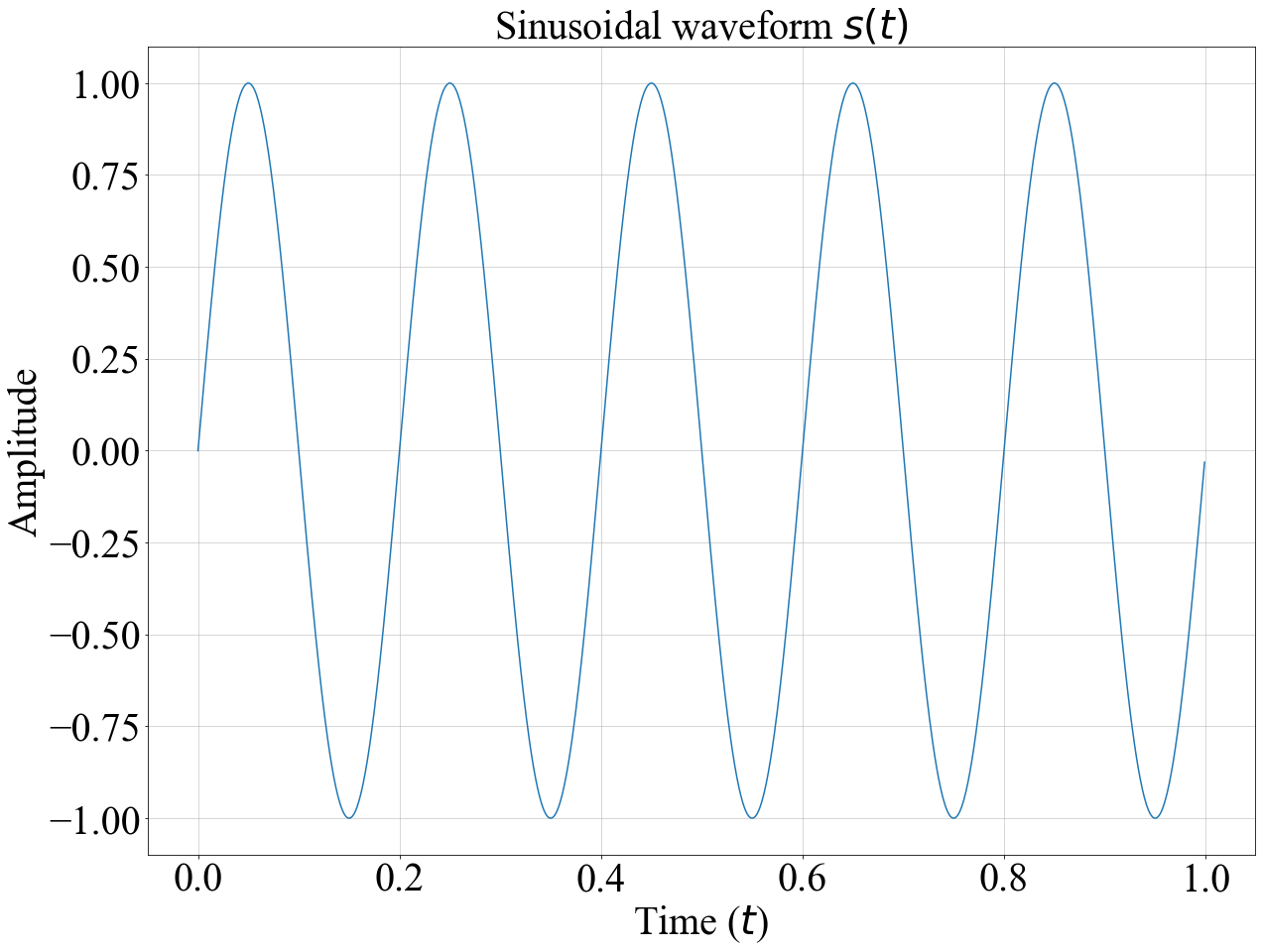}}
\caption{Sinusoidal waveform $s(t)$.}
\label{sin}
\end{figure}

\begin{figure}[!htbp]
\centerline{\includegraphics[width=0.5\textwidth]{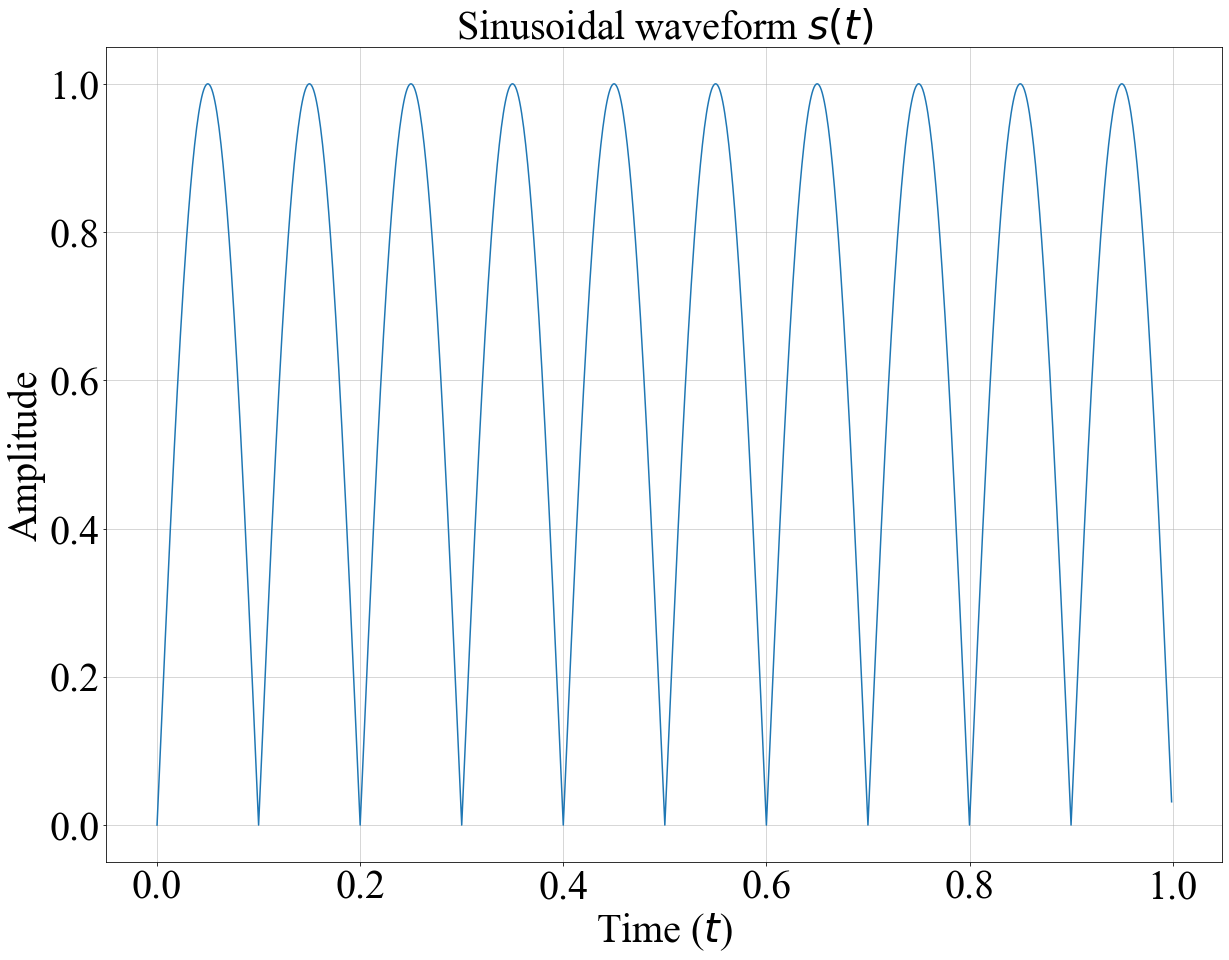}}
\caption{Magnitude of the sinusoidal waveform $s(t)$.}
\label{sin_mag}
\end{figure}

Where $A$ is the signal’s peak amplitude, $f$ is frequency, and $\phi$ is the phase.  With respect to zero, signal amplitude can be either positive or negative, as shown in Fig.\ref{sin}. 
On the other hand, the magnitude of the signal $|s(t)|$, shown in Fig. \ref{sin_mag}, 
is the absolute value of the amplitude, it shows the change in the signal regardless of its direction since it is always a positive quantity. Amplitude values indicate the strength of the signal, reflecting the intensity of the measured physical phenomenon.\

As explained earlier, a signal represents a physical entity or phenomenon that changes over time. A signal that remains unchanged over time when measured again under the same conditions is referred to as a deterministic signal. Such a signal has no uncertainty  about its value at any given point in time and can be modeled using mathematical expressions. The sinusoidal function is an example of a deterministic function. However, signals that are completely free of unknown and uncontrollable factors and have true deterministic properties are extremely rare  \cite{ch12_sd03}. In practical scenarios, signals are non-deterministic as they are random in nature and exhibit various degrees of uncertainty. Since non-deterministic signals are stochastic, they are described in terms of their statistical properties. From this perspective, random signals can be classified into two categories: stationary and nonstationary. Stationary signals generally maintain constant statistical properties over time, while nonstationary signals exhibit changing statistical properties over time.\

From the preceding discussion, it is apparent that signal variation describes the changes in the measured physical entity or phenomenon as it evolves over time. This variation is subject to the influence of uncontrollable factors during the measurement period, which manifest themselves as noise  or outliers in the signal. Other influencing factors include subsequent processing and operations imposed on the signal, commonly referred to as "systems" in the context of signal processing. Specifically, a system in signal processing is any process or operation that produces an output signal in response to an input signal. In practical situations, real-world systems exhibit mostly nonlinear behaviors\footnote[2]{A linear system in signal theory is defined by two key properties: homogeneity and additivity. Homogeneity means if an input signal $x(t)$ results in an output $y(t)$, then any scaled input $a   x(t)$ leads to a scaled output $a   y(t)$. Additivity implies that the response to a sum of input signals $x_1(t) + x_2(t)$ is the sum of their responses, $y_1(t) + y_2(t)$.} and operate under transient, nonstationary conditions \cite{ns17}. Therefore, real-world signals in practical situations exhibit nonlinear, time-varying, and nonstationary characteristics. Understanding the unique characteristics of a signal holds significant importance, as it directly influences the selection of the proper signal processing tool, ensuring precise and effective analysis of the signal and, consequently, extracting discriminative features.\

\begin{figure}[!htbp]
\centerline{\includegraphics[width=0.5\textwidth]{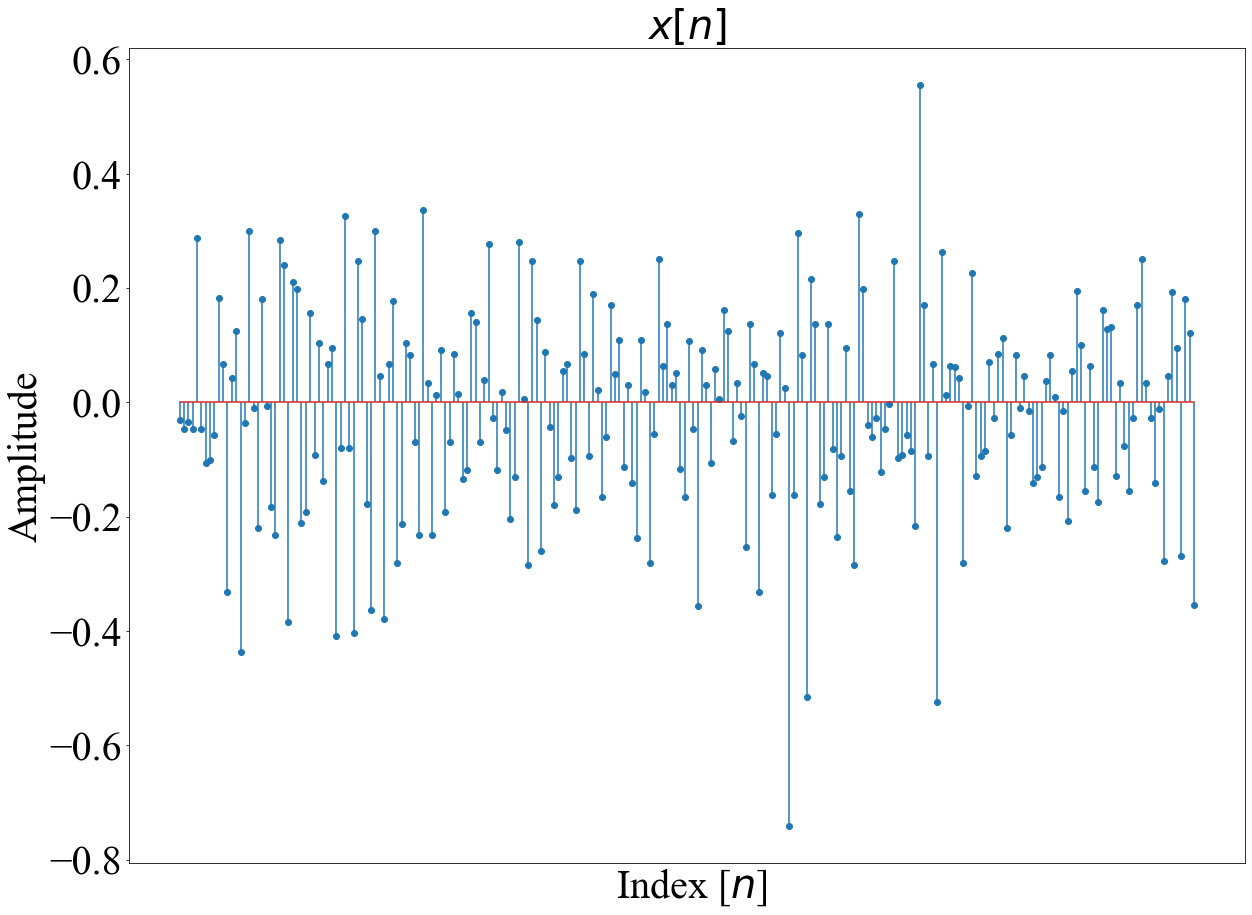}}
\caption{$x[n]$, a discrete-time version of signal $x(t)$.}
\label{x_n}
\end{figure}

Another fundamental aspect of signal processing is signal acquisition and the time representation of the signal. From a time perspective, signals are classified as continuous or discrete-time signals. The continuous-time signal $x(t)$ is the signal that has a continuous value over the observation time range, as depicted in Fig. \ref{wavefrom}. A discrete-time version of the signal $x(t)$ is displayed in Fig. \ref{x_n}, denoted by $x[n]$. In the discrete-time representation, the independent variable $n$ is discrete in type and takes integer values only. That's it; in contrast to a continuous-time signal, the discrete signal is defined only over discrete time intervals. 
The discrete-time representation $x[n]$ is obtained by sampling the continuous-time signal $x(t)$ at time instants separated by $T_s$. \textit{i.e.}
\begin{equation}
x[n]=x(nT_s),\; n =0,1,2,\dots
\end{equation}
where $T_s$ represents the sampling interval or sampling period in seconds. Accordingly, the sampling frequency or sampling rate $f_s$ is the number of samples obtained in $1$ second; it is given by:
\begin{equation}
f_s=\frac{1}{T_s}
\end{equation}
The unit of $f_s$ is samples per second or hertz. In practice, a measured signal is a sampled version of the actual physical signal since it is acquired by recording a specific number of measurements (samples) every second. The sampling rate plays a crucial role in signal measurement and processing. A higher sampling rate means more measured samples per second and, hence, a higher signal resolution. However, processing time and memory requirements increase as the sampling rate increases. A fundamental concept in signal sampling is the Nyquist theorem \cite{sl67}, which states that a signal can be correctly reconstructed from its sampled discrete-time sequence provided that the sampling rate is equal to or greater than twice the highest frequency (effective bandwidth) of the signal. This implies that the highest frequency, represented as $f_{max}$,  that can be represented accurately in an acquired signal is one-half of its sampling rate $f_s$; \textit{i.e.}
\begin{equation}
\label{nr}
f_s=2f_{max}
\end{equation}
This formula is called the Nyquist rate; it provides a lower bound on the sampling rate that is required to reconstruct a signal from its sampled version accurately. Accordingly, a sampled signal is categorized as either undersampled, critically sampled, or oversampled:
\begin{itemize}
    \item Undersampled signal: is one where the sampling rate is below the Nyquist sampling rate. Undersampling leads to a distortion in the signal known as aliasing, causing signal components at higher frequencies than the sampling frequency to appear at a lower, aliased frequency \cite{te12}.
    \item Critically sampled signal: is one where the sampling rate equals the Nyquist sampling rate. In this case, the signal is sampled at the lowest rate, which still allows for a complete reconstruction of the continuous-time signal, assuming the signal bandwidth is limited to half of the sampling frequency. While perfect reconstruction is theoretically still possible, it is less common in practical applications.
    \item Oversampled signal: is one where the sampling rate exceeds the Nyquist rate. Oversampling is widely used in signal measurements because it improves the resolution of the acquired signal. Additionally, with oversampling, the quantization noise (an error that represents the difference between the value of the measured sample and the closest mapped digital value) is spread over a broader frequency range compared to the effective bandwidth of the signal. Accordingly, filtering out high frequencies outside the effective bandwidth effectively reduces noise in the signal, improving the signal-to-noise ratio (SNR).
\end{itemize}

\subsection{Signal Energy and Power}
As noted above,  signal amplitude reflects the intensity of the measured physical phenomenon, making it significant in identifying critical features of the signal, such as energy and power. Energy quantifies the total "work" done by the signal; mathematically, the energy of a signal is the area under the squared magnitude of the signal over a given time interval. for a continuous-time signal $x(t)$, its energy $E$ is defined as:
\begin{equation}
    E = \int_{-\infty}^{\infty} |x(t)|^2 \, dt
\end{equation}
For a discrete-time signal $x[n]$, the energy is given by:
\begin{equation}
    E = \sum_{n=-\infty}^{\infty} |x[n]|^2
\end{equation}
This definition assumes the signal is energy-limited, meaning that its energy is finite over the entire time domain. Signal power measures the average rate at which the signal transmits energy. It is calculated as the average power over a period of time. For a continuous-time signal, its average power $P$ is defined as:
\begin{equation}
    P = \lim_{T \to \infty} \frac{1}{2T} \int_{-T}^{T} |x(t)|^2 \, dt
\end{equation}
Similarly, for discrete signals:
\begin{equation}
    P = \lim_{N \to \infty} \frac{1}{2N+1} \sum_{n=-N}^{N} |x[n]|^2
\end{equation}
If the signals' energy $E$ converges to a finite but non-zero value, the signal is classified as an energy signal. Examples include non-zero signals over a limited period but zero elsewhere, such as pulse signals. The average power of an energy signal is zero because the calculation of power involves averaging the finite energy over an infinite period of time. Power signals, on the other hand, are signals that have finite, non-zero average power. A common example is a periodic sinusoidal signal that continues indefinitely; such signals have infinite energy. For signals of finite duration, energy and power can be evaluated over their active time intervals. For periodic signals, due to the repetitive nature of the signal, the computation is simplified by considering one single period $T$ as follows:
\begin{equation}
    P = \frac{1}{T} \int_{-\frac{T}{2}}^{\frac{T}{2}} |x(t)|^2 dt
\end{equation}
For a discrete-time periodic signal $x[n]$ with period $N$, the power is:
\begin{equation}
   P = \frac{1}{N} \sum_{n=0}^{N - 1} |x[n]|^2
\end{equation}\
Instantaneous power refers to the power of a signal at a specific moment in time. Unlike average power, which is calculated over a period of time, instantaneous power gives a moment-to-moment view of the power level of the signal. For a continuous-time signal $x(t)$, the instantaneous power $P(t)$ at time $t$ is defined as the square of the signal's magnitude at that particular time. Mathematically, it is expressed as:
\begin{equation}
   P(t) = |x(t)|^2
\end{equation}
For a discrete signal $x[n]$, where $n$ represents discrete time indices, the instantaneous power $P[n]$ at point $n$ is defined as the square of the magnitude of the signal at that point. Mathematically, this is expressed as:
\begin{equation}
   P[n] = |x[n]|^2
\end{equation}
For a given discrete time signal $x[n]$, represented by $N$ samples, the total energy $E$ is the sum of its instantaneous power values over its discrete time interval: 
\begin{equation}
E=\sum_{n=0}^{N-1}|x[n]|^2
\end{equation}
Accordingly, the average power $P$ is obtained by dividing the energy $E$ by the number of samples $N$:
\begin{equation}
P=\frac{\sum_{n=0}^{N-1}|x[n]|^2}{N}
\end{equation}
along with power and energy, statistical properties of the signal, such as mean and variance, represent important aspects of signal characteristics and play a major role in signal processing. For the discrete time signal $x[n]$, its mean $\mu$  is equal to:
\begin{equation}
\mu = \frac{1}{N} \sum_{n=0}^{N-1} x[n]
\end{equation}
The mean represents the central tendency or the average value of the signal over its time period. The signal's variance $\sigma^2$ is given by:
\begin{equation}
\sigma^2 = \frac{1}{N} \sum_{n=0}^{N-1} (x[n] - \mu)^2
\end{equation}
The variance measures the spread of the signal's amplitude around its mean value, quantifying the deviation of the signal's samples from its mean. Note that for a zero-mean signal, The variance equals the signal's average power. For such signals, the root-mean-square (RMS) is a useful measure that represents the effective value of the signal because it provides a meaningful measure of the average magnitude of the signal, even though its mean is zero. The RMS is expressed as:
\begin{equation}
x_{RMS} = \sqrt{\frac{1}{N} \sum_{n=0}^{N-1} [x[n]]^2}
\end{equation}

\subsection{Signal Correlation}
\begin{figure}[!htbp]
\centerline{\includegraphics[width=0.5\textwidth]{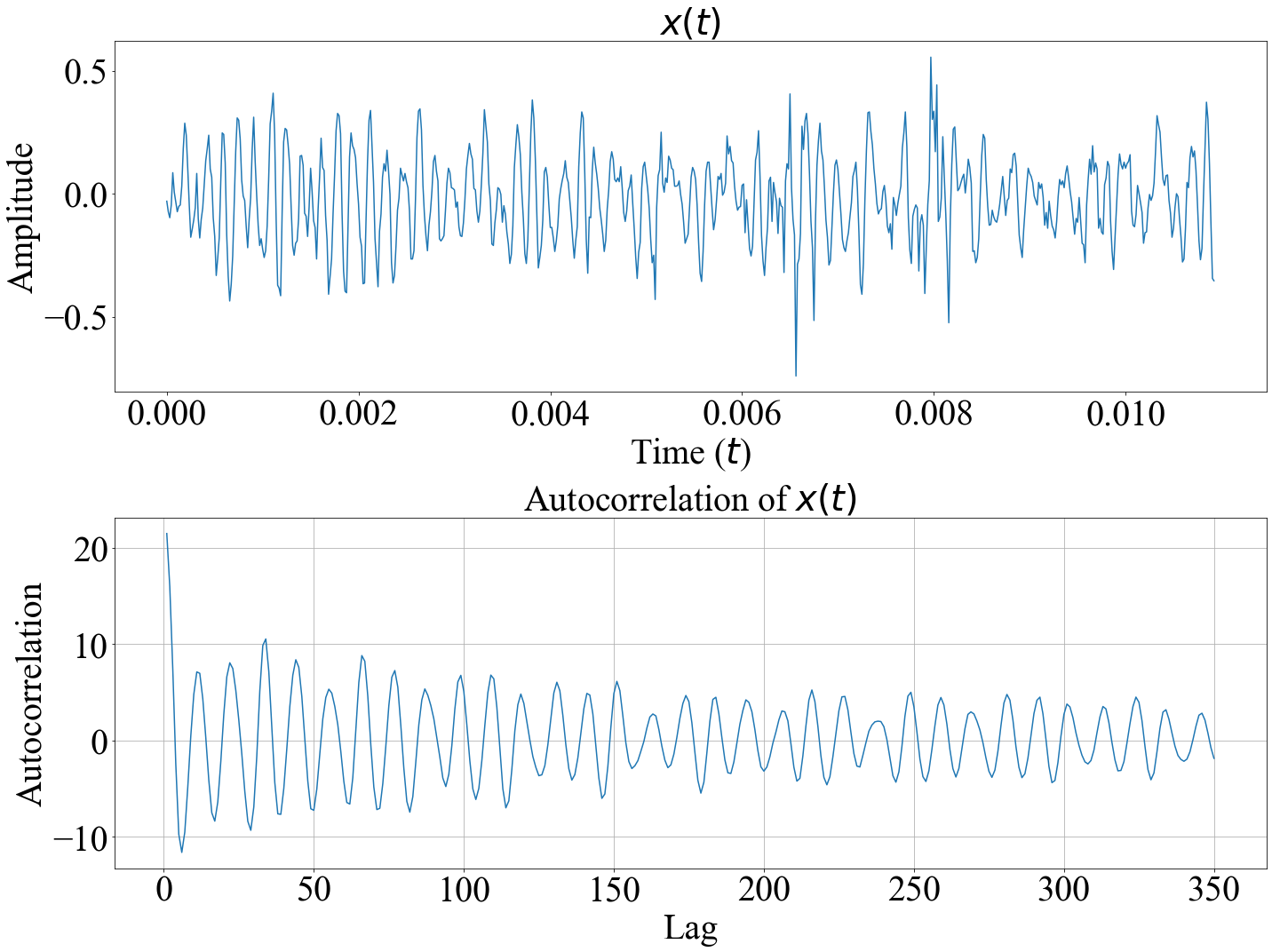}}
\caption{Signal $x(t)$ and its autocorrelation.}
\label{auto_corr}
\end{figure}

Correlation can be viewed as a similarity measure; it quantifies the similarity or the degree of dependence between two signals, providing insights into their common characteristics and patterns. Auto-correlation refers to the correlation of a signal with a delayed version of itself. It represents the degree of similarity between a signal and a time-shifted (lagged) version of itself over a given time interval. Auto-correlation helps to detect repeating patterns in the signal and assess its stability or predictability over time. For a continuous-time signal $x(t)$, the auto-correlation function is expressed as:
\begin{equation}
    R_{xx}(\tau) = \int_{-\infty}^{\infty} x(t)  x(t - \tau) dt
    \label{acf}
\end{equation}
where $\tau$ represents time shift. For a discrete-time signal $x[n]$, autocorrelation $R_{xx}[k]$, at a lag $k$ is given by defined as:
\begin{equation}
\label{ac}
R_{xx}[k]=\sum_{n=-\infty}^{+\infty}x[n]   x[n-k], \; k=0,\pm 1,\pm 2,\dots
\end{equation}
Note that for zero lag, there is no time shift; hence, the autocorrelation $R_{xx}$ will be at its maximum value,--- which equals the signal's total energy--- representing the maximum similarity as depicted in Fig. \ref{auto_corr} which shows the autocorrelation of the signal $x(t)$. 
\

Cross-correlation extends the concept of auto-correlation to two different signals, measuring the degree of similarity or dependence between them. It quantifies the degree of similarity between one signal and a time-shifted (lagged) version of another signal over a given time interval.  For two continuous-time signals $x(t)$ and $y(t)$, their cross-correlation is expressed as:
\begin{equation}
    R_{xy}(\tau) = \int_{-\infty}^{\infty} x(t) y(t - \tau) dt
\end{equation}
For discrete-time signals $x[n]$ and $y[n]$, the cross-correlation $R_{xy}$  at a given time lag $k$ is defined as:
\begin{equation}
\label{cc}
R_{xy}[k]=\sum_{n=-\infty}^{+\infty}x[n]y[n-k], \; k=0,\pm 1,\pm 2,\dots
\end{equation}
The value of $R_{xy}[k]$ indicates how much of one signal is present in the other signal at a given delay $k$. Hence, it describes the mutual relationship between the two signals as a function of the displacement of one relative to the other. Cross-correlation is often used to search for desired features or patterns within a signal.\

The correlation coefficient between two signals is a statistical measure that quantifies the degree to which the two signals are linearly related. It is a normalized form of the cross-correlation function and is widely used in signal processing and statistics to assess the strength of the linear relationship between two signals \cite{dn21}. For discrete-time signals $x[n]$ and $y[n]$, each of length $N$, the correlation coefficient $\rho_{xy}$ is defined as:
\begin{equation}
\label{cc}
\rho_{xy} = \frac{\sum_{n=0}^{N-1} (x[n] - \mu_x)   (y[n] -  \mu_y)}{\sqrt{\sum_{n=0}^{N-1} (x[n] -  \mu_x)^2   \sum_{n=0}^{N-1} (y[n] - \mu_y)^2}}
\end{equation}
where $\mu_x$ and $\mu_y$ are the means of the signals $x[n]$ and $y[n]$, respectively.\newline
The correlation coefficient lies between $-1$ and $+1$. Values close to $+1$ indicate a strong positive linear relationship, i.e. as one signal increases, so does the other. Values close to $-1$ indicate a strong negative linear relationship, where one signal increases as the other decreases. Values close to $0$ suggest a weak or no linear relationship between the signals.

\subsection{Signal Convolution}
Convolution is an integral operation on two signals that produces a third signal; it expresses the amount of overlap of one signal as it is shifted over another signal. The convolution of two continuous-time signals $x(t)$ and $h(t)$ is given by:
\begin{equation}
x(t) \ast h(t) = \int_{-\infty}^{\infty} x(\tau) h(t - \tau) d\tau
\end{equation}
where $\ast$ denotes the convolution operator, which is defined as the integral of the product of $x(t)$ with a time-reversed and shifted version of $h(t)$. The time reversal of $h(\tau)$, represented by $h(-\tau)$, involves reflecting the signal about the vertical axis that represents time origin ($\tau = 0$). Accordingly, the integral is evaluated for all shift values $t$, producing the convolution signal $x(t) \ast h(t)$. For discrete-time signals $x[n]$ and $h[n]$ the convolution is expressed as:
\begin{equation}
x[n] \ast h[n] = \sum_{k=-\infty}^{\infty} x[k] h[n - k]
\end{equation}
The convolution process can be visualized as one signal sliding over the other signal, computing the area of overlap as a function of the degree of overlap. In contrast to correlation, which measures the similarity between two signals as a function of the displacement of one relative to the other, the convolution represents the amount of overlap between two signals as one is inverted and shifted over the other signal. Therefore, while correlation measures the similarity between two signals, convolution can be viewed as a measure of the effect of one signal on the other.\

Convolution is particularly useful for describing how a linear time-invariant (LTI) system or filter responds over time to an input signal since the output of an LTI system or a filter for any arbitrary input can be determined by convolving the input signal with the system's impulse response. LTI Systems are systems that are characterized by linearity and time-invariance. Linearity, as explained earlier, implies that the response (output) to a weighted sum of inputs is the weighted sum of the responses to each input. Time-invariance means the system's response to a given input does not change over time. In signal theory, the impulse response of an LTI system or a filter, denoted as $h(t)$ for continuous-time representation or $h[n]$ in discrete-time case, represents the system's output when subjected to an impulse function. In continuous-time representation, the impulse function is denoted as $\delta(t)$ and is defined as a function that is zero at all times except at $t=0$, where it is infinitely high in such a way that its integral over time is $1$. In discrete time, the impulse function is denoted as $\delta[n]$ and is defined as a sequence that is zero at all samples except at $n=0$, where its value is $1$. The impulse response fully characterizes the system's or filter's behavior. Hence, given the impulse response of an LTI system or a filter for any arbitrary input, its output can be determined by convolving the input signal with the impulse response.

\subsection{Signal Entropy}
\begin{figure}[!htbp]
\centerline{\includegraphics[width=0.5\textwidth]{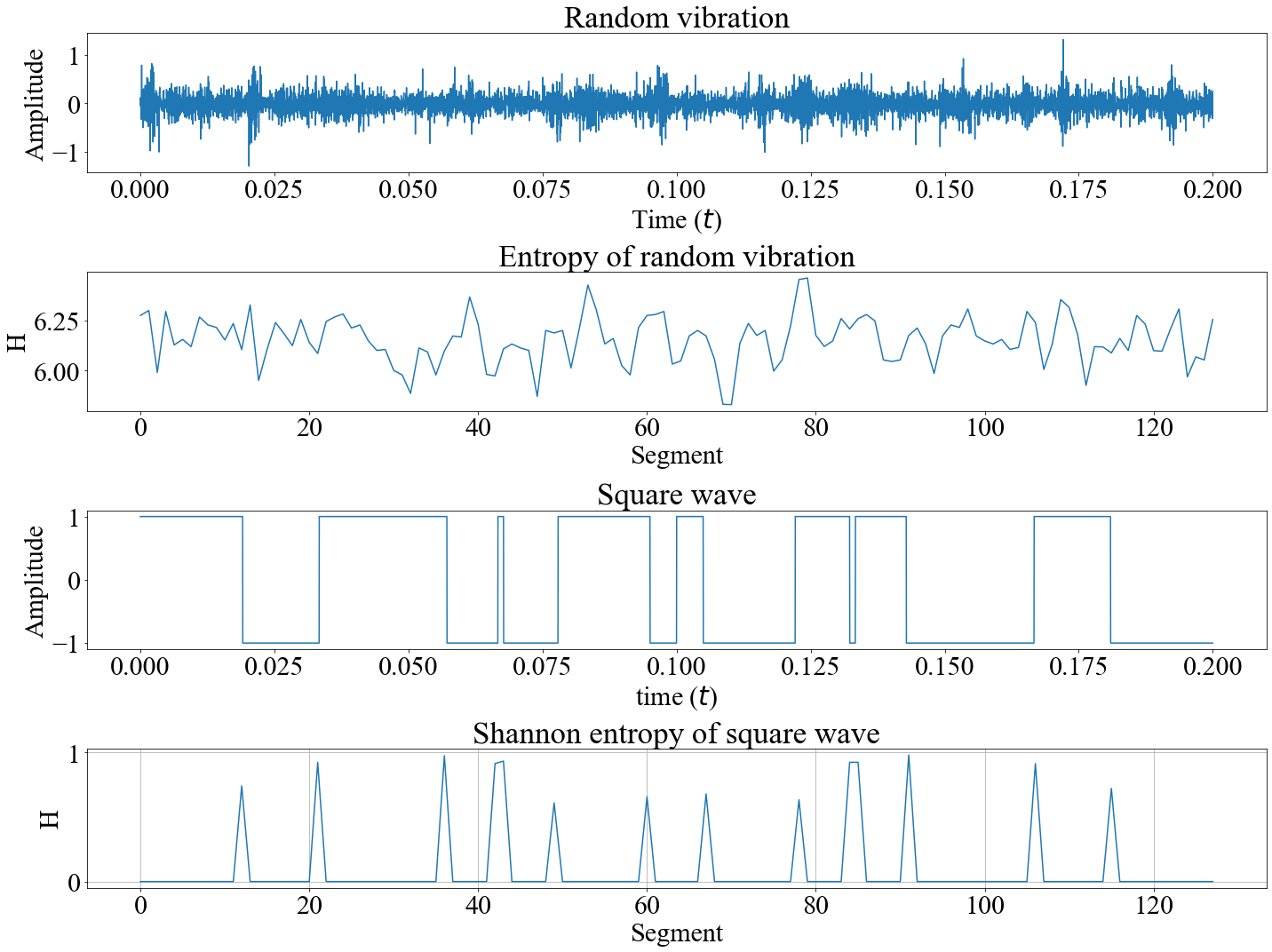}}
\caption{A comparison between Shannon entropy values of a random vibration $v(t)$, and a square wave.}
\label{entropy}
\end{figure}

Entropy is an important concept that originated in information theory as a measure of "information" in a random variable. In signal processing, entropy quantifies the degree of unpredictability or randomness in a signal, thereby reflecting its structure and offering a quantitative measure of various signal characteristics. The fundamental concept of entropy-based applications in signal processing is to treat the signal as a random stochastic process. In this context, the signal samples represent a collection of possible outcomes of the process; each outcome is associated with a probability of occurrence. Hence, for a discrete signal where its samples are represented as a set of possible outcomes $x_1, x_2,\dots, x_n$ with associated probabilities $p(x_1), p(x_2),\dots, p(x_n)$, the Shannon entropy is calculated as follows:
\begin{equation}
    H(X) = -\sum_{i=1}^{n} p(x_i) \log p(x_i)
    \label{SE}
\end{equation}
This formula calculates the average "information," "surprise," or "uncertainty" inherent in the signal's possible outcomes. In information theory, high entropy suggests that a signal contains a wide range of information. In signal processing, high entropy values are associated with high degrees of unpredictability, uncertainty, irregularity, and randomness, indicating a complex signal with more varying structures compared to predictable or repetitive patterns within the signal. On the other hand,  low entropy values often imply a high degree of uniformity and predictability within a signal. This is illustrated in Fig. \ref{entropy}, which compares the entropy values of two signals: the random vibration signal $v(t)$ and a square wave. Each signal contains $12800$ samples. The signals are segmented into $128$ segments so that each segment contains $100$ samples. Accordingly,  the Shannon entropy ($H$) is calculated for each segment to quantify uncertainty within each segment and to assess signal irregularity in terms of the variation in entropy values between signal segments.
The comparison shows that segments of the random vibration have higher entropy values than those of the square wave, indicating a higher degree of uncertainty. Moreover, the high variation of entropy values among different segments of the random vibration reflects the complex and irregular patterns in the signal. These concepts find useful applications in signal processing, including but not limited to:
\begin{itemize}
    \item Measurement of complexity: Entropy serves as a metric for signal complexity, with higher entropy values indicating more complex signals. Such signals are typically characterized by their low compressibility.
    \item Uncertainty Quantification: Entropy is a direct measure of the uncertainty in a signal. A signal with high entropy suggests less predictability and greater uncertainty about its current state or future values.
    \item Regularity Analysis: Lower entropy values indicate a higher degree of regularity in the signal, which could indicate a fault or abnormal behavior in applications such as condition monitoring and fault detection.
    \item Randomness Evaluation: The degree of randomness in a signal can be assessed using entropy. High entropy signals have a high degree of randomness, whereas low entropy signals involve minimal randomness.
\end{itemize}\

The Shannon entropy formula in (\ref{SE}), introduced by Claude Shannon in 1948 \cite{ac48}, is the fundamental form of entropy that is widely used in signal processing. Spectral Entropy is another type of entropy that applies Shannon entropy to the power spectrum of a signal, providing insight into its spectral characteristics. Other types of entropy include approximate entropy and sample entropy, commonly used in time-series analysis to measure the regularity and unpredictability of fluctuations in data series. More information on the different types of entropy can be found in \cite{el20}.

\section{Signal Transformation and Analysis}
Signal transformation is a fundamental concept in signal theory; it plays an important role in many signal processing applications, such as signal analysis, filtering, denoising, compression, and feature extraction. Transforming signals from their time-domain representations into other-domain representations allows the signal to be analyzed in a new paradigm, revealing new aspects of the signal that are not apparent in the time domain. The process of signal transformation involves mapping a signal from its original domain (usually time) to a new domain. This is achieved by computing the inner product of the signal with a set of basic functions (kernels) that serve as the building blocks for representing the signal in the new domain. These functions are chosen based on the properties desired in the transformed domain, for example, sinusoidals in Fourier transforms and wavelets in wavelet transforms. The inner product measures how much of the base function is present in the signal. A large value of the inner product indicates a high degree of similarity or a strong presence of the characteristics represented by the base function in the signal. This section reviews common frequency-domain and time-frequency domain transformations widely used in signal processing.

\subsection{Fourier Transform}
The Fourier transform (FT) is one of the most widely used transforms in signal processing. It provides a frequency-domain representation of the signal, revealing its spectral contents and enabling frequency-based analysis. The basic principle underlying the FT is that any waveform can be expressed as a combination of sinusoidal components with different amplitudes, frequencies, and phases. Accordingly, FT attempts to decompose signals into their constituent sinusoidal components. FT is mathematically expressed as:
\begin{equation}
 F(\omega) = \int_{-\infty}^{\infty} f(t) e^{-j\omega t} \, dt   
 \label{FT}
\end{equation}
where,\newline
$F(\omega)$ represents the FT the function f(t),\newline
$f(t)$ is the original time-domain signal,\newline
and $e^{-j\omega t}$ is the complex exponential function used in the transform.\newline
The discrete Fourier transform (DFT) is the discrete version of the continuous FT, which is used for analyzing discrete signals; it is defined as:
\begin{equation}
X[k] = \sum_{n=0}^{N-1} x[n] e^{-\frac{j2\pi}{N}kn} 
\end{equation}
where,\newline
$X[k]$ represents the DFT of the discrete time-domain signal, $x[n]$,\newline
$N$ is the total number of samples in $x[n]$,\newline
$k = 0,1,...,N-1$, is the index in the frequency domain,\newline
and $n = 0, 1,...,N-1$, is the index in the time domain.\newline
Both the FT and the DFT are reversible operations where the frequency domain representation of a signal can be transformed back into its original time-domain form by applying the inverse transform operation. This property is fundamental in signal processing and has various applications, such as signal compression, filtering, and noise reduction. \

The fast Fourier transform (FFT) and its inverse, IFFT, are computational algorithms that efficiently implement the DFT and its inverse, IDFT. The key advantage of FFT and IFFT is their high computational efficiency. NumPy and SciPy Python libraries offer reliable functions for FFT and IFFT computation. When computing the FFT of a signal, two critical parameters need to be considered: the sampling rate, $f_s$, and the FFT length, $n$. The values of these parameters determine the frequency range and frequency resolution of the resulting spectrum. It is essential to choose an appropriate FFT length, usually a power of two (\textit{e.g.}, 256, 512, 1024), as the FFT algorithm performs efficiently when the number of points is a power of two. The FFT length also determines the number of frequency bins in the spectrum, where each bin corresponds to a specific range of frequencies within the total frequency spectrum analyzed by the FFT. The number of unique frequency bins in the FFT output is:
\begin{equation}
 \frac{n}{2}+1
 \label{fft_length}
\end{equation}
The bin width defines the frequency resolution, $\Delta f$, of the FFT representing the smallest distinguishable frequency in the spectrum. It is defined as:
\begin{equation}
    \Delta f = \frac{f_s}{N}
    \label{freq_res}
\end{equation}
Frequency resolution is critical in frequency analysis because it affects the accuracy with which closely spaced frequency components can be resolved in the resulting spectrum. If the FFT length is set equal to the number of samples $N$ in the input signal, the FFT operates directly on the signal as it is. If the FFT length is set greater than $N$, the signal is typically zero-padded to the desired length. Zero padding does not alter the actual frequency content of the signal but increases the number of frequency bins in the FFT result, leading to a finer frequency resolution in the spectrum. On the other hand, setting the FFT length to less than $N$ truncates the signal, resulting in loss of information and reduced frequency resolution. The maximum frequency that can be accurately represented in the spectrum is the Nyquist frequency of the signal ($\frac{f_s}{2}$, which equals half of the sampling rate. Frequencies higher than the Nyquist frequency will not be correctly resolved in the spectrum. The output of the FFT function is an array that contains complex Fourier coefficients representing the amplitude and phase information of each frequency bin. The function \textit{fftfreq} in Python can be used to obtain the frequency bins; it takes two inputs, the FFT length $n$ and the sampling period, $T_s=\frac{1}{f_s}$, and returns an array containing frequency bins. Accordingly, the frequency spectrum is constructed by mapping the coefficients to the $x$-axis and the corresponding frequency bins to the $y$-axis. Since Fourier coefficients are complex numbers, each bin is expressed either by the magnitude or power of the corresponding Fourier coefficients, resulting in amplitude spectrum or power spectrum, respectively. This gives a two-sided frequency spectrum symmetric around the $y$ axis, showing positive and "negative" frequencies.  The negative frequency spectrum is inherent to the Fourier analysis of signals; it is a mirror of the positive "real" spectrum around the $y$-axis. Thus, it represents redundant frequency information. Reconstructing a positive spectrum involves discarding the second half of coefficients and frequency bins arrays. Thus, to accurately represent the signal's energy  in the frequency domain, it's necessary to multiply the magnitudes of Fourier coefficients by $2$ to restore the energy distribution in the positive frequency since the signal's energy was equally divided between the positive and negative frequencies. It is worth noting that the number of Fourier coefficients is directly proportional to the length of the input signal, as stated in (\ref{fft_length}). Consequently, the magnitude of the FFT coefficients increases with the length of the signal. Therefore, when longer signal segments are used, the frequency spectrum displays larger values as more components contribute to the sum. Therefore, to reflect the true amplitudes of the signal components, the magnitudes of the Fourier coefficients are usually normalised by the signal length.\

\begin{figure}[!htbp]
\centerline{\includegraphics[width=0.5\textwidth]{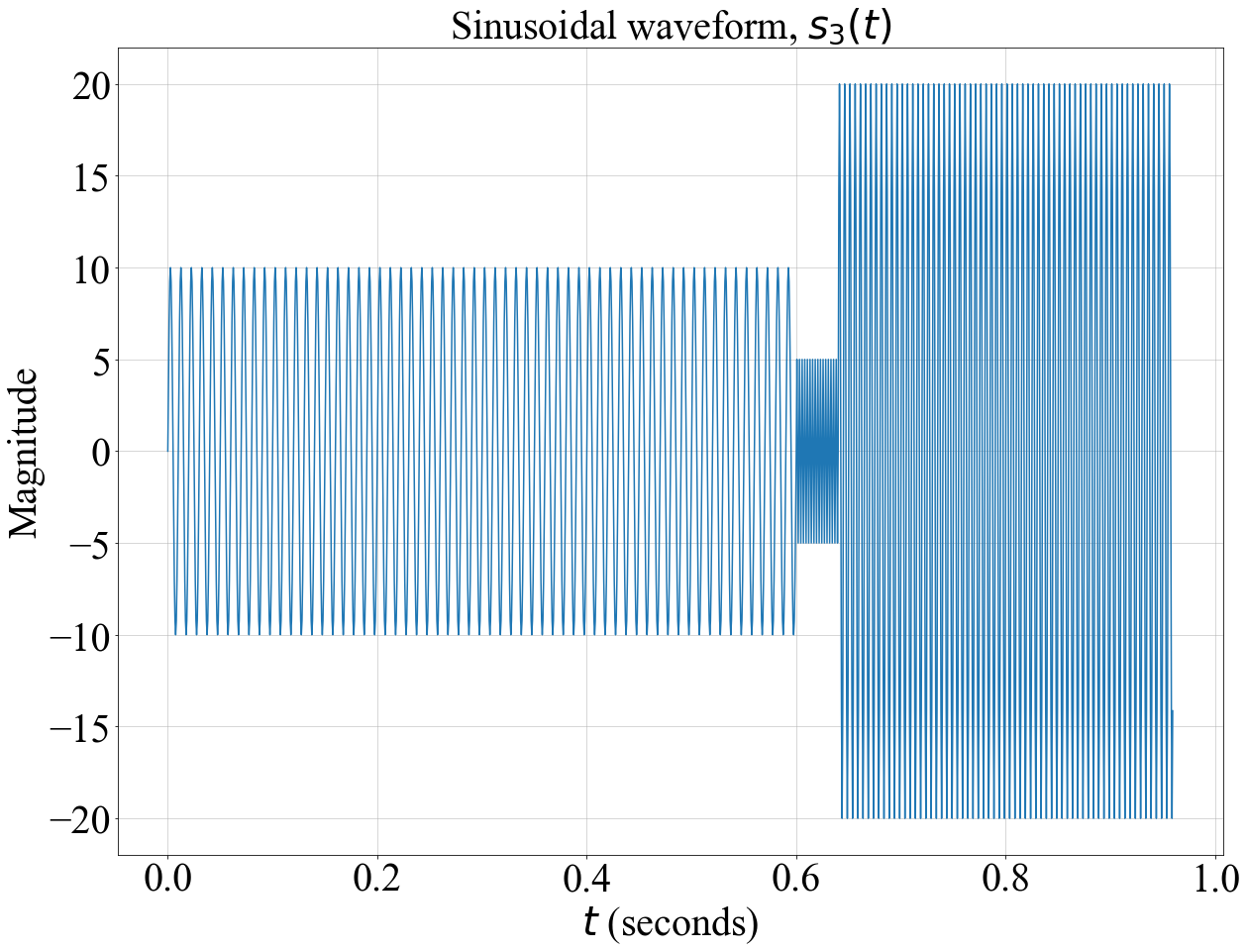}}
\caption{ Composite sinusoidal signal $s_3(t)$.}
\label{sin_3}
\end{figure}

\begin{figure}[!htbp]
\centerline{\includegraphics[width=0.5\textwidth]{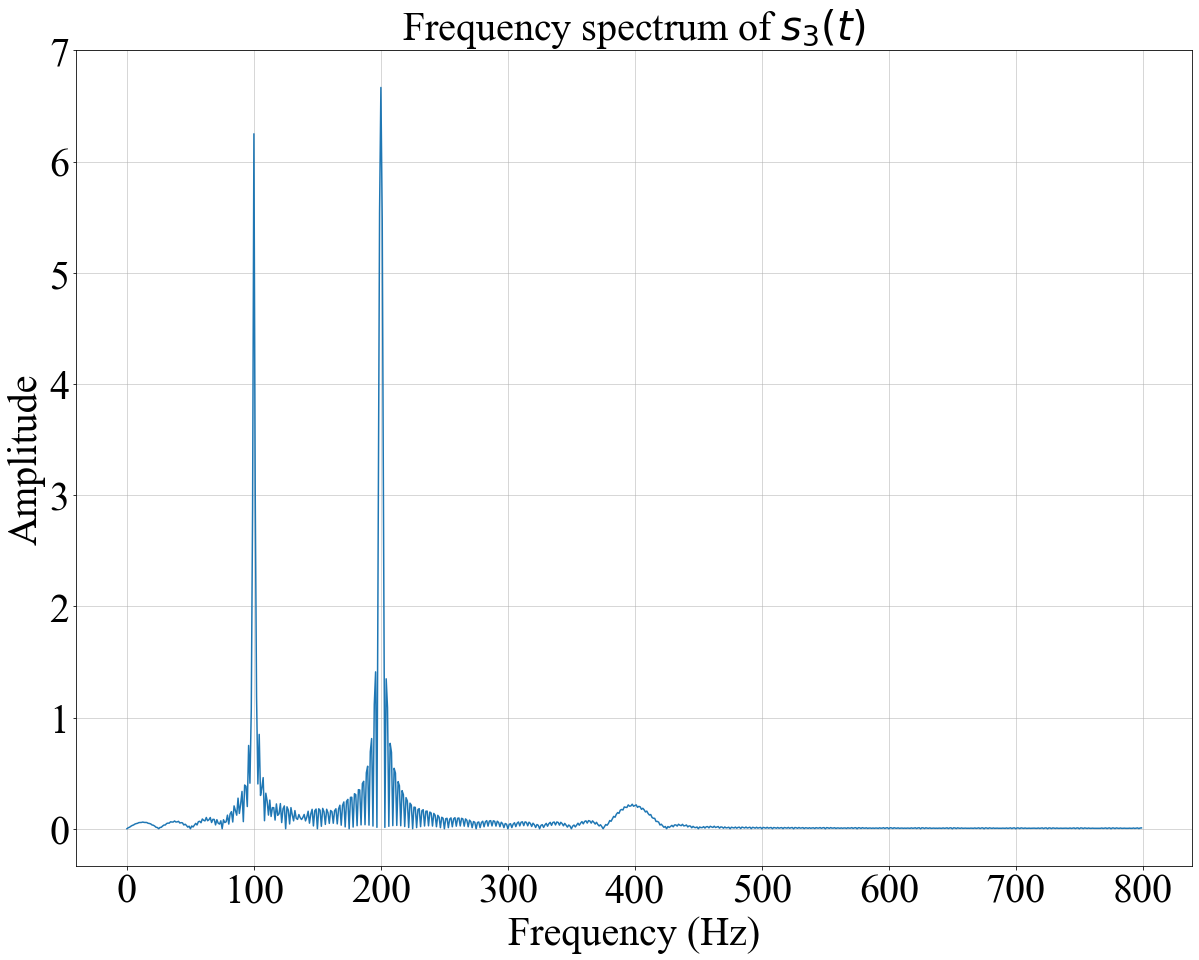}}
\caption{ Frequency spectrum of $s_3(t)$.}
\label{sin_3_fft}
\end{figure}

\begin{figure}[!htbp]
\centerline{\includegraphics[width=0.5\textwidth]{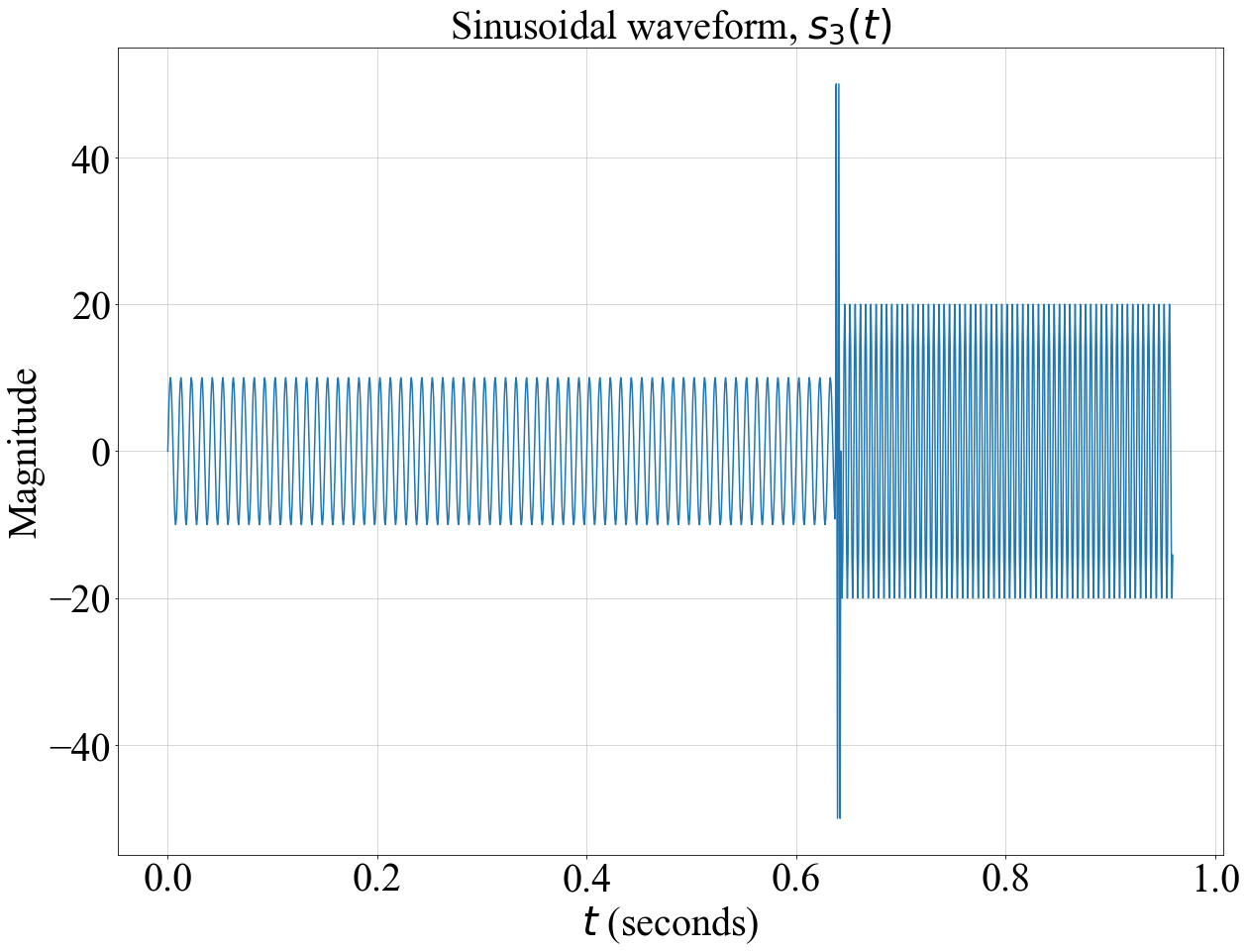}}
\caption{Composite sinusoidal signal $s_{m3}(t)$.}
\label{sin_3_m}
\end{figure}

\begin{figure}[!htbp]
\centerline{\includegraphics[width=0.5\textwidth]{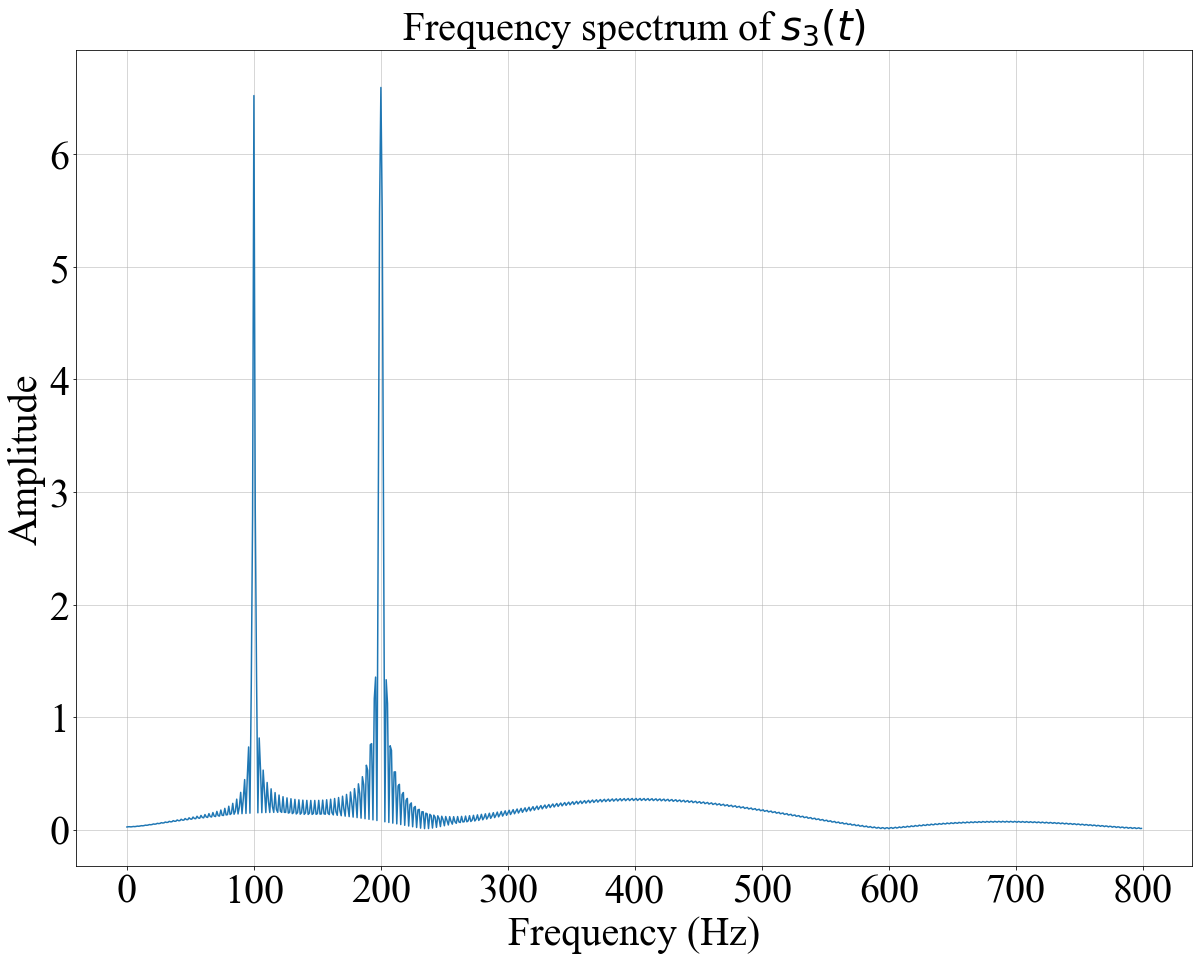}}
\caption{Frequency spectrum of $s_{m3}(t)$.}
\label{sin_3_fft_narrow}
\end{figure}

To demonstrate the concept of FT in resolving various frequency components, a sinusoidal signal, denoted as $s_3(t)$ composed of three components, is created as shown in Fig. \ref{sin_3}. 
The used sampling rate is $1600$ Hz, and the length of the signal is $1536$ data points; accordingly, this corresponds to $0.96$ seconds of the signal's time duration. The signal consists of three sinusoidal components with different durations, amplitudes and frequencies. The first component consists of $960$ data points and oscillates at $100$ Hz with an amplitude of $10$. The second component contains $64$ data points with a frequency of $400$ Hz and an amplitude of $5$. The third component consists of $512$ data points oscillating at a dominant frequency of $200$ Hz and an amplitude of $20$. Consequently, the time duration which corresponds to each of the components are $0.6$ seconds, $0.04$ seconds, and  $0.32$ seconds, respectively. Compared to the first and third components, the second component represents weak, rapid oscillations of very short duration, simulating a transient component of particular importance in the composite sinusoidal. The frequency spectrum of the signal is displayed in Fig. \ref{sin_3_fft}, which clearly shows that the three components in the composite signal at $100$ HZ, $200$ Hz, and $400$ Hz are accurately resolved and represented in the spectrum. The peaks correspond to the amplitudes of the components and therefore convey how much energy is being dissipated within each component. In practical terms, the presence of a high peak in the spectrum indicates the presence of a dominant (energetic) oscillatory component in the time domain signal. The frequency value associated with this peak reflects the oscillation rate of the component.
While the FT can identify the three frequency components within the signal, its analysis spans the signal's entire duration, which is evident from the infinite integral in the transform equation (\ref{FT}). Thus, although it accurately distinguishes between different frequency components, it cannot provide temporal information about the timing of the occurrence of these frequencies within the signal. This lack of time resolution in the FT limits its applications to pure-frequency analysis only. Additionally, the FT is not ideal for analyzing signals with highly time-localized components, like short bursts with high energy concentration. Such components produce a broad range of frequencies in the frequency spectrum of the signal due to the inherent uncertainty principle associated with Fourier analysis \cite{ar14}, which states that a signal cannot simultaneously have an arbitrarily small duration in time and an arbitrarily narrow bandwidth in frequency. This means that if a signal is highly localized in time, it must spread widely in the frequency domain and vice versa. This concept is demonstrated in Fig. \ref{sin_3_m} and Fig. \ref{sin_3_fft_narrow}, which show a composite sinusoidal signal, denoted as $s_{m3}(t)$, and its frequency spectrum, respectively. The signal $s_{m3}(t)$ is a modified version of the sinusoidal signal $s_3(t)$ where the duration of the second component is reduced to $0.05$ (The duration of the first component is increased accordingly), and its amplitude is increased to $30$ to simulate localized oscillations of high energy concentration. 
As depicted in Fig. \ref{sin_3_fft_narrow}, the frequency of the localized component in $s_{m3}(t)$ is broadly spread across the frequency spectrum, which poses a challenge in distinguishing subtle frequency components in practical applications.

\subsection{Short-Time Fourier Transform}
\begin{figure}[!htbp]
\centerline{\includegraphics[width=0.5\textwidth]{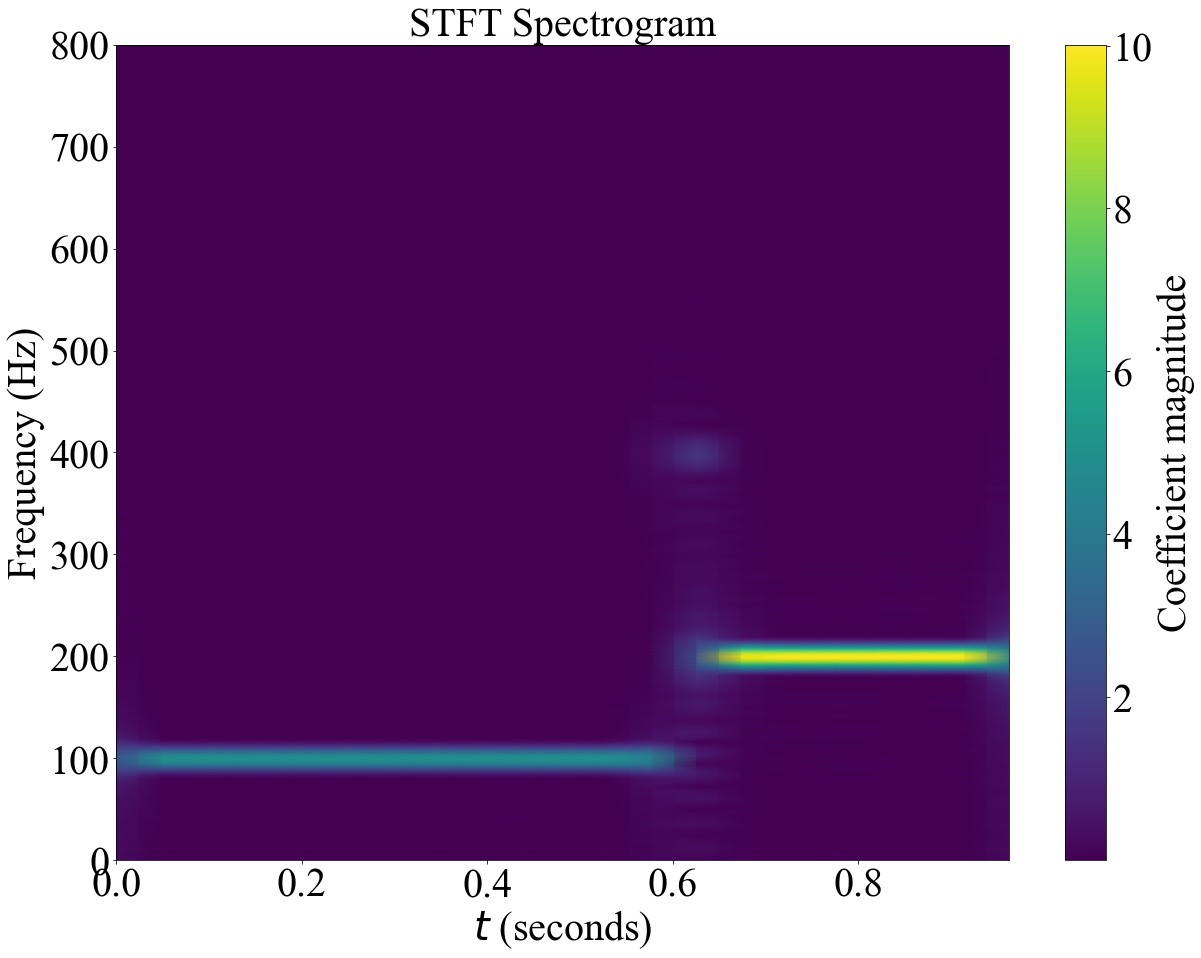}}
\caption{STFT spectrogram of the composite sinusoidal signal $s_3(t)$.}
\label{stft_s3}
\end{figure}

\begin{figure}[!htbp]
\centerline{\includegraphics[width=0.5\textwidth]{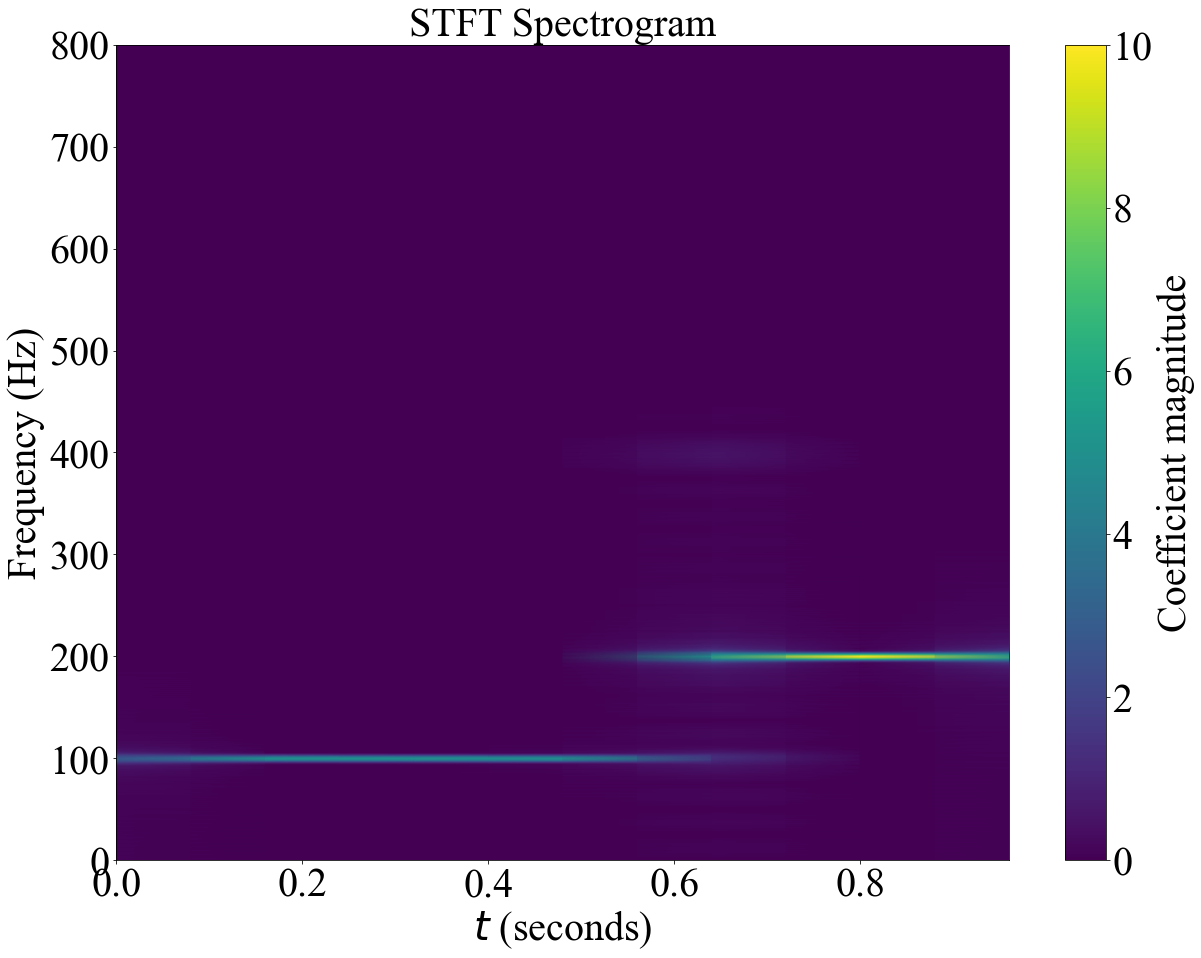}}
\caption{STFT spectrogram of the composite sinusoidal signal $s_3(t)$.}
\label{stft_s3_3_seg}
\end{figure}

The short-time Fourier transform (STFT) bridges the gap between time-domain and frequency-domain analysis, making it particularly useful for analyzing nonstationary signals whose spectral characteristics evolve over time. Unlike FT, which spans the signal's entire duration for frequency analysis, STFT analyzes the signal over finite, overlapping time windows, thereby preserving temporal information. The computation of the STFT involves dividing the time-domain signal into segments of equal time length, which is achieved by multiplying the signal with a sliding window function. The FFT is then computed for each segment, providing frequency bins and corresponding Fourier coefficients for each segment. For a continuous signal, $x(t)$, the STFT is mathematically defined as:
\begin{equation}
STFT\{x(t)\}(\tau, \omega) = \int x(t) w(t-\tau) e^{-j\omega t} \, dt
\end{equation}
where $w(t-\tau)$ is the window function centered at time $\tau$, and $\omega$ is the angular frequency. For discrete signal $x[n]$, the discrete STFT is defined as:
\begin{equation}
  DSTFT[n, k] = \sum_{m=0}^{N-1} x[n + m]   w[m]   e^{-j2\pi\frac{km}{N}}
\end{equation}
where,\newline
$N$ is the size of the window, defining the number of signal samples included in each signal's segment for analysis,\newline
$n$ is the time index around which the window is centered,\newline
$k$ is the frequency bin index,\newline
$w[n + m]$ represents the signal samples within the window,\newline
$w[m]$ is the window function of size $N$,\newline
and $e^{-j2\pi \frac{km}{N}}$ is is the DFT kernel.\newline
The SciPy Python library provides an efficient implementation of SFTF and its inverse transformation. The output of the STFT function contains three sets: 
\begin{itemize}
    \item The temporal set contains the duration of each segment in seconds. Its size equals the number of segments.
    \item The frequency set that contains groups of frequency bins corresponding to each time segment. The size of the set is equal to $\frac{n}{2}+1$.
    \item Coefficients set: A matrix containing Fourier coefficients in each frequency bin for every segment. Thus, its size is $d_1\times d_2$, where $d1$ and $d2$ are sizes of frequency and temporal sets, respectively. STFT features are usually obtained by examining the coefficients matrix. However, since Fourier coefficients are complex numbers, extracting features in the STFT commonly involves analyzing the magnitudes of these coefficients.
\end{itemize}
Heatmaps serve as an effective tool for visualizing the STFT output, providing a clear representation of the variations in the frequency content of the signal over time. These heatmaps, commonly referred to as spectrograms, are generated by mapping the temporal set to the $x$-axis and frequency set to the $y$-axis.  The "intensity" of each time-frequency pair in the $x-y$ plane is represented by the magnitude of the corresponding Fourier coefficients. Fig. \ref{stft_s3} shows the STFT of the composite signal $s_3(t)$ introduced earlier. The signal is segmented into ten segments using a Hamming window with an overlap of $50\%$ of the segment length. The Fourier coefficients are computed for each segment using an FFT length of $2048$. Compared to the signal's frequency spectrum depicted in \ref{sin_3_fft}, the STFT spectrogram shows the spectral contents and reveals their temporal characteristics simultaneously, providing time-frequency analysis of various frequency components in the signal. Specifically, The duration of each frequency component in the signal is clearly shown over the timeline ($X$-axis).  Furthermore, the temporal edges, represented by the start and end times of each component, are well-defined, thereby reflecting high temporal resolution that allows  the identification of the transient component at $400$ Hz.
In real-world scenarios, the efficiency of the STFT in time-frequency analysis is significantly influenced by the choice of its parameters, particularly the segment length and the percentage of overlap between adjacent segments. Segment overlap provides continuity and reduces information loss between adjacent segments. A high degree of overlap introduces more redundancy into the segments and improves time resolution, which is particularly useful when a large segment length is used. The segment length determines the resolution of the time-frequency analysis. A smaller segment length enables better time resolution, which helps identify rapid transient patterns in the signal, as demonstrated in Figure \ref{stft_s3}. On the other hand, a larger segment length enhances frequency resolution by allowing more frequency cycles to fit within the segment. However, this comes at the cost of reduced temporal resolution, as transient changes in the signal over short time durations may be overlooked. Therefore, there is a trade-off between time and frequency resolutions in the STFT analysis. This concept is demonstrated in Fig. \ref {stft_s3_3_seg}, which shows the STFT spectrogram of $s_3(t)$ re-computed by dividing the signal into three segments only instead of ten segments. The frequency resolution is represented by the "thickness" of the timeline. A thinner timeline--- reflecting a finer frequency resolution--- allows precise identification of the corresponding frequency bin on the $y$-axis compared to a thick timeline that spans several frequency bins.  A comparison of the two spectrograms shows that the use of a longer segment improves the frequency resolution but reduces the temporal resolution, resulting in overlapping temporal edges that obscure the identification of the transient component.
The fixed segment length is a major limitation in STFT that results in a uniform resolution analysis of the signal and, hence, leads to an inherent compromise between time and frequency resolutions when applying STFT for time-frequency analysis.

\subsection{Power Sepctral Density}
Power spectral density (PSD) describes how the power of a signal is distributed across different frequencies, providing a measure of the power per unit frequency. This is achieved through the normalization of the power spectrum by frequency resolution (frequency bin width), converting the power spectrum into a power spectral density that shows the power distribution per unit frequency. Mathematically, the PSD $S_{xx}(f)$ of a continuous-time signal $x(t)$ is defined as the FT of the autocorrelation function $R_{xx}(\tau)$; it is expressed as:
\begin{equation}
    S_{xx}(f) = \int_{-\infty}^{\infty} R_{xx}(\tau) e^{-j2\pi f\tau} \, d\tau
\end{equation}
where,
\begin{equation}
   R_{xx}(\tau) = E[x(t)x(t+\tau)]
\end{equation} 
and $E[ ]$ denotes the expected value.
For discrete signals, the discrete PSD (DPSD) is utilized, involving the DFT of the autocorrelation sequence. The DPSD 
$S_{xx}[k]$ of a discrete-time signal $x[n]$ is given by:
\begin{equation}
S_{xx}[k] = \sum_{n=-\infty}^{\infty} r_{xx}[n] e^{-j\frac{2\pi}{N}kn}
\end{equation} 
where $r_{xx}[n]$ is the autocorrelation sequence of $x[n]$.\

\begin{figure}[!htbp]
\centerline{\includegraphics[width=0.5\textwidth]{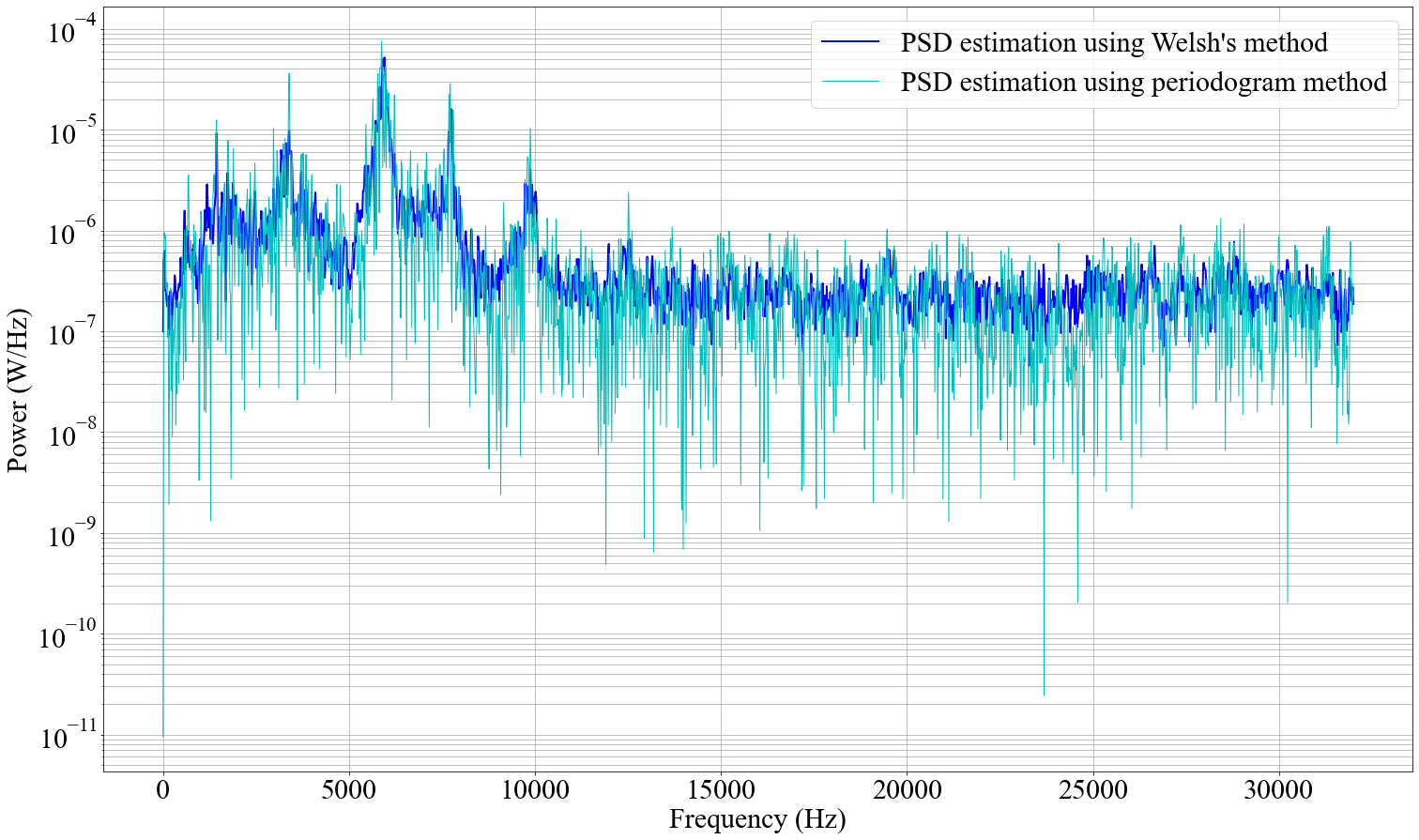}}
\caption{Periodogram and Welsh's  PSD estimations of vibration signal $v(t)$.}
\label{psd_w_vs_p}
\end{figure}

The SciPy Python library provides two methods for PSD estimation: the periodogram method (\textit{scipy.signal.periodogram}) and the Welsh method \textit{scipy.signal.welch} \cite{tw67}. The periodogram estimation involves obtaining the spectral power of the signal by applying FFT and squaring the magnitude of the resultant FFT coefficients. The PSD estimate is obtained by normalizing the spectral power by the frequency resolution of the FFT. One of the limitations of the periodogram method is its high variance, especially when the number of samples in the signal is limited. The Welsh method overcomes this limitation by segmenting the signal into overlapping segments using a sliding window, computing the periodogram for each segment, and averaging the periodograms to obtain the PSD estimate. This effectively reduces noise and provides a smoother estimate than the periodogram method. Fig. \ref{psd_w_vs_p} compares the two methods in estimating PSD of the random vibration signal $v(t)$ introduced earlier. In the Welsh method, the signal is divided into four segments with $50\%$ of overlap. It is evident that averaging the estimate over several segments greatly reduces the variance of the estimate, resulting in a smoother PSD. The smoothness of the PSD is of significant importance in power spectral analysis and feature extraction since high estimation variance can hinder major characteristics in the PSD, such as peaks. A smooth PSD allows precise identification of such characteristics, improving the sensitivity of the extracted features.
\

Similar to the STFT, the choice of parameters in the Welch method, such as window type, segment length and overlap, has a significant effect on the PSD estimate. While STFT and the PSD possess similar aspects, they serve different purposes and are based on distinct theoretical foundations. Understanding their differences is crucial, especially in machine learning and feature extraction applications. Table \ref{sftf_vs_psd} compares STFT and PSD and summarises their main characteristics. 
\begin{table*}[!htbp]
    \centering
\caption{Comparison between STFT and PSD}
\begin{tabularx}{1\textwidth} 
{| m{10em} | X | X | }
 \hline
 \textbf{Properties} & \textbf{STFT} & \textbf{PSD}\\
 \hline
 \textit{Purpose and Concept}&  STFT is used for time-frequency analysis of signals, especially nonstationary signals where frequency components vary over time&  PSD is used to describe the power distribution of a signal across frequency. It is particularly useful in understanding the energy content of different frequency components of a signal \\
\hline
\textit{Representation}&  The result of STFT is a two-dimensional representation of the signal, showing both time and frequency dimensions. This is crucial in analyzing signals whose spectral properties evolve over time&  PSD provides a one-dimensional frequency domain representation, indicating how power is distributed across frequencies. Unlike STFT, it does not offer time-domain information\\
\hline
\textit{Characteristics}&   STFT involves a trade-off between time and frequency resolution controlled by the window size. A larger window offers better frequency resolution but poorer time resolution, and vice versa.& PSD assumes signal stationary, making it less suitable for analyzing signals with time-varying characteristics  \\
\hline
\textit{Application} &  STFT is used for analyzing signals with time-varying spectral properties. & PSD is more suited for stationary signals or for assessing overall frequency content \\
\hline
\end{tabularx}
 \label{sftf_vs_psd}
\end{table*}

\subsection{Wavelet Transform}
\begin{figure}[!htbp]
\centerline{\includegraphics[width=0.5\textwidth]{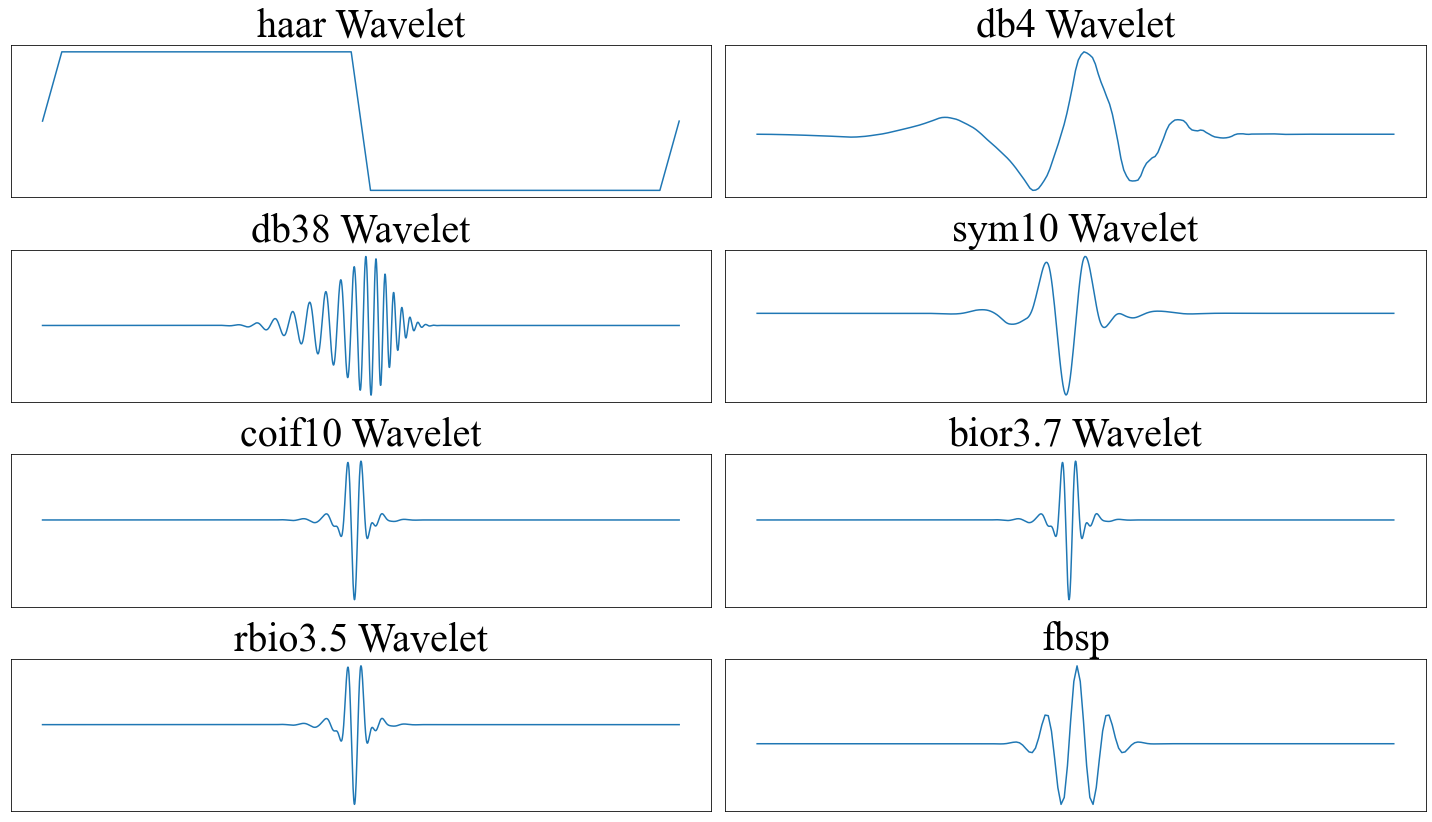}}
\caption{Some of the commonly used wavelet functions.}
\label{wavelet_func}
\end{figure}

The accuracy of time-frequency analysis depends heavily on the localization of the base function in time and frequency domains. In Fourier analysis, the basis functions are sinusoidals (complex exponentials) that are perfectly localized in the frequency domain since each corresponds to a single frequency component with no frequency spread. In the time domain, these sinusoidal functions extend infinitely, lacking time localization due to their infinite duration. In the context of wavelets, signal analysis is based on a family of wavelets that are localized in both time and frequency domains.  These wavelets are generated from a single mother wavelet through scaling and translation, facilitating signal analysis at different frequencies and time intervals. A wavelet is a mathematical function that satisfies certain conditions, such as finite energy and a zero mean. The term "wavelet" itself means a small wave, which captures the essence of these functions and how they behave. The main characteristics of a wavelet function include the following:
\begin{itemize}
    \item Zero Mean: This ensures the wavelet oscillates around the zero level; this is crucial for detecting changes in the signal since it allows it to capture both positive and negative variations in the signal effectively.
    \item Finite Energy: This implies that the total energy of a wavelet is finite, ensuring integrability and existence of the inner product with the signal. The energy of a given wavelet function is calculated by integrating the squared amplitude of the wavelet over its entire duration. Mathematically, the total energy of a  wavelet function, $\psi(t)$, is expressed as:
    \begin{equation}
         \int_{-\infty}^{\infty} |\psi(t)|^2 \, dt < \infty
    \end{equation}
A wavelet with finite energy is inherently localized in time and has compact support, meaning it is non-zero only over a limited range, and its oscillations are confined to a small region. This time localization allows wavelets to effectively represent signals that have nonstationary or transient components. A wavelet's integrability implies that it won’t produce infinite results when integrated over its domain. This is crucial for signal analysis since it is essentially carried out by computing the inner product of the signal with scaled and transformed versions of the wavelet.
\end{itemize} \
Wavelet analysis is conducted through two operations, scaling and translation:
\begin{itemize}
    \item Scaling (Dilation): This involves stretching or compressing the wavelet. Scaling changes the frequency content of the wavelet, allowing the signal to be analysed in different frequency bands. Specifically, the oscillation frequency of a wavelet is directly affected by the scale. At lower scales, the wavelet compresses and oscillates more rapidly, which allows it to detect high-frequency components or rapid changes in the signal. On the other hand, at higher scales, the wavelet is stretched, resulting in fewer oscillations for the same amount of time, making it more sensitive to low-frequency components or slow-changing features in the signal.
    \item Translation (Shifting): This refers to moving the wavelet along the time axis, enabling the wavelet to analyze different signal parts at a given scale.
\end{itemize} \
These aspects allow high time-frequency localization and provide a multi-resolution analysis through varying scales and translations. The choice of the mother wavelet and scale values is derived by signal characteristics and application-specific requirements. Attributes such as orthogonality, compact support, vanishing moments, and the ability to capture signal discontinuities and singularities play a crucial role in this choice. The wavelet family includes various mother wavelets; each with unique properties that cater to various signal types and application requirements. Fig. \ref{wavelet_func} displays the time-domain waveforms of some of the commonly used wavelet functions.

\subsubsection{The Continuous Wavelet Transform (CWT)} 
\begin{figure}[!htbp]
\centerline{\includegraphics[width=0.5\textwidth]{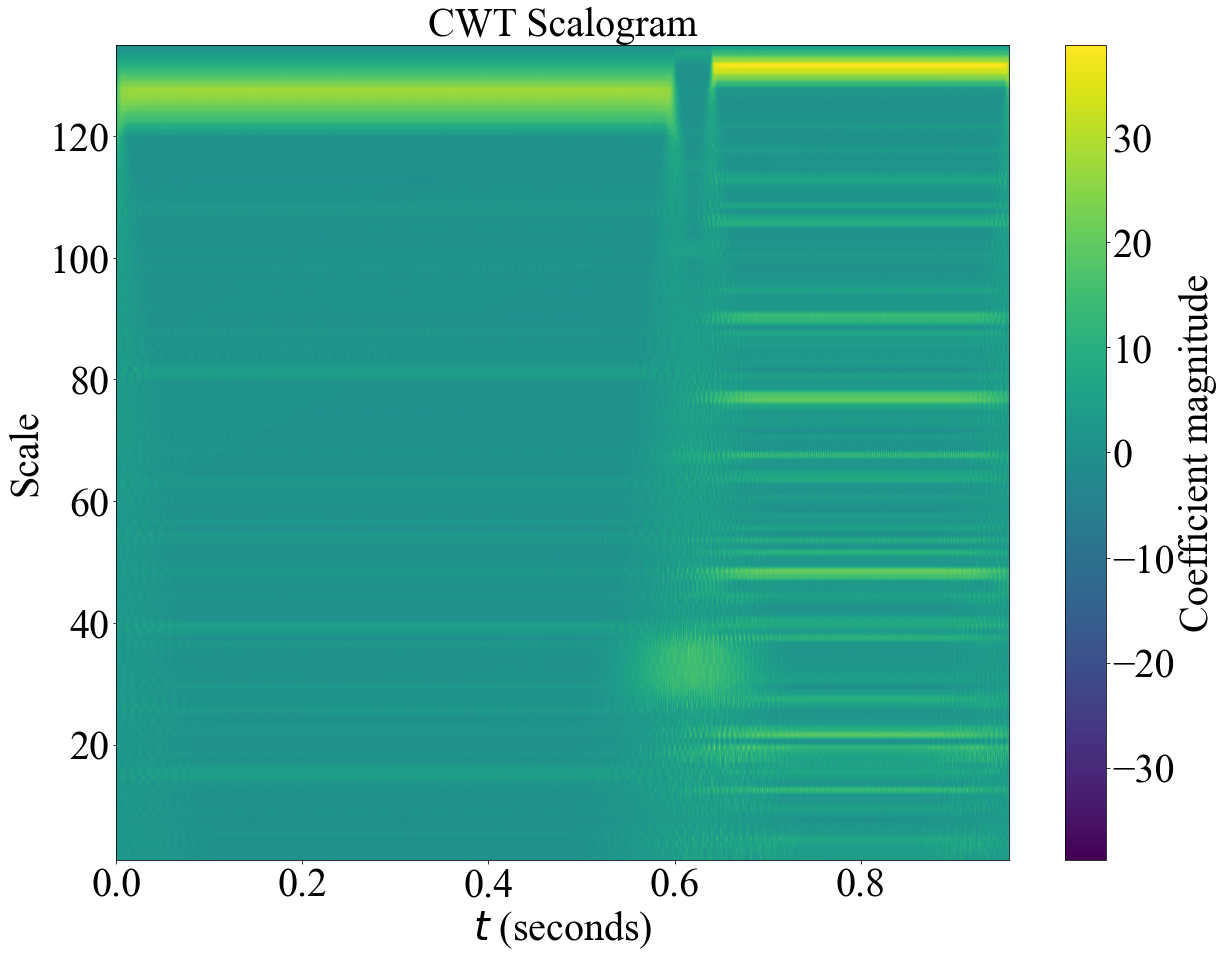}}
\caption{ CWT (scalogram) of the composite signal $s_3(t)$.}
\label{sin_3_cwt}
\end{figure}

The CWT of a signal $x(t)$ can be expressed mathematically as follows:
\begin{equation}
    CWT_x(a, b) = \frac{1}{\sqrt{|a|}} \int_{-\infty}^{\infty} x(t)   \psi^*\left(\frac{t - b}{a}\right) dt
\end{equation}
where, \newline
$a$ is the scale parameter, $b$ is the translation parameter, \newline
$CWT_x(a, b)$ represents the wavelet coefficient at scale $a$ and translation $b$,  \newline
$\psi(t)$ is is the mother wavelet, and $\psi^*(t)$ is its complex conjugate. \newline
The output of CWT is a matrix of wavelet coefficients, where each row contains the coefficients of the corresponding scale. These coefficients represent the correlation between the signal and a scaled and translated wavelet, indicating how closely the signal matches the wavelet at those specific scales and translations. This emphasizes the importance of selecting a proper wavelet function for signal analysis. The CWT output is commonly visualized through 2D heatmaps, known as scalograms, where the $x$-axis and $y$-axis represent time and scale, respectively. The intensity of the heatmap is represented by the magnitude of the wavelet coefficients, which reflect the signal wavelet energy at different scales and translations. The scale parameter inversely relates to the signal's frequency content. High scales correspond to low frequencies, capturing the slower-varying components of the signal. On the other hand, low scales, corresponding to high frequencies, capture rapidly changing components within the signals. Hence, scalograms show how the spectral content varies over time, enabling the identification of transient or nonstationary components within the signal. Fig.\ref{sin_3_cwt} displays the scalogram of the composite signal $s_3(t)$ using a scale range from 1 to 135. The \textit{fbsp}, displayed in the bottom right corner of Fig. \ref{wavelet_func}, is employed in the analysis; \textit{fbsp} is a spline wavelet constructed using a spline function. The selection of this wavelet is based on its sinusoidal-like waveform, resembling the waveform of the composite signal $s_3(t)$. The scalogram demonstrates a multi-resolution signal analysis at various scale values. In particular, scales greater than $120$ effectively capture the slowly varying components (components $1$ and $3$) of the signal. On the other hand, the raid change in the signal, represented by component $2$, can only be detected at small-scale values (scales between $30$ and $40$).

\subsubsection{Discrete Wavelet Transform (DWT)}
\begin{figure}[!htbp]
\centerline{\includegraphics[width=0.5\textwidth]{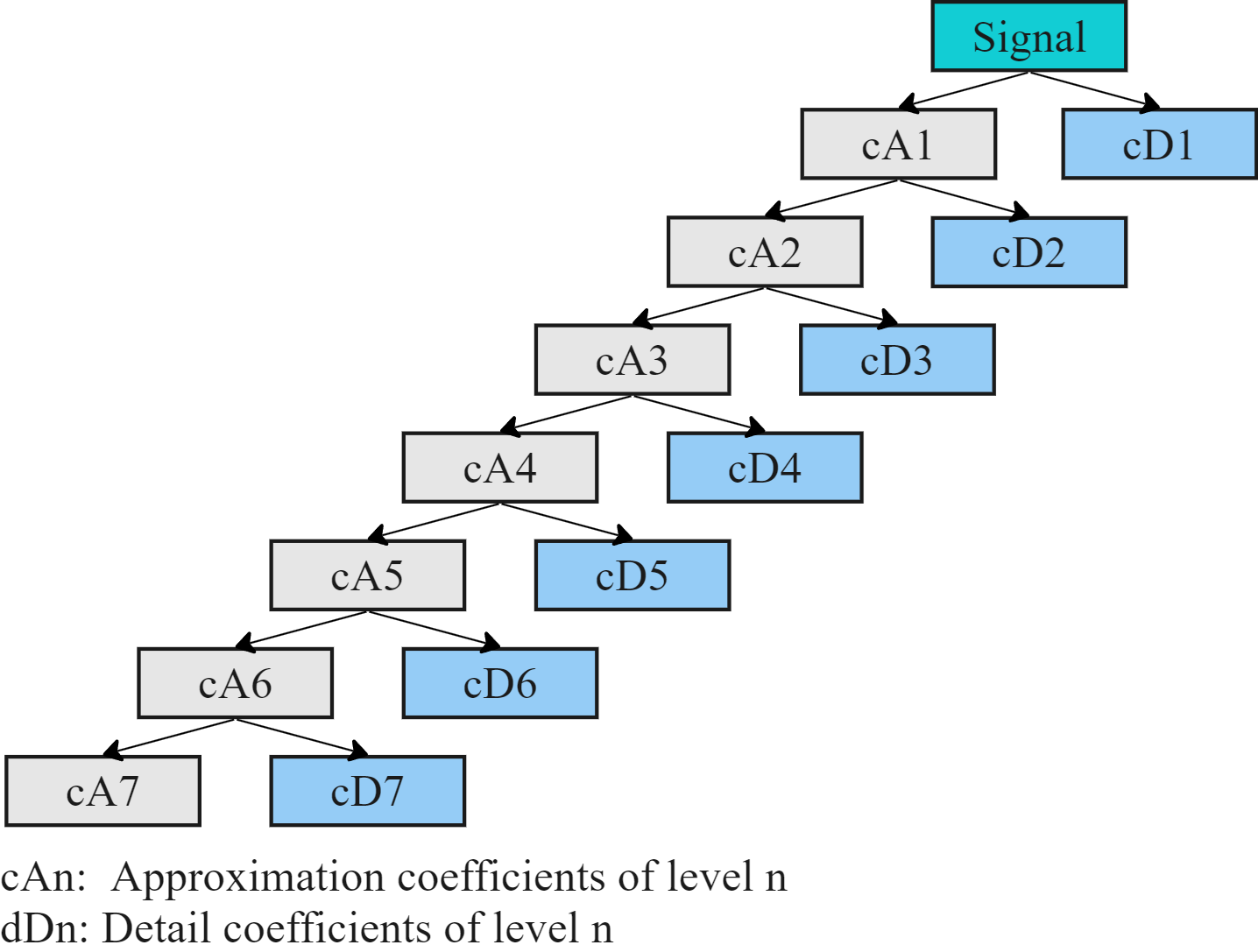}}
\caption{$7$-level DWT decomposition tree.}
\label{dwt_tree}
\end{figure}

\begin{figure}[!htbp]
\centerline{\includegraphics[width=0.5\textwidth]{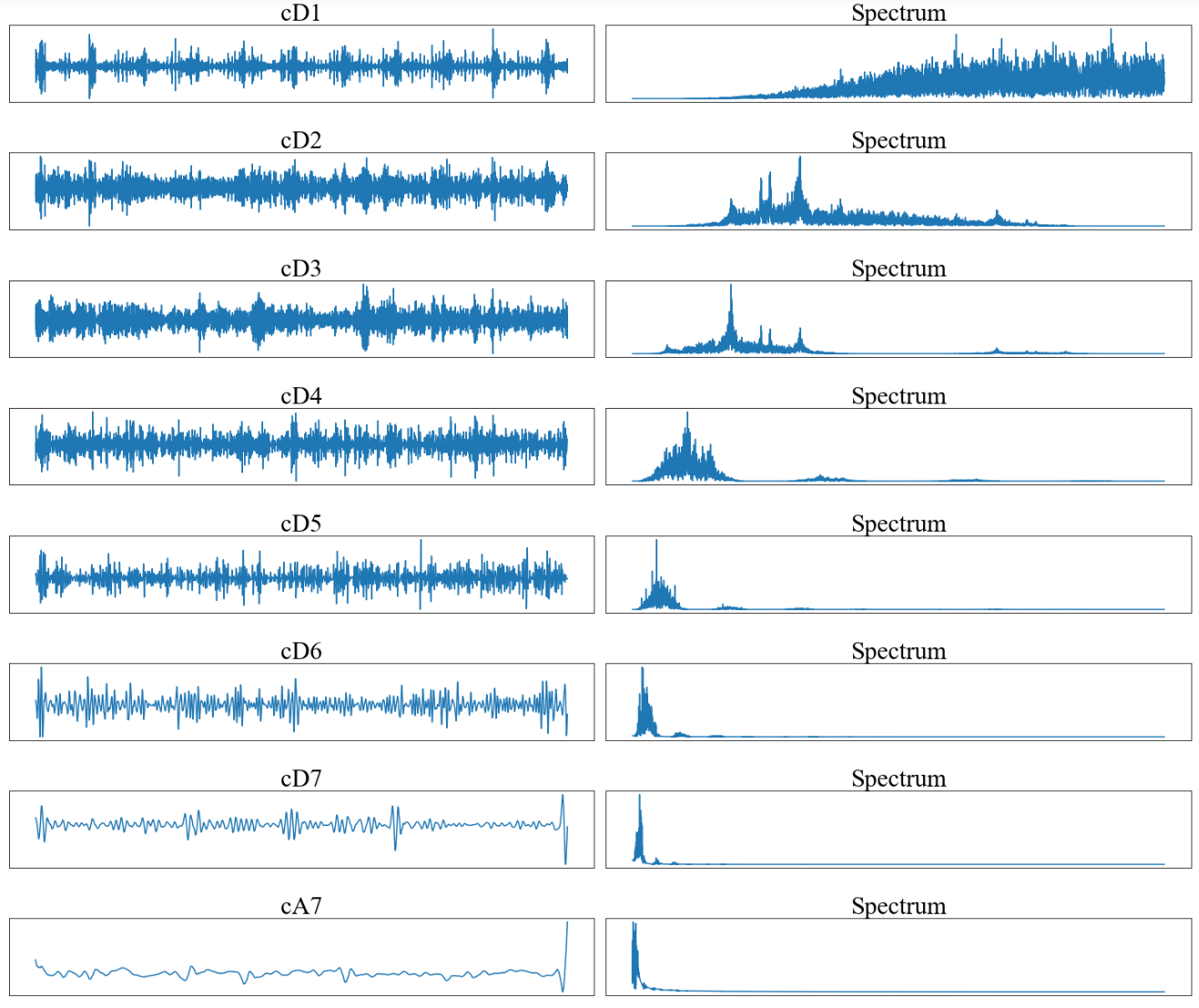}}
\caption{$7$-level DWT decomposition of random vibration signal $v(t)$: Elementary modes and their corresponding frequency spectra.}
\label{dwt_vib}
\end{figure}

DWT is a wavelet-based signal decompression tool that uses discrete values for scale and translation, reducing redundancy and computational complexity. In DWT, wavelet multi-resolution analysis is achieved by decomposing the signal into high and low-frequency components, commonly known as elementary modes, using digital filter banks and down-sampling operations. The Key parameters include the type of mother wavelet and the degree of decomposition. Mathematically, the DWT of a discrete signal $x[n]$ can be expressed as:
\begin{equation}
    DWT_x[j, k] = \sum_{n=-\infty}^{\infty} x[n]   \psi_{jk}[n]
\end{equation}
where, \newline
$DWT_x[j, k]$ represents the wavelet coefficient at the $j$-th level of decomposition and the $k$-th position, \newline
$\psi_{jk}[n]$ are the discrete wavelets, defined as scaled and shifted versions of the mother wavelet.\newline
In the decomposition process, the signal undergoes convolution with consecutive low-pass and high-pass filters followed by down-sampling. The main aspects and steps of the decomposition process include:\newline
Mother Wavelet: The first step involves selecting an appropriate wavelet base function, often called the "mother wavelet." \newline
Scaling and Translation: The decomposition process starts by scaling and translating the mother wavelet to generate a family of wavelet functions that serve as bases to decompose the signal. In the DWT, the mother wavelet is used in a discretized form where scaling and translation are done in discrete steps, typically in powers of two for scaling and integer multiples of that scale for translations. For each level of decomposition, the wavelet is scaled by a factor of $2^j$, where $j$ indicates the level of decomposition. This scaling effectively compresses the wavelet to capture signal features at different frequencies. Consequently, the wavelet is translated along the signal in steps that are multiples of the current scale factor $2^j$. This stepwise translation allows the wavelet to cover the entire signal, ensuring that all signal parts are analyzed. \newline
High-Pass Filter:  This filter represents the mother wavelet since its coefficients are essentially a discrete representation of the mother wavelet. As a high-pass filter, it extracts high-frequency \textit {detail} components from the signal, capturing rapid changes within the signal. \newline
Low-Pass Filter: This filter represents the scaling function, as its coefficients are derived from the scaled wavelets. As a low-pass frequency filter, it extracts low-frequency \textit {approximate} components that vary slowly over time. \newline
Down-Sampling: After separating the signal into high and low-frequency components, each component is down-sampled by a factor of two. Down-sampling involves keeping every second sample of the filtered signals. This process reduces the sample rate by half, effectively decreasing the data size and ensuring the transform is computationally efficient. \newline
Approximate and Detail Coefficients: Approximation and detail coefficients are outputs of down-sampling operations performed on approximate and detail components, respectively. The approximate coefficients represent a smoothed and down-scaled version of the original signal, while the detail coefficients describe high-frequency content in the signal. \newline
Iterative Analysis: The process of high and low pass filtering followed by down-sampling is performed iteratively on the resultant \textit{approximation coefficients only} until reaching the specified number of levels ($j$) or until reaching maximum decomposition constrained by signal length. This process creates a $j$-level tree structure of wavelet coefficients, shown in Fig.\ref{dwt_tree}.\newline
Signal reconstruction:  The original signal can be perfectly reconstructed by applying the inverse DWT, which processes the resultant coefficients through reconstruction filters in an inverse manner to the decomposition process. \newline
Elementary modes:  These modes are obtained through the individual reconstruction of the resultant coefficients (coefficients cD1 to cDj and cAj as shown in Fig. \ref{dwt_tree}), allowing the processing of the composite signal at the level of its constituent modes. Since these modes form the underlying structure that constitutes the original signal, perfect reconstruction of the original signal is possible through direct summation of these modes.\newline
Fig. \ref{dwt_vib} displays elementary modes of the random vibration signal $v(t)$ along with their corresponding spectra. These modes are obtained through $7$-level DWT decomposition of the signal using Daubechies \textit{db4} wavelet (shown in the top right corner of Fig. \ref{wavelet_func}). 
The Daubechies family of wavelets, especially those with higher orders (e.g., \textit{db4}, \textit{db5}), are compactly supported and well-localized in time, making them highly effective for analyzing signals that contain transient patterns or nonstationary components. This is particularly useful in vibration analysis \cite{aat22}, where detecting sudden changes or faults is essential for condition monitoring and diagnostics. The DWT decomposition of $v(t)$ iteratively extracts slowly varying modes of the signal, corresponding to lower frequencies in the signal at each level. This progressively reduces the resolution of high-frequency details with each level of decomposition, making DWT well-suited for signals where the information of interest is concentrated in lower frequencies.

\subsubsection{Stationary Wavelet Transform (SWT)}
A major limitation of the DWT is the lack of time invariance, which stems primarily from the down-sampling process. The down-sampling performed at each level makes the output of DWT sensitive to shifts in the input signal. Even small shifts can cause significant changes in the resulting wavelet coefficients, which is problematic in applications where the precise timing of signal features is essential. Furthermore, down-sampling reduces the signal length by half at each level of decomposition, which may lead to the loss of specific signal details, especially at higher decomposition levels. SWT is an extension of the DWT that addresses these limitations by omitting the down-sampling step, thus maintaining the original size of the signal throughout the decomposition levels. Unlike DWT, which provides a critically sampled (non-redundant) representation, SWT provides a more detailed but redundant representation of the signal. SWT is implemented by applying high-pass and low-pass filters without downsampling the filtered signals. Instead, the filters are up-sampled at each level to match the size of the original signal. The up-sampling of filters is done by inserting zeros between the filter coefficients, effectively increasing the size of the filters as the decomposition level increases. This leads to an overcomplete signal representation and increases computational burden compared to DWT.

\subsubsection{Wavelet Packet Transform (WPT)} 
\begin{figure}[!htbp]
\centerline{\includegraphics[width=0.5\textwidth]{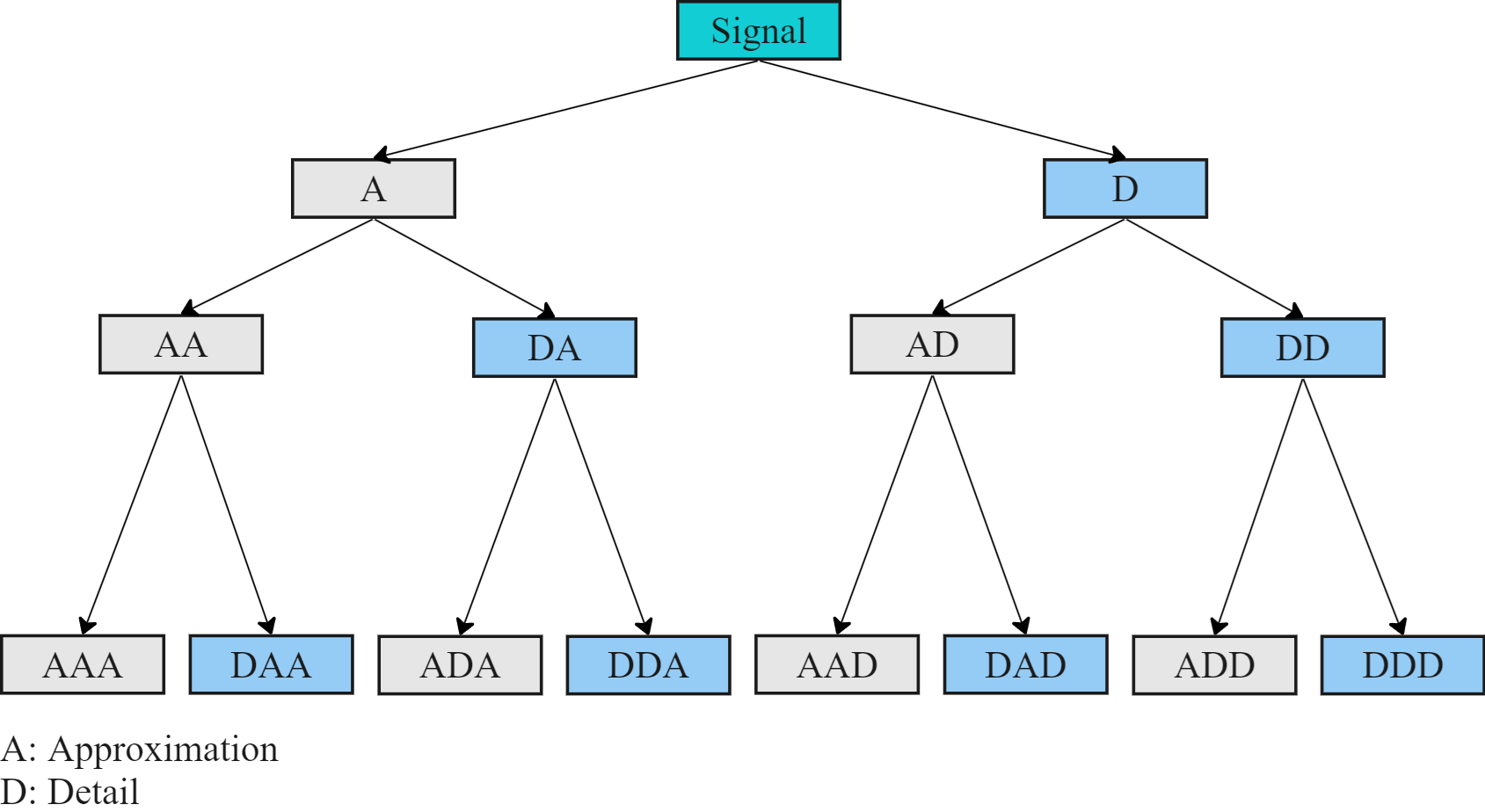}}
\caption{$3$-level WPT decomposition tree.}
\label{wpt_tree}
\end{figure}

\begin{figure}[!htbp]
\centerline{\includegraphics[width=0.5\textwidth]{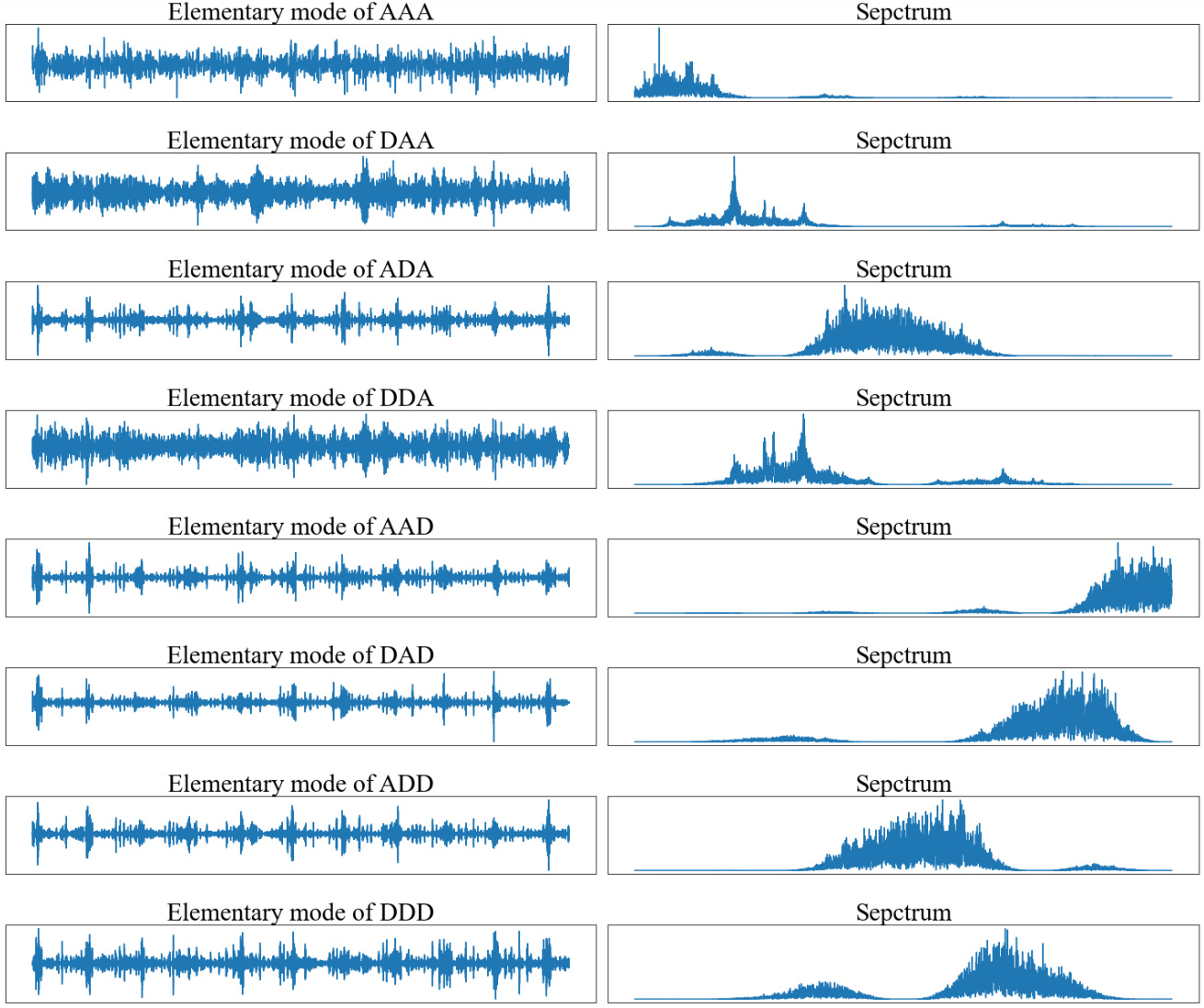}}
\caption{$3$-level WPT decomposition of random vibration signal $v(t)$: Elementary modes and their corresponding frequency spectra.}
\label{wpt_vib}
\end{figure}

The WPT extends the capabilities of classical wavelet-based decomposition. Unlike the DWT, which limits its decomposition to approximation coefficients, the WPT performs a full decomposition  at each level by processing detail and approximation coefficients, as depicted in Fig. \ref{wpt_tree}, thereby capturing both low-frequency and high-frequency components. While this process is more computationally intensive, it enhances frequency resolution in the analysis, enabling the detection of subtle signal features that DWT may overlook. 
Fig. \ref{wpt_vib} displays elementary modes of the vibration signal $v(t)$, obtained through $3$-level WPT decomposition using the \textit{db4} wavelet, along with their corresponding spectral contents. 
In contrast to DWT, which offers a hierarchical frequency decomposition with good frequency resolution at lower frequencies, WPT provides a more uniform time-frequency localization across the spectrum due to its symmetric decomposition of detail and approximation coefficients at each level. Hence, it maintains a balance between time and frequency resolution throughout the frequency spectrum, allowing for a more precise localization across a wider range of frequencies. 
Additionally, the spectrum in DWT is skewed towards lower frequencies with each successive level of decomposition. While this can be advantageous for analyzing signals characterized by low-frequency components, it inadequately represents higher-frequency details. WPT, on the other hand,  provides a balanced spectrum with a more equitable representation of both high and low-frequency components.\

Wavelet decomposition offers an effective tool for analyzing complex real-world signals that involve transit and nonstationary components, opening the door for a wide range of applications such as: 
\begin{itemize}
    \item Multi-resolution analysis: Wavelet decomposition methods attempt to decompose the signal into elementary modes characterized by their high time-frequency localization, offering the capacity for conducting a multi-resolution analysis of the original signal.
    \item Feature Extraction: The obtained modes encompass meaningful information as they exhibit distinct amplitude and frequency characteristics. Accordingly, distinctive features directly related to the inherent structure of the signal can be extracted from these modes or their corresponding coefficients.
    \item Noise Reduction: Noise and signal components often have different characteristics in the wavelet domain. By thresholding certain wavelet coefficients before signal reconstruction \cite{ch65}, noise can be effectively reduced or eliminated.
    \item Signal Enhancement: Wavelet reconstruction can be utilized to enhance signals, where certain characteristics can be amplified or smoothed out as required, improving the quality of the reconstructed signal.
    \item Signal Compression: By keeping only the most significant wavelet coefficients for reconstruction, signals can be compactly represented with minimal loss of quality. 
\end{itemize} \
As previously stated, selecting the appropriate mother wavelet and decomposition level are crucial steps in wavelet analysis. Different wavelets have different shapes and properties, such as smoothness, symmetry, and the number of vanishing moments, making the selection mainly dependent on application requirements and the shape similarity between the wavelet and the signal. The decomposition level determines the resolution at which the signal is analyzed. Higher decomposition levels allow for the analysis of lower-frequency components but also increase computational complexity and could dilute the significance of higher-frequency components. The maximum level of decomposition in DWT and WPT is typically limited by the length of the signal, $N$. Generally, each level of decomposition reduces the number of data points by half. Thus, the number of maximum possible decomposition levels equals $log_2(N)$. In SWT, the maximum decomposition level is determined by the signal's length and the filters' length at each decomposition level, which increases with each level due to the up-sampling operation. Further decomposition becomes impractical when the filters become too long relative to the signal.\

The Python PyWavelets package provides comprehensive modules for wavelet analysis. It supports custom wavelets and provides various wavelet functions such as CWT, DWT, SWT, WPT and over 100 built-in wavelet filters. Other libraries supporting wavelet analysis include PyCWT, a Python module for continuous wavelet spectral analysis, and scipy.signal, which provides CWT and other wavelet functions, particularly for signal filtering.
 
\subsection{Hilbert Transfrom}
The Hilbert transform (HT) is a fundamental operator in signal theory; it is particularly useful in obtaining the analytic signal representation of real-valued signals. An analytic signal is a complex-valued representation of the signal that provides a comprehensive way to describe both its amplitude and phase characteristics. 
The HT of a signal $x(t)$ is defined as:
\begin{equation}
    H\{x(t)\} = \frac{1}{\pi} \int_{-\infty}^{\infty} \frac{x(\tau)}{t-\tau} d\tau
\end{equation}
The transform essentially modifies the phase of each frequency component of the signal by $\pm 90^\circ$. The $X_a(t)$ of $x(t)$ is formed by augmenting the signal with its HT $H\{x(t)\}$ as the imaginary part. Mathematically, it is expressed as:
\begin{equation}
    X_a(t) = x(t)+jH\{x(t)\}
\end{equation}
Represented in its polar form, the analytical is expressed as:
\begin{equation}
    x_a(t) = A(t)   e^{j \theta(t)}
\end{equation}
where, $A(t)$ is the instantaneous amplitude (also known as the amplitude envelope), it is given by:\newline
\begin{equation}
    A(t) = |x_a(t)| = \sqrt{x(t)^2 + H\{x(t)\}^2},
\end{equation}
 and $\theta(t)$ is  the instantaneous phase; given by:
\begin{equation}
    \theta(t) = \arctan\left(\frac{H\{x(t)\}}{x(t)}\right)
\end{equation}
Accordingly,  the instantaneous frequency $f(t)$ can be obtained by taking the derivative of  the instantaneous phase $\theta(t)$:
\begin{equation}
    f(t) = \frac{1}{2\pi} \frac{d\theta(t)}{dt}
\end{equation}\

\begin{figure}[!htbp]
\centerline{\includegraphics[width=0.5\textwidth]{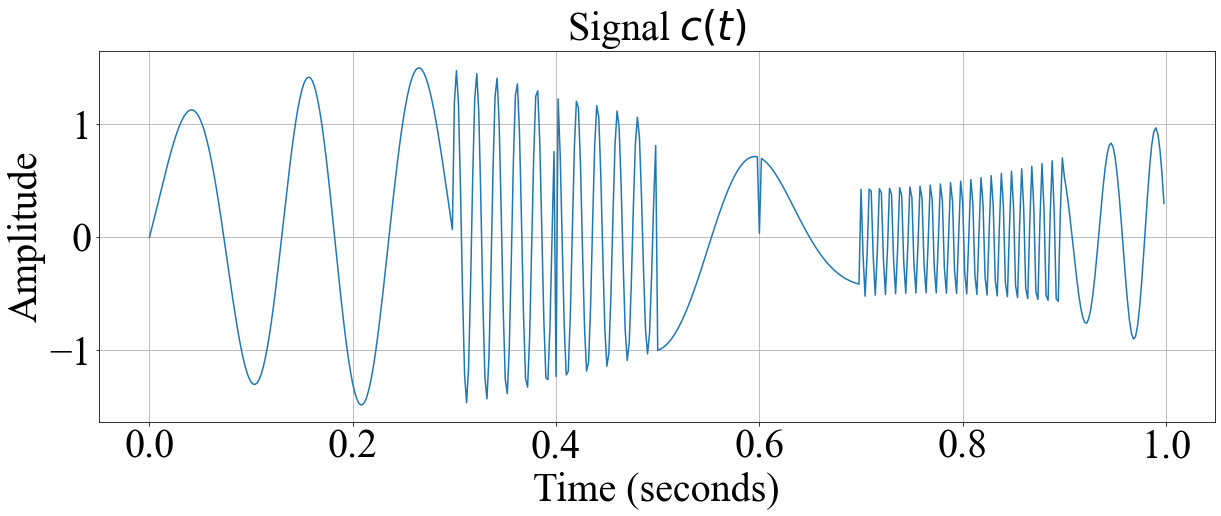}}
\caption{Composite signal $c(t)$.}
\label{signal_c}
\end{figure}

\begin{figure}[!htbp]
\centerline{\includegraphics[width=0.5\textwidth]{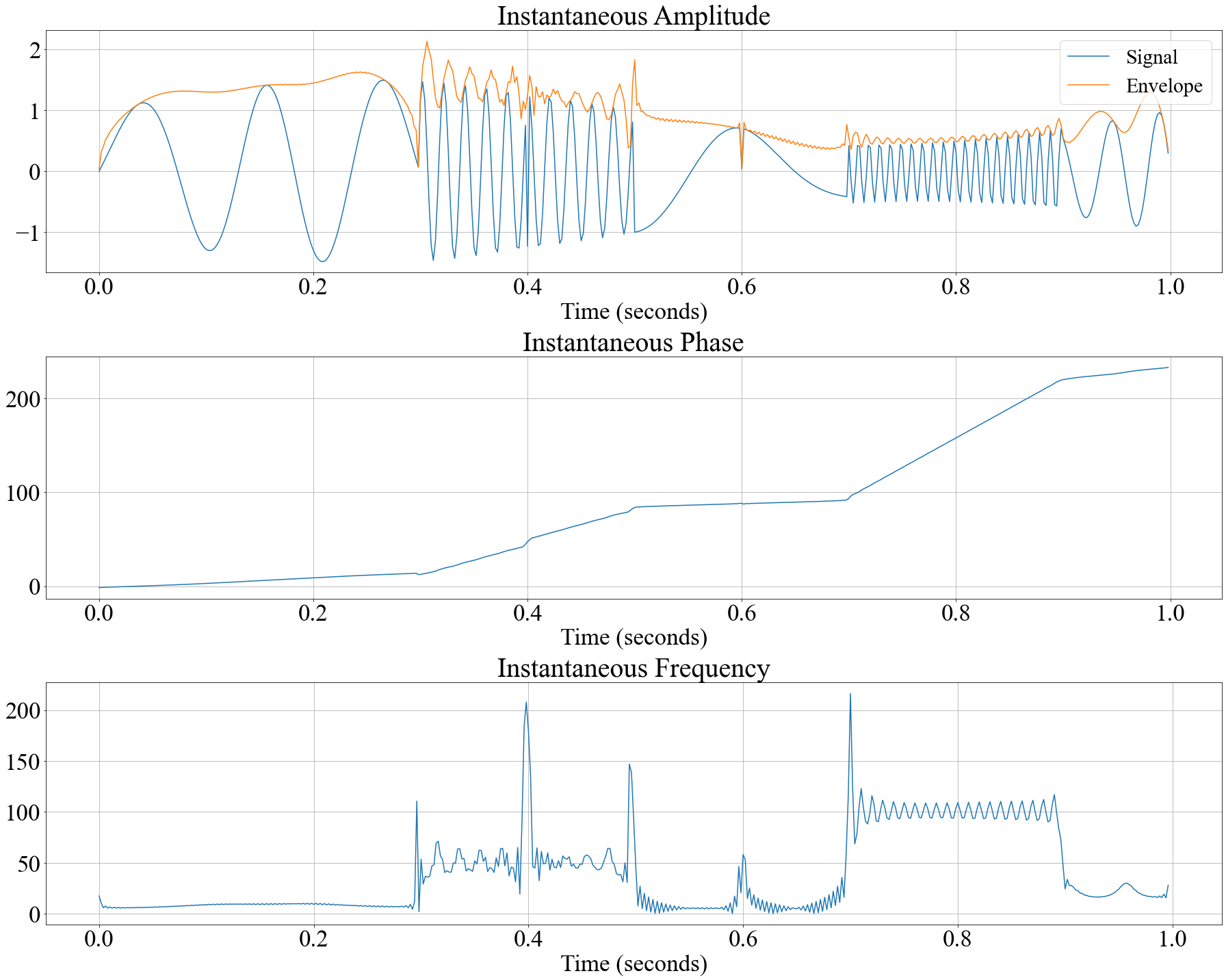}}
\caption{Hilbert Transform of the signal $c(t)$.}
\label{HT_signal}
\end{figure}

The instantaneous amplitude of a signal shows how the signal strength varies with time. The instantaneous phase provides the phase angle of the signal as a function of time. The instantaneous frequency reveals how the frequency content of a signal evolves over time, providing a dynamic view of the signal's spectral properties. By obtaining instantaneous amplitude, phase, and frequency information, the HT serves as an effective tool to identify distinct characteristics, such as common patterns and sudden changes in phase and frequency where relevant features can be extracted accordingly. For instance, HT can be used to obtain instantaneous amplitudes of amplitude-modulated signals such as fault vibration signals. In these signals, vibrations of damaged bearings are manifested as modulations in the amplitude of the generated vibration signal.  Hence, obtaining the signal envelope provides an efficient approach for extracting fault signature frequencies from the envelope's spectrum. Fig. \ref{signal_c} displays a composite signal, $c(t)$, that has a time duration of $1$ second. The signal is composed of various components so that it  exhibits four sudden changes as follows:
\begin{itemize}
    \item Rapid oscillations of $50$ Hz from $0.3$ time instant to $0.5$ time instant.
    \item Rapid oscillations of $100$ Hz from $0.7$ time instant to $0.9$ time instant.
    \item Abrupt phase change of $180$ degrees at $0.4$ time instant.
    \item Abrupt phase change of $90$ degrees at $0.6$ time instant.
\end{itemize}
The function \textit{scipy.signal.hilbert} in the scipy.signal library is used to compute the HT of $c(t)$. The function returns the analytical signal $c_a(t)$ as an output. The instantaneous amplitude, phase, and frequency are calculated accordingly and displayed in Fig. \Ref{HT_signal}. 
As shown in the plots, the obtained instantaneous information provides critical insights into the signal's behavior, highlighting common patterns and sudden changes. Specifically, the envelope shows the main pattern of the signal, reflecting its energy content over time. The abrupt phase changes at the $0.4$ and $0.6$ time instants are manifested as small step changes at these time instants in the instantaneous phase. Moreover, the phase is directly impacted by temporal changes in the signal frequency, as shown in the plot. The instantaneous frequency provides vital information about frequency variations and sudden changes in the signal.  Particularly, the instantaneous frequency plot shows the rapid oscillations of $50$ Hz and $100$ Hz between $0.3-0.5$ and $0.7-0.9 $ time instants, respectively. Additionally, the sudden changes in frequency and phase are manifested as a spike at the corresponding time instant in the plot. In real-world applications, such spikes represent significant features in the signal that would be of particular interest.

\subsection{Hilbert–Huang transform}
The Hilbert–Huang transform (HHT) \cite{th98} is a well-known adaptive method for analyzing nonlinear and nonstationary signals. Unlike traditional transforms, HHT does not impose a fixed basis function in signal analysis. Instead, it utilizes an adaptive approach to analyze the signal, making it highly responsive to variations in signal. The HHT is a two-stage process that involves empirical mode decomposition (EMD) of the signal followed by the Hilbert Spectral Analysis (HSA).\newline

\subsubsection{Empirical Mode Decomposition (EMD)}
\begin{figure}[!htbp]
\centerline{\includegraphics[width=0.5\textwidth]{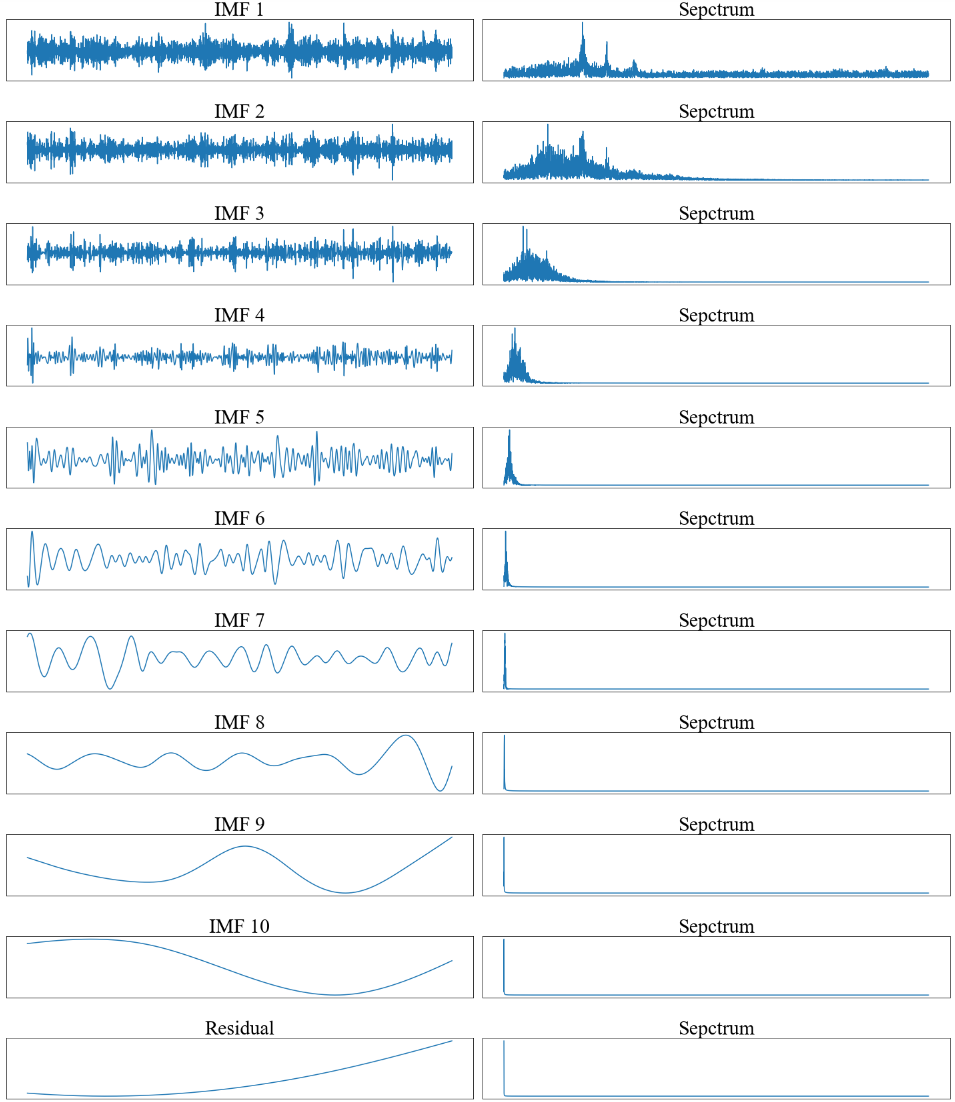}}
\caption{EMD of the random vibration signal $v(t)$: IMFs, residual, and their respective frequency spectra.}
\label{emd_vib}
\end{figure}

EMD is an adaptive decomposition method that decomposes the signal into a set of simpler functions known as intrinsic mode functions (IMFs). EMD forms the basis of adaptive mode decomposition (AMD) methods introduced over the last three decades. In contrast to wavelet-based decomposition, AMD uses adaptive approaches rather than priori basis functions to decompose the signal. Hence, the obtained IMFs are not influenced by a priori basis function \cite{ag17}. This makes AMD particularly effective when dealing with nonlinear and nonstationary signals due to its ability to adapt to varying signal characteristics. However, this adaptability comes at the cost of higher computational complexity. The EMD algorithm, introduced in 1998 as part of the HHT, represents the core of all AMD methods. The process of EMD, commonly known as sifting, involves the following  steps: 
\begin{enumerate}
    \item Identification of Extrema: Start by identifying all the local maxima and minima of the original signal.
    \item Envelope Creation: Construct the upper and lower envelopes of the signal by interpolation between the local maxima and minima, respectively. These envelopes essentially outline the signal's oscillatory amplitude.
    \item Mean Envelope Calculation: Compute the mean of the upper and lower envelopes.
    \item Extraction of Detail: Subtract the mean envelope from the original signal. This step isolates a component of the signal that potentially qualifies as an IMF.
    \item IMF Check: Verify if the extracted component meets the criteria for being an IMF:
          \begin{itemize}
              \item  The number of extrema (local maxima and minima) and the number of zero crossings in the component must either be equal or differ at most by one. This condition ensures that the IMF captures a well-defined oscillatory mode without bias toward upward or downward trends.
              \item The mean value of the envelope defined by the local maxima and the envelope defined by the local minima must be zero at any point in the component. This condition guarantees that the IMFs have well-balanced oscillations around zero, reflecting true oscillatory modes rather than trends or biases in the signal.
          \end{itemize}
          If the component doesn't qualify for an IMF, return to step 2 and use this component as the new signal.
    \item Completion of one IMF: Once an IMF is identified, subtract it from the original signal. This leaves a residue signal.
    \item Repetition: Repeat steps 1-6 on the residue signal. This process is iterated, yielding a new IMF at each iteration, and the residual becomes the input for the next iteration.
    \item Termination (Stopping Criterion): The process terminates when the residual signal becomes a monotonic function from which no more IMFs can be extracted, or it becomes sufficiently small in amplitude (based on a predefined threshold)
\end{enumerate}
The output of the EMD process of the random vibration signal $v(t)$ is displayed in Fig. \ref{emd_vib}, including IMFs, residual, and their respective spectra.
As shown, the EMD adaptively decomposed the signal into ten IMFs of distinct frequency spectra where the frequency contents of IMFs decrease with each successive IMF. While EMD is well-known to be effective in analyzing complex signals, the decomposition process is sensitive to noise since it works directly on the signal's extrema and minima. Additionally, the process requires a complete envelope to decompose the signal into IMFs accurately, which can be difficult to define at the boundaries due to the absence of neighboring data points. This can lead to distortions of IMFs near the ends of the signal, commonly known as the end effect \cite{ez23}, impacting the accuracy and reliability of EMD for signal analysis, particularly when the signal's behavior at the boundaries is critical. Another challenge in EMD is mode mixing \cite{ag08}; ideally, each IMF should represent a unique frequency component. However, in signals with close spectral proximity or exhibit intermittency characteristics such as abrupt amplitude and frequency changes, the standard EMD algorithm may not be able to separate these features accurately. As a result, a single IMF may contain information from multiple modes, which can lead to a loss of significant physical interpretation of the IMFs. Further improvements to the standard EMD algorithm have been introduced to address these limitations. For example, the ensemble empirical mode decomposition (EEMD) algorithm \cite{ew09} was developed by incorporating an ensemble approach and adding finite amplitude white noise to alleviate problems of mode mixing and end effects. Accordingly, the natural oscillatory modes are obtained by averaging the corresponding IMFs obtained from an ensemble of the signal and the added noise. Other common AMD algorithms include variational mode decomposition (VMD)\cite{vd14},  local mean decomposition \cite{ts05}, and empirical wavelet transform (EWT) \cite{eg13}. More details on AMD techniques and their applications can be found in \cite{de23}\cite{ac21}\cite{af17}\cite{al18}.

\subsubsection{Hilbert Spectral Analysis}
\begin{figure}[!htbp]
\centerline{\includegraphics[width=0.5\textwidth]{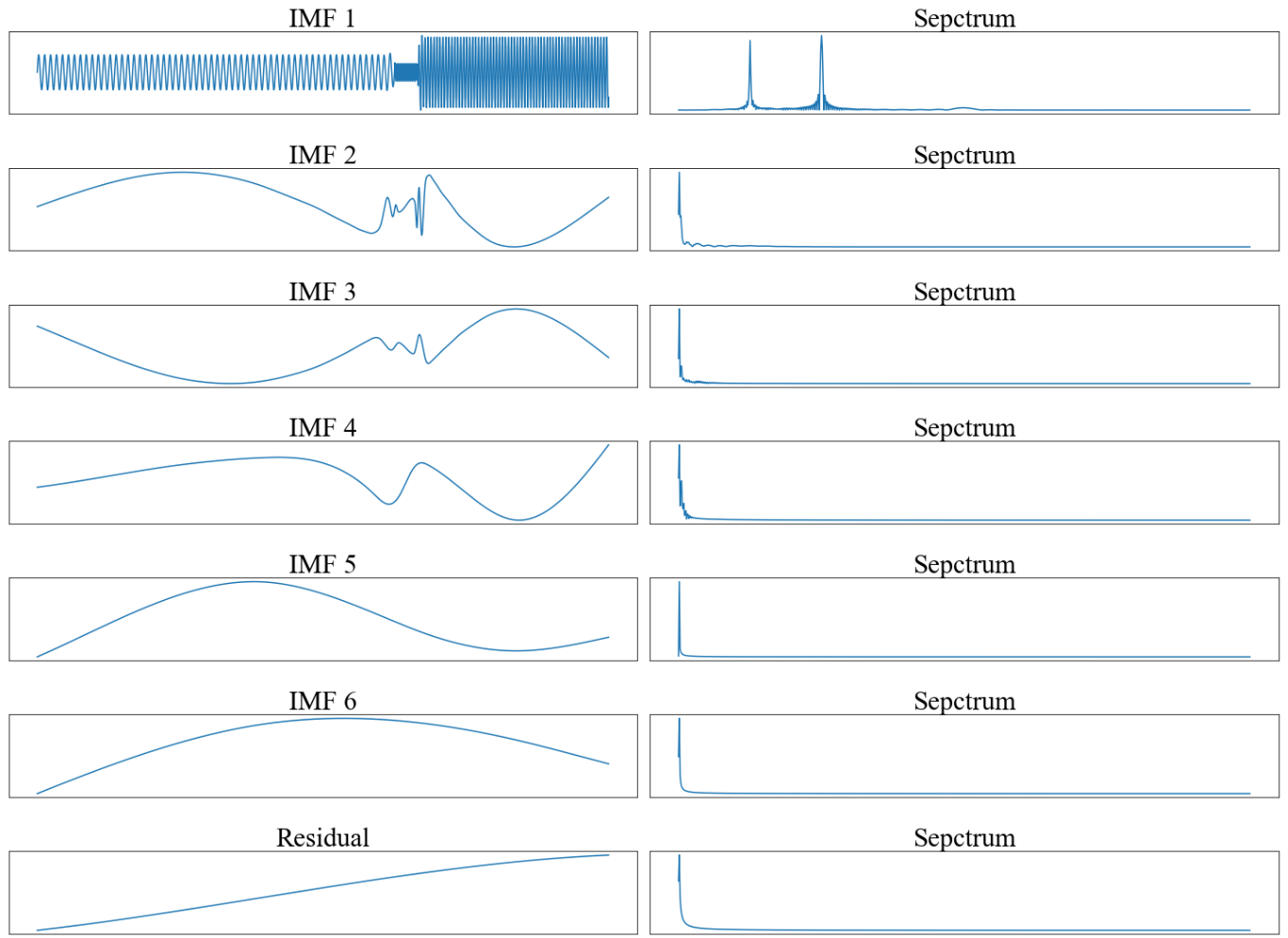}}
\caption{EMD of the composite sinusoidal signal $s_3(t)$: IMFs and residual alongside  with their respective frequency spectra.}
\label{emd_sin3}
\end{figure}

\begin{figure}[!htbp]
\centerline{\includegraphics[width=0.5\textwidth]{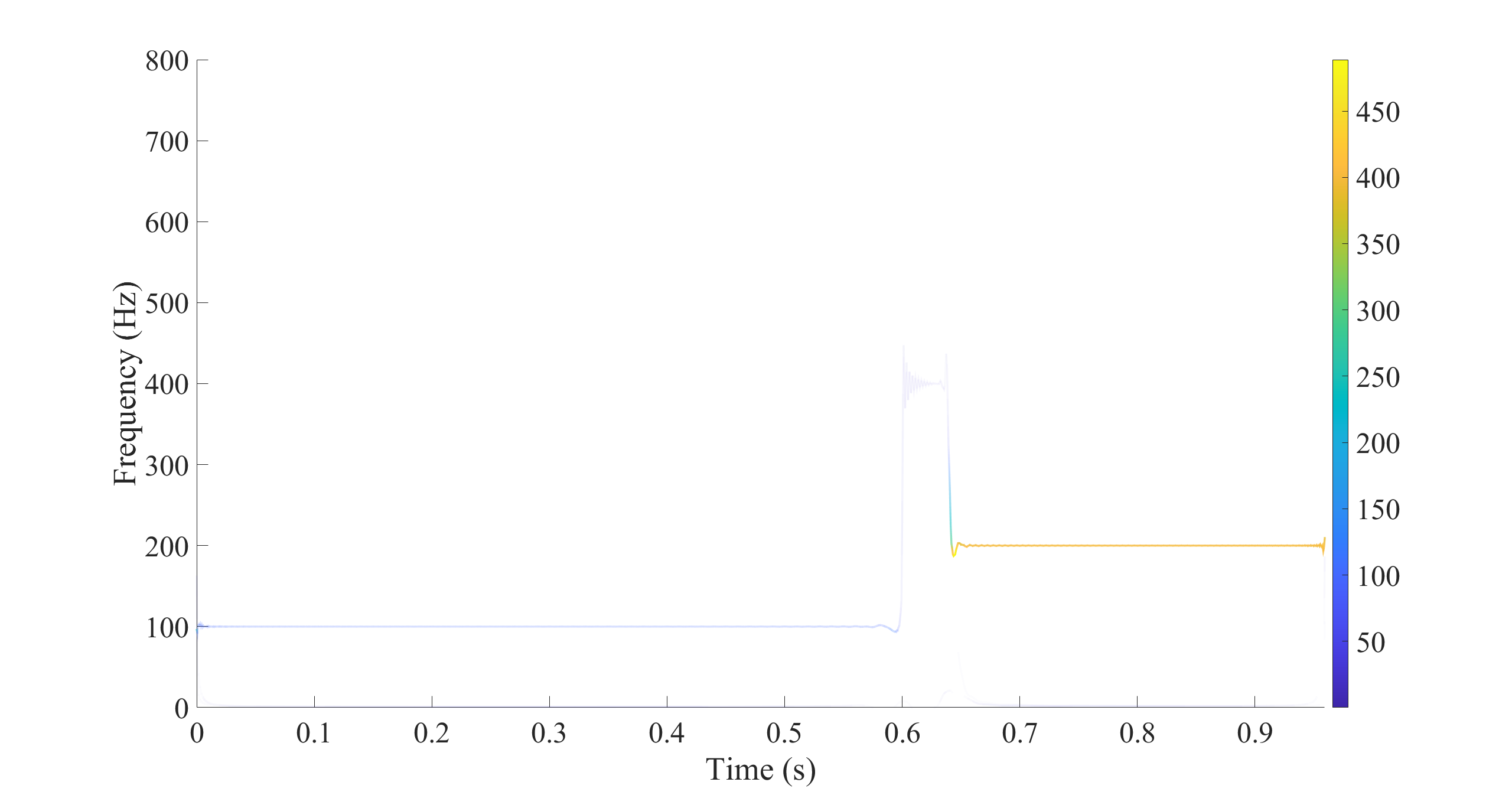}}
\caption{2D heatmap visualization of the HTT of $s_3(t)$.}
\label{hht_sin3}
\end{figure}

The second step of HHT involves applying the HT to each IMF to obtain its analytic signal and, accordingly, its insinuations amplitude, phase, and frequency. An IMF, due to its defining conditions, ensures that its HT is well-behaved and meaningful for signal analysis \cite{al19}. Specifically, due to the equal number of extrema and zero crossings, an IMF exhibits a consistent oscillatory pattern, which is essential for a coherent HT. Further, the symmetry of the IMF's envelope about the zero line guarantees that the IMF doesn’t have a bias towards positive or negative values, facilitating a more accurate HT.\newline
Mathematically, the HSA can be expressed as follows:
Given a signal $x(t)$, decomposed into $n$ IMFs $c_i(t)$, where $i=1,2,  ,n$, the analytic signal $Z_i(t)$ for each IMF is formed as follows:
\begin{equation}
    Z_i(t) = c_i(t) + jH\{c_i(t)\}
\end{equation}
where $H\{c_i(t)\}$ is the HT of $c_i(t)$. The analytical signal in the polar form is expressed as:
\begin{equation}
    Z_i(t) = A_i(t)   e^{j \theta_i(t)}
\end{equation}
Accordingly, the instantaneous amplitude $A_i(t)$ is:\newline
\begin{equation}
    A_i(t) = |Z_i(t)| = \sqrt{c_i(t)^2 + H\{c_i(t)\}^2},
\end{equation}
 and the instantaneous phase $\theta_i(t)$ is:
\begin{equation}
    \theta_i(t) = \arctan\left(\frac{H\{c_i(t)\}}{c_i(t)}\right)
\end{equation}
Accordingly,  the instantaneous frequency $f_i(t)$ is obtained by taking the derivative of $\theta_i(t)$:
\begin{equation}
    f_i(t) = \frac{1}{2\pi} \frac{d\theta_i(t)}{dt}
\end{equation}
The instantaneous frequencies and amplitudes obtained from the HT, can be used to construct a time-frequency distribution of the signal. This is accomplished by plotting the instantaneous amplitude or energy against the instantaneous frequencies of the IMFs for each point in time. The result is a distribution that shows how the signal's frequencies evolve over time, highlighting the signal's nonlinear and nonstationary characteristics. The time-frequency-energy distribution is commonly visualized as a 2D heatmap, where the $x$-axis represents time, the $y$-axis represents frequency, and the color intensity represents the amplitude or energy level of the signal at each time-frequency point. Considering the composite sinusoidal signal $s_3(t)$, Fig. \ref{emd_sin3} displays the resulting IMFs and residuals from the EMD process, as well as their spectral contents. Consequently, Fig. \ref{hht_sin3} shows the 2D heatmap visualization of the output from the HTT. These results are obtained using textit{emd}, \textit{vmd}, and \textit{hht} built-in MATLAB functions, which, in contrast to Python, provide a mature and user-friendly environment for EMD, VMD, and HHT computations and visualization.
It is obvious that EMD and, consequently, HHT effectively capture all components in the signal, including the transit osculations of $400$ Hz at $0.6$ seconds, and effectively shows the temporal energy distribution of the signal at different frequencies. This demonstrates the advantage of the HHT that lies in its adaptability to nonlinear and nonstationary signals.\

However, the performance of the HHT is tied to the choice of AMD method, which is fundamental to the accurate decomposition of the signal into its IMFs. Furthermore, the employed stopping criterion for the sifting process \cite{it14} and the used interpolation method for envelope estimation \cite{aa23}, have direct impacts on the effectiveness of the HHT. A proper stopping criterion ensures that the decomposition process neither overfits nor underfits the signal, thereby preserving the signal's essential characteristics without introducing artifacts. The interpolation method for envelope estimation influences the HHT's performance since it determines how well the upper and lower envelopes capture the true oscillatory modes of the signal.

\subsection{Wigner–Ville Distribution}
The Wigner–Ville distribution (WVD), as a time-frequency analysis tool is characterized by its ability  to provide highly resolved energy-time-frequency representations of signals. The WVD of a continuous-time signal 
$x(t)$ is expressed as:
\begin{equation}
     WVD(t, f) = \int_{-\infty}^{\infty} x(t + \frac{\tau}{2}) x^*(t - \frac{\tau}{2}) e^{-j2\pi f\tau} \, d\tau
\end{equation}
where, $x^*(t)$ is the complex conjugate of $x(t)$, $\tau$ and $f$ are variables representing time-shift and frequency, respectively. \newline
According to this expression, the kernel function of WVD is given by:
\begin{equation}
    x(t + \frac{\tau}{2}) x^*(t - \frac{\tau}{2}),
\end{equation}
which is essentially the instantaneous autocorrelation function (IACF) of $x(t)$, commonly known as Wigner auto-correlation function \cite{tc95}. The use of the complex conjugate of the signal in the autocorrelation accounts for the magnitude and phase parts of the signal. Note that the traditional autocorrelation function of (\ref{acf}) integrates the autocorrelation over time $t$, providing a global measure of the correlation. In contrast, the IACF provides a local measure of the correlation, which is better suited to reflect local and time-varying features of the signal. Hence, facilitating a more accurate representation of the signal's characteristics \cite{dt10}. The Fourier kernel $e^{-j2\pi f\tau}$ in the WVD expression transforms the IACF into the frequency domain with respect to the time-shift variable $\tau$, thereby providing a joint time-frequency representation of the signal. For a discrete signal $x[n]$ with $N$ samples,The discrete WVD is given by:
\begin{equation}
   WVD[n, k] = \sum_{m=-N}^{N} x[n + \frac{m}{2}] x^*[n - \frac{m}{2}] e^{-j2\pi \frac{km}{N}}
\end{equation}
Here, $x^*[n]$is the complex conjugate of $x[n]$, and $k$ represents the discrete frequency index. The term,\newline
\begin{equation}
  x[n + \frac{m}{2}] x^*[n - \frac{m}{2}], 
\end{equation}
captures the auto-correlation of the signal at different lags.\

\begin{figure}[!htbp]
\centerline{\includegraphics[width=0.5\textwidth]{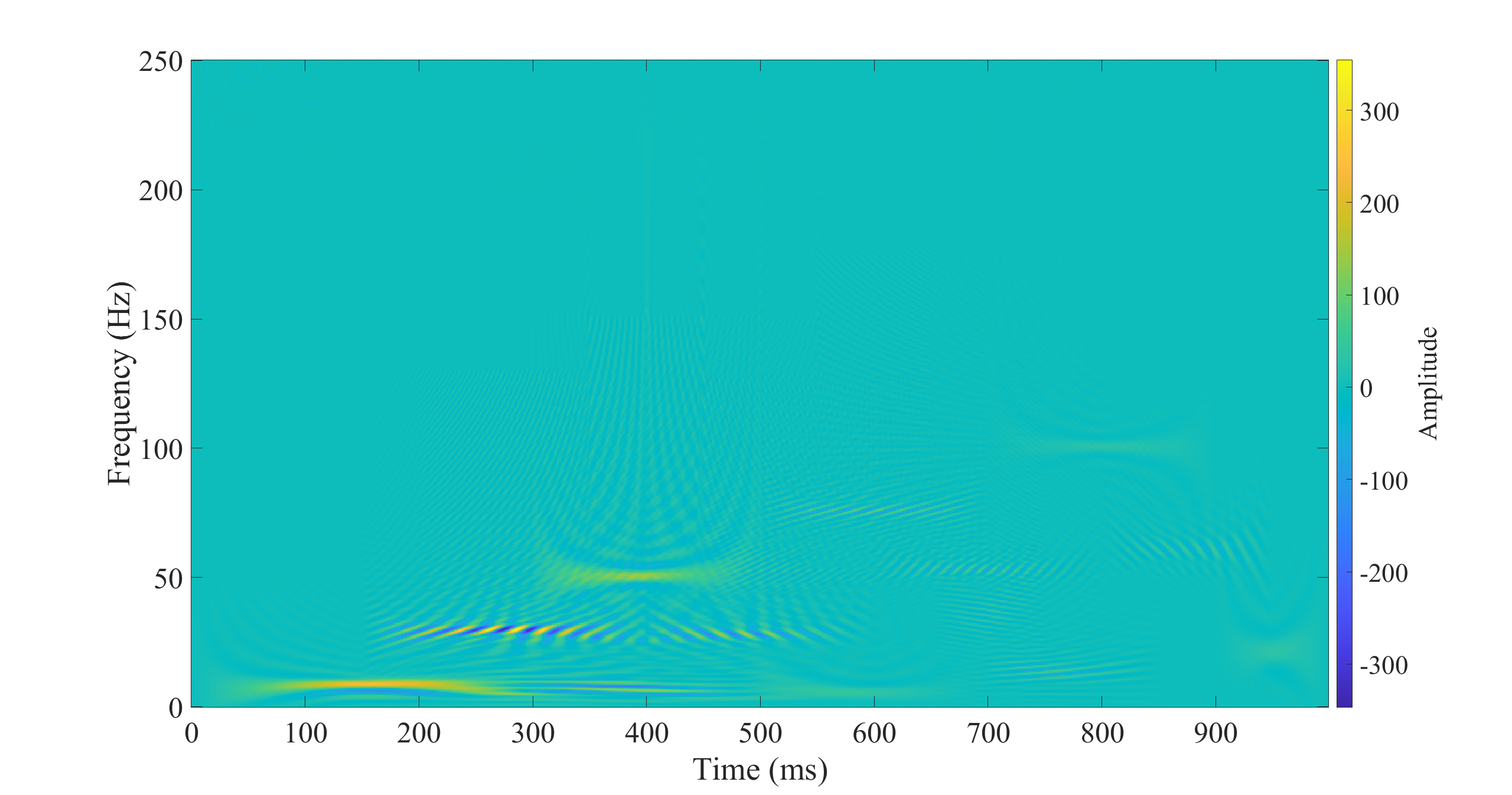}}
\caption{WVD of the composite sinusoidal signal $c(t)$.}
\label{wvd_sig_c}
\end{figure}

\begin{figure}[!htbp]
\centerline{\includegraphics[width=0.5\textwidth]{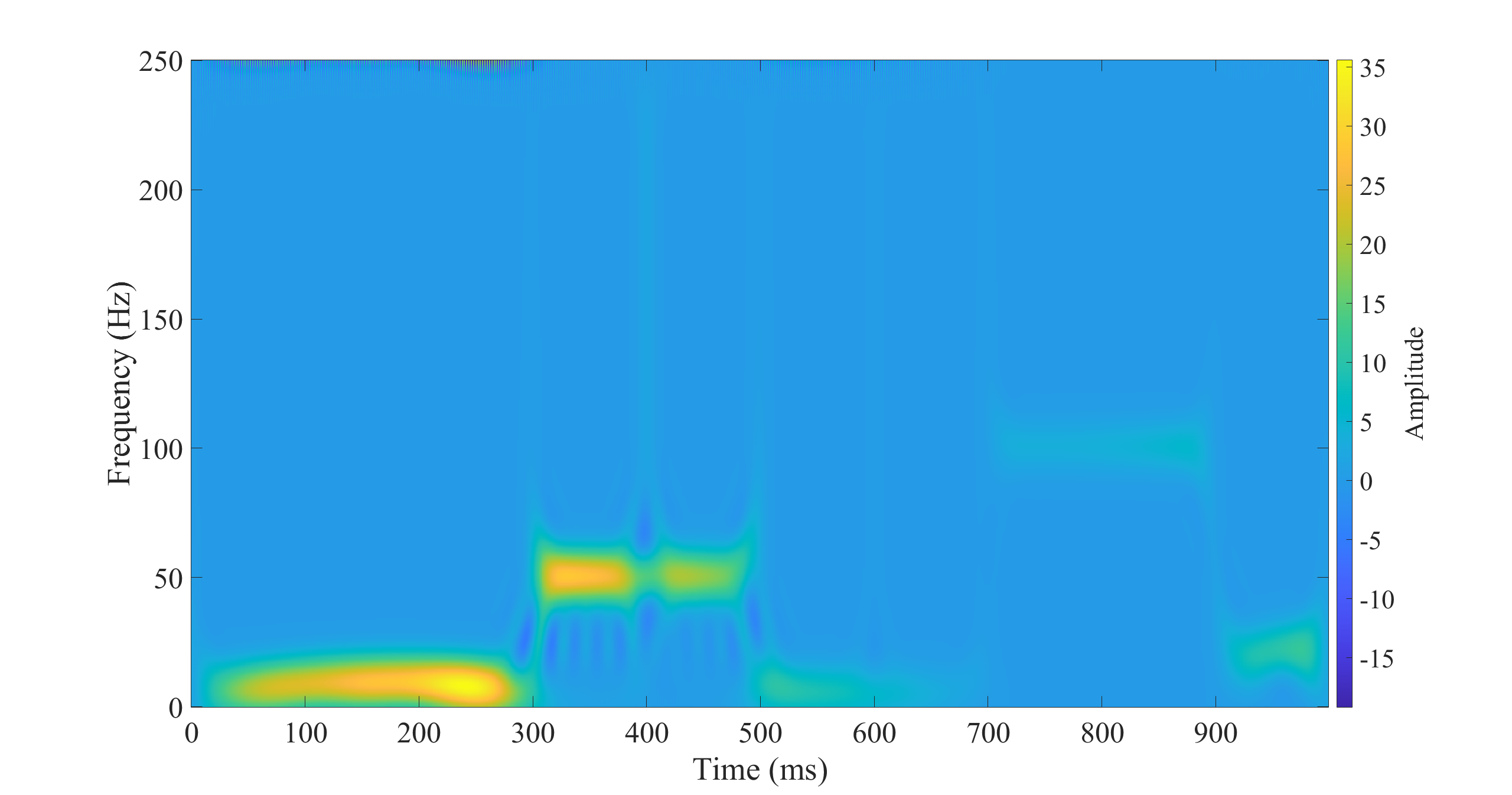}}
\caption{PWVD of the composite sinusoidal signal $c(t)$.}
\label{pwvd_sig_c}
\end{figure}

The output of WVD, $WVD(t, f)$, is essentially a 2D function of time and frequency that is commonly visualized as a 2D heatmap where the value of $WVD(t, f)$ at any point $(t, f)$ reflects signal's energy at that particular time and frequency. The high energy concentration of WVD results from its inherent quadratic (nonlinear) structure since the energy itself is a quadratic representation of the signal \cite{lh92}. In contrast to FT, STFT, and CWT, which are liner transforms\footnote[3]{In signal theory, a linear transform satisfies the superposition or linearity condition, which states that if the input signal $x(t)$ is a linear combination of some signal components, then,  the transom of $x(t)$ is also a linear combination of the transforms of each signal component.}, the WVD is nonlinear since it involves a product of the signal with a time-shifted version of itself (correlation). This energetic and correlative nature of the WVD \cite{lh92} makes it a unique tool for energy-time-frequency analysis compared to other transforms, however the quadratic nature of the WVD presents a major challenge when analysing multi-component signals due to the high level of cross-term interference \cite{ew17} that appears in the time-frequency representations. Cross-terms are caused by undesired cross-correlation between various signal components, leading to a false indication of signal components between the desired auto-correlation terms in the WVD representation of the signal. To overcome this problem, window-based approaches \cite{sf84, tf98} and kernel-based approaches \cite{ic89, tz90}, such as the pseudo-Wigner distribution (PWVD), are commonly utilized to suppress cross-terms. PWVD is a windowed version of the WVD that applies a smoothing kernel to WVD in either time-domain or frequency domain to reduce the cross-terms at the cost of a certain loss in resolution. The PWVD of a signal $x(t)$ can be expressed as:
\begin{equation}
PWVD(t, f) = \int_{-\infty}^{\infty} h(\tau) x\left(t + \frac{\tau}{2}\right) x^*\left(t - \frac{\tau}{2}\right) e^{-j2\pi f\tau} d\tau
\end{equation}
where $h(\tau)$ is the smoothing kernel in the time domain. In the frequency domain, the smoothing operation becomes the convolution operation between the smoothing kernel and the WVD. The smoothing operation essentially averages the signal's energy over the time and/or frequency domains, reducing the presence of cross-terms arising from the quadratic nature of the WVD. However, this comes at the cost of blurring the signal's time-frequency representation. The limitation of WVD resulting from undesired cross-correlation and the advantage of PWVD are demonstrated by comparing the WVD and PWVD of the composite signal, $c(t)$, which, as illustrated earlier, comprises multiple time-varying components.  The MATLAB function \textit{wvd} is used to compute WVD and PWVD and visualize the outputs as 2D heatmaps. This built-in function provides straightforward computations and visualizations of various WVD types. The WVD and PWVD results are displayed in Fig.\ref{wvd_sig_c} and Fig.\ref{pwvd_sig_c}, respectively. 
The comparison reveals that the WVD representation is impaired by false indications of non-existent signal components due to the cross-correlation between the actual signal components. In contrast, the PWVD, through its smoothing operation, eliminated the interference of cross-terms and accurately resolved all the time-varying frequency components. However, it can be seen that the improvement achieved with the PWVD led to a reduction in frequency resolution, manifested as wider lines in the PWVD's heatmap compared to the WVD's heatmap, thereby affecting the precise estimation of the corresponding frequency values.

\subsection{Conclusive Comparison}
Aimed at providing interested readers a thorough understanding of signal processing fundamentals, this tutorial offered a comprehensive introduction to signal characteristics with an in-depth discussion of prevalent signal transformation and analysis tools. The discussion focused on elucidating the main concepts, mathematical foundations, essential characteristics, advantages, and limitations. Additionally, the tutorial addressed implementation considerations by underscoring the use of programming libraries and built-in functions to facilitate efficient implementations of these tools. Furthermore, the codes employed for generating the illustrative plots throughout this section have been made publicly accessible as previously mentioned.\

Throughout the section, it became evident that selecting the proper tool depends entirely on signal nature, application-specific requirements, and available computational resources. For stationary signals, where the primary interest lies in frequency analysis, The FT efficiently obtains the frequency spectrum, thereby enabling the identification of predominant frequency components in the signal. The PSD is particularly advantageous in scenarios involving comparative spectral analysis of multiple signals varying in length and/or bandwidth. It provides a normalised spectral density measure that facilitates effective comparison. The HT is particularly effective in applications where the signal envelope is of primary importance. Moreover, instantaneous phase and frequency information can be efficiently obtained from the analytical signal of the HT.\

For signals with time-varying spectral characteristics, the STFT provides a computationally efficient tool for the time-frequency representation of the signal. However, its fixed window size results in a uniform resolution across all frequencies, reducing its effectiveness for applications requiring variable-resolution analysis. In such contexts, Wavelet analysis presents a viable solution due to its ability to perform multi-resolution analysis through scaled and shifted versions of a wavelet base function.\

For the analysis of nonstationary and nonlinear signals characterized by rapidly changing components, the HHT provides an adaptive and more robust approach than CWT but with more computational requirements.  The WVD offers a powerful method for providing a joint energy-time-frequency representation of the signal, although it is more computationally intensive than the other tools. It is particularly useful in high-resolution analysis scenarios or where energy concentration is a focal requirement. \

Signal decomposition techniques allow the analysis of the elementary components of a signal. This makes it possible to extract unique manageable-sized features closely related to the signal's inherent structure. Regarding wavelet decomposition, its performance is highly dependent on the base wavelet function and the level of decomposition. In contrast, AMD methods use an adaptive mechanism rather than an a priori basis function to decompose the signal, thereby generating elementary modes that are not influenced by an a priori basis function. Moreover, AMD methods dynamically adapt to varying signal characteristics, making them effective for analyzing nonlinear and nonstationary signals. However, AMD methods are more computationally intensive than wavelet decomposition. Additionally, AMD methods are sensitive to noise and their efficiency highly depends on the employed stopping criterion and the used interpolation method for envelope estimation. Table \ref{tools_comp} summarizes the comparison, highlighting the main aspects of each tool regarding signal nature, application requirements, and computational complexity.
\begin{table*}[!htbp]
    \centering
\caption{Comparison between common signal analysis and transformation tools}
\label{tools_comp}
    \begin{tabular}
    {| c| c | c | c | } \hline 
         \textbf{Tool} & \textbf{Signal Nature} & \textbf{Application} & \textbf{Computational Complexity} \\ \hline 
         Fourier Transform (FT)&  Stationary&  Frequency analysis& Low\\ \hline 
         Power Spectral Density (PSD)&  Stationary&  Spectral analysis& low\\ \hline 
         Hilbert Transform (HT)&  Stationary/nonstationary&  Envelope and  instantaneous phase/frequency analysis& low\\ \hline 
         Short-Time-Frequency Transform (STFT)&  nonstationary&  Fixed-resolution time-frequency analysis& moderate\\ \hline 
         Wavelet Transform (WT)&  nonstationary&  Multi-resolution time-frequency analysis& moderate\\ \hline 
         Hilbert-Huang Transform (HHT)&  nonstationary and nonlinear &  Adaptive time-frequency analysis& High\\ \hline
         Wigner-Ville Distribution (WVD)& nonstationary& High-resolution energy-time-frequency analysis&High\\
         \hline
         Wavelet Decomposition (WD)&  nonstationary&  Signal decomposition& moderate\\ \hline 
         Adaptive Mode Decomposition (AMD)&  nonstationary and nonlinear&  Signal decomposition& High\\ \hline    
    \end{tabular}
\end{table*}

\section{A Typical Signal-Based ML Pipeline}
\begin{figure*}[!htbp]
\centerline{\includegraphics[width=0.65\textwidth]{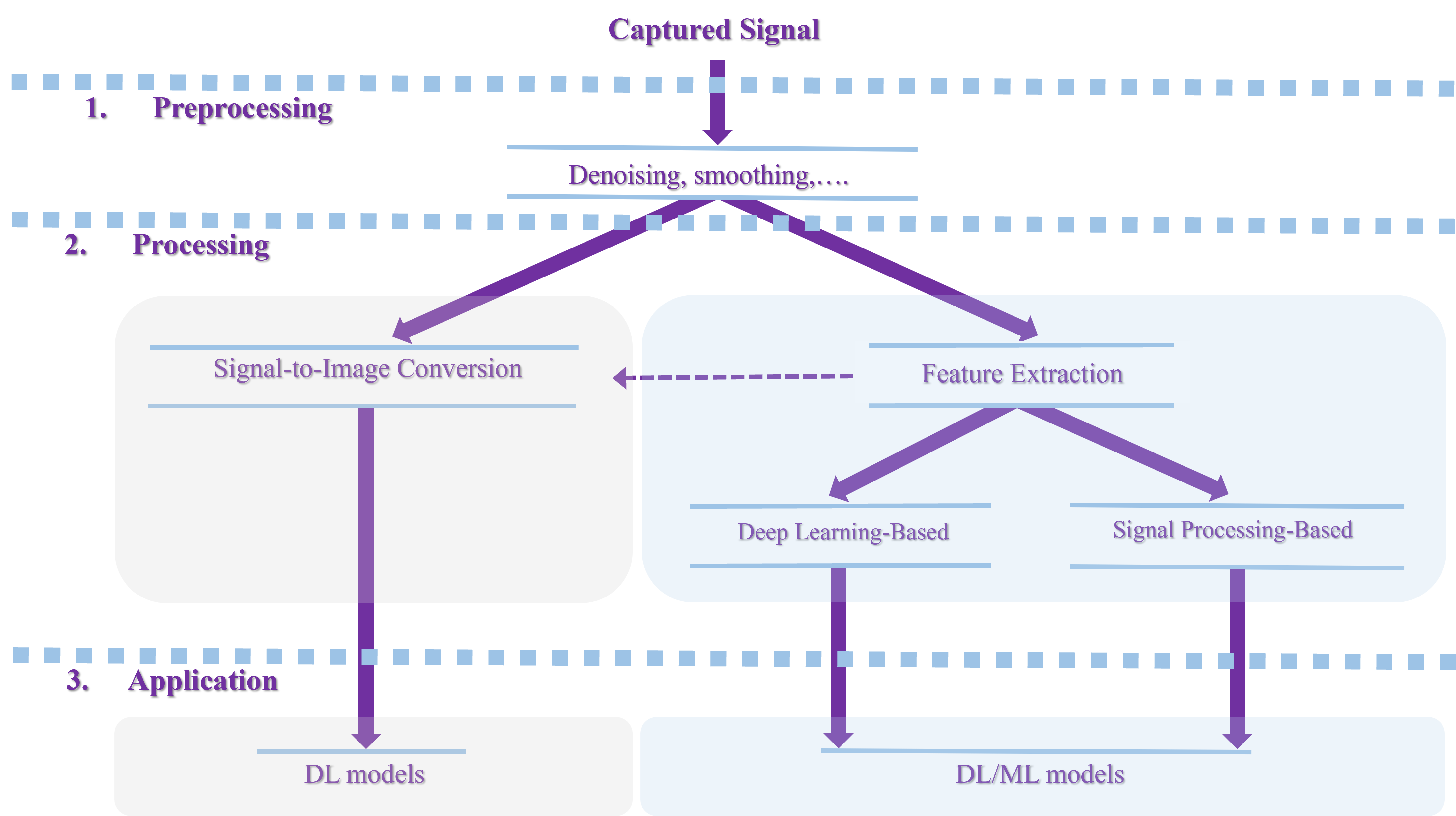}}
\caption{Overview of a typical signal-based ML framework.}
\label{framework}
\end{figure*}

In signal-based ML applications, the pipeline of signal processing is typically divided into three core stages: preprocessing, processing, and application, as illustrated in Fig. \ref{framework}. This section addresses each stage, highlighting its purpose and main aspects.

\subsection{Preprocessing} 
The first stage encompasses the preprocessing of acquired signals. Here, it is essential to distinguish between data preprocessing techniques\cite{ta21}, commonly used in ML models, and signal preprocessing, which involves the tasks of signal smoothing, signal denoising, and signal segmentation. Signal preprocessing is performed on the raw signals, whereas data preprocessing techniques, such as normalization and scaling, are applied to the extracted features.

\subsubsection{Signal Smoothing}
The acquired signal may exhibit rapid amplitude changes between successive samples. Such random changes in amplitude can negatively impact application-level performance. Furthermore, during measurements, the signal is susceptible to outlier samples caused by external factors such as instrument malfunction \cite{rm21}. Signal smoothing attempts to remove such impairments in the signal by adjusting the amplitudes of individual samples with respect to the amplitudes of adjacent samples. Smoothing acts as an approximation filter that works on $N$-successive samples at a time and outputs $N$ smoothed samples. Common smoothing methods involve the well-known moving average (MA) filter and the Savitzky–Golay (SG) filter \cite{ss64}. MA filter offers a simple method to smooth the signal by averaging neighboring data points while preserving the signal's shape. It operates by averaging a set of signal data points (window ) and sliding this window across the signal to smooth out short-term fluctuations.
The SG filter uses a different approach to smoothing the signal by performing a local polynomial regression on the signal data points present in the moving window to provide more nuanced smoothing that preserves the original shape and characteristics of the signal. The design of an SG filter involves properly selecting the window size and the order of the polynomial \cite{ch20, ch21, ch22, ch23}. Signal smoothing helps refine the signal's waveform shape and captures its main patterns, particularly useful in many signals-based applications. However, as a lossy operation,  smoothing could cause a notable distortion in applications where peak-related features are of main interest. More information on signal smoothing can be found in \cite{ca22, ar22, ck20, rk18}.  

\subsubsection{Signal Denoising}
Denoising refers to the process of removing noise from the acquired signal or, in other words, reconstructing the desired signal from its acquired noisy representation. This can be expressed mathematically as follows:
\begin{equation}
w\left ( t \right )=s\left ( t \right )+n\left ( t \right )    
\end{equation}
where $w\left ( t \right )$ denotes the noisy signal, $s\left ( t \right )$ is the desired signal, and $n\left ( t \right )$ is the noise component, Noise can be defined as any unwanted signal present in the measurement process other than the desired signal \cite{nh88}. Signal denoising and signal smoothing are often treated synonymously in the literature \cite{ab97}, as both aim to refine the signal's waveform while preserving meaningful patterns within the signal. However, they have different impacts on the noise and frequency contents of the signal. Smoothing is typically applied in the time domain on a sample-by-sample basis while denoising attempts to remove the whole noise component, $n\left ( t \right )$. 

Furthermore, in smoothing, the rapid fluctuations can be viewed as highly oscillating components of small amplitudes compared to the whole signal over a relatively long duration. From a frequency point of view, such oscillations represent high-frequency components of low amplitudes in the signal's spectrum. Smoothing, therefore, acts as a low-pass filter that removes these high-frequency components from the signal. On the other hand, signal denoising typically involves the use of various signal processing methods to develop more advanced noise filtering techniques. Signal denoising is addressed with more details in Section \ref{nosie}.

\subsubsection{Signal Segmentation}
Depending on the length of the acquired signal, expressed in terms of the number of samples, it might be necessary to divide the signal into smaller segments for the following practical considerations:
\begin{itemize}
    \item Online deployment: Online applications require working on a pre-defined length of the input signal.
    \item Real-time and delay-sensitive applications: Given the sampling frequency, $f_s$ in samples per second, the time duration, $t_s$, of the segment, expressed in \textit{seconds} equals:
    \begin{equation}
        t_s = \frac{N_o}{f_s}
    \end{equation}
    Hence, in real-time and delay-sensitive applications, the number of samples, $N_o$, in the input segment is of paramount importance since time delay increases as a function of $N_o$ \cite{aat22}.
    \item Computation burden: The computational complexity of signal-based ML systems increases as a function of $N_o$ \cite{aat22}.
    \item Effectiveness of extracted features: The number of samples, $N_o$, in the input segment significantly shapes the discriminative characteristics of the extracted features. Shorter segments may not provide an adequate representation of the signal's characteristics, reducing the reliability of extracted features. Conversely, very long segments could incorporate a high redundancy level, weakening the discriminative capability of the extracted features. 
\end{itemize}
Constant segmentation is commonly used in various applications where the segment length, $N_o$, is kept constant. Consequently, signals are divided into smaller segments of equal lengths. Determining the value of $N_o$ is application-specific and typically relies on achieving good trade-offs among the aforementioned practical considerations. Constant segmentation is commonly implemented using a sliding window of size $N_o$. Another aspect of constant segmentation is the percentage of overlap between successive segments. Introducing overlap between segments increases the number of segments extracted from the signal. The more overlap, the more segments are extracted. However, increasing the overlap increases correlation or dependence among adjunct segments. In general, the extent to which overlap can be used and its percentage depends on the nature of the application and the signal processing techniques used \cite{daa20, ads19, vfm15} \

In some applications, the acquired signal exhibits random intervals of activity and inactivity, as observed in various sound and biomedical signals. In such scenarios, constant segmentation becomes unreliable, as it does not count for these inherent intervals. In such situations, adaptive segmentation \cite{nva23, ams22, acs21, ajn18, anp17, sbp16, mpi10} is commonly utilized to dynamically adjust the segment length, $N_o$. Adaptive segmentation exploits the underlying characteristics of the signal and adapts $N_o$ accordingly. Unlike fixed length segments, adaptive length segments are more discriminative because the signal's characteristics and inherent patterns change from segment to segment, allowing more discriminative features to be extracted. However, adaptive segmentation involves more computational burden compared to constant segmentation.

\subsection{Processing} 
The processing stage involves processing the resultant segments to extract appropriate features that serve as inputs to ML models. The methods and algorithms available in the literature can be broadly categorized into two main approaches: feature extraction and signal-to-image conversion. In feature extraction, signal processing or deep learning (DL) can be used to extract relevant features from the input segment. The next section comprehensively covers the topic of feature extraction. Signal-to-image conversion approaches \cite{vam22, sys19, ass22, fgc20, esa23, rgc23, aye23, sve20, cym19, oc2_8} transform 1-dimensional (1D) signal segments into 2-dimensional (2D) spatial representations, allowing to leverage DL models in signal-based applications. Gramian Angular Field (GAF) and Markov Transition Field (MTF) \cite{vam22} are two well-known techniques that are commonly used to transform time series into images. Advanced techniques involve the use of signal processing to generate 2D time-frequency representations of the signal segments. The essence of these techniques is to represent time-frequency-energy characteristics of the signal and features as width-height-color intensity, mimicking image features. However, to serve training and testing purposes of DL models, a large number of transformed images should be available. Nevertheless, signal-to-image conversion is especially feasible in integrated systems such as unmanned aerial vehicles (UAVs) and autonomous vehicles (AVs) \cite{oc2}. In these systems, computer vision applications, such as object detection and navigation, and signal-based applications are deployed within the same environment. Therefore, it is deemed computationally efficient to convert signals into images and use transfer learning approaches that utilize the trained deep-learning infrastructure for inference tasks of signal-based applications. Moreover, signal-to-image conversion and transfer learning offer a practical solution when the size and type of available signal data are insufficient to fulfill training and testing requirements.

\subsection{Application} 
The application is the final stage of a typical signal-based ML pipeline. As a task-specific stage, it uses appropriate ML models \cite{oc2_0, oc2_1, oc2_2} to address tasks, including classification, clustering, and anomaly detection. Feature preprocessing, hyperparameter tuning, and relevant metrics \cite{oc2_3, oc2_4} are often used to improve and evaluate model performance.\
Table \ref{table1} serves as a high-level overview of a typical signal-based ML framework, highlighting each stage's purpose, inputs, functions, and outputs.
\begin{table*}[!htbp]
    \centering
\caption{Inputs, functions, and outputs of each stage of typical signal-based ML framework}
\label{table1}
    \begin{tabular}{|c|c|c|c|} \hline 
         \textbf{Stage}&  \textbf{Inputs}&  \textbf{Functions}& \textbf{Outputs}\\ \hline 
         \textbf{\textit{Preprocessing}}&  Acquired signal+impairments (noise / outliers)&  Smoothing / Denoising / Segmentation& Segmented signal\\ \hline 
         \textbf{\textit{Processing}}&  Segmented signal&  Feature extraction / Signal-to-image conversion& Features/ Images\\ \hline 
         \textbf{\textit{Application}}&  Features/ Images&  Classification / Clustering / Anomaly detection& Results \\ \hline
    \end{tabular}
\end{table*}

\begin{figure*}[!htbp]
\centerline{\includegraphics[width=0.9\textwidth]{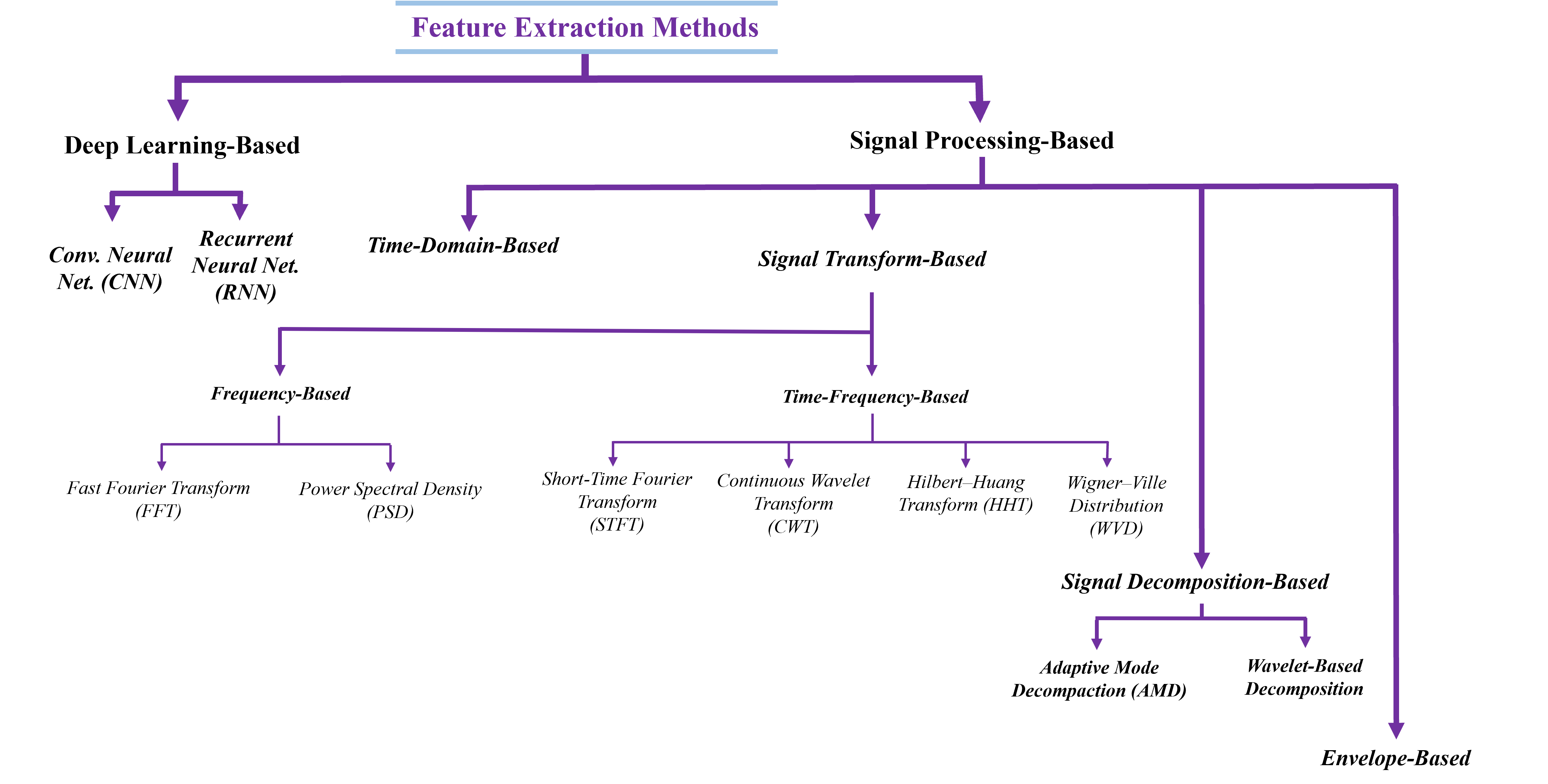}}
\caption{Taxonomy of feature extraction techniques.}
\label{taxonomy}
\end{figure*}

\section{Feature Extraction Methods}
This section provides a comprehensive review of existing feature extraction methods. The review utilizes a new hierarchical taxonomy, depicted in Fig. \ref{taxonomy}, to categorize and structure various methods that exist in the literature. The taxonomy starts by grouping the existing methods under two main categories: DL-based and signal processing-based. This categorization distinguishes between two mechanisms that exist for feature extraction: feature \textit{learning} and feature \textit{engineering}. The first mechanism relies on DL models--- such as convolutions neural networks (CNN), recurrent neural networks (RNN), and auto-encoders--- to \textit{learn} high-level feature representations from signal data \cite{dl1,dl2, dl3, dl5, dl6, dl7, dl8, dl9, dl10, dl11}. The second mechanism, on the other hand, employs signal processing to extract distinctive features from the signal, deliberately \textit{engineered} to emphasize specific attributes and reveal certain characteristics. This deliberate engineering allows meaningful connections to be made between the resulting features and various conditions or classes that exist within the signals. In contrast, DL features are not interpretable since DL models are black-box models, making establishing such meaningful connections impossible. In addition, features extracted by DL models are model-dependent, whereas features based on signal processing are solely signal-dependent. This implies that variations in the model's architecture, parameters, training, or tuning lead to different feature representations and, consequently, inconsistent performance results. Thus, it can be deduced that in DL-based approaches, performance is model-dependent, while in signal processing-based approaches, performance is mainly feature-dependent. Moreover, DL features would exhibit higher dimensionality and redundancy than signal processing-based features. This requires the use of appropriate  dimensionality reduction and feature selection techniques\cite{dl5, dr1,dr2}, thereby increasing the online processing time and computational burden. Table \ref{table2} summarizes the aforementioned comparison between signal processing-based and DL-based approaches.
\begin{table*}
    \centering
\caption{A comparison between deep learning-based and signal processing-based feature extraction}
\label{table2}
    \begin{tabular}{|r|c|c|} \hline 
         \textbf{\textit{Aspect}}&  \textbf{Deep Learning Approaches}& \textbf{Signal Processing Approaches}\\ \hline 
         \textbf{\textit{Feature Extraction Mechanism}}&  Feature learning& Feature engineering\\ \hline 
 \textbf{\textit{Interpretability of Extracted Features}}& Not interpretabile&Interpretabile \\\hline
 \textbf{\textit{Size of  Required Dataset}}& Large&Small-to-medium\\ \hline 
         \textbf{\textit{Data Dimensionality and Redundancy  in Extracted Features}}&  High, depends on DL model& Low-to-medium\\ \hline 
 \textbf{\textit{Computational Cost}}& High&Moderate\\ \hline
    \end{tabular}    
\end{table*}\
A considerable amount of work in the literature uses hybrid approaches where signal processing-based features are fed into DL models to further extract highly distinctive representations from these features \cite{dl&sp1, dl&sp2, dl&sp3, dl&sp4, dl&sp5, dl&sp6,dl&sp7, dl&sp8, dl&sp9, dl&sp10, dl&sp11, dl&sp12, dl&sp13, dl&sp14}. Compared to DL-based and signal processing-based features, hybrid features have higher discriminative power, which generally leads to better performance. However, employing hybrid approaches involves increased online processing time and higher computational complexity, making practical deployment subject to trade-offs between system complexity and performance requirements. In general, hybrid approaches prove feasible in applications that involve complex signals, such as radio modulation recognition \cite{dl&sp9, dl&sp10, dl&sp11, dl&sp12, dl&sp13, dl&sp14}. In these applications, signals are randomly modulated, corrupted with high noise levels, and subject to various types of interference.\

Focusing on signal processing-based feature extraction, the taxonomy of Fig. \ref{taxonomy}, from a computational point-of-view, divides the existing methods into four main categories: time domain-based methods, transform-based methods, decomposition-based methods, and envelope-based methods. Time domain-based methods extract features directly from time domain waveforms of the signal without further processing. On the other hand, transform-based, decomposition-based, and envelope-based methods use additional steps to process the signal and, consequently, additional computations to extract features. The remainder of this section reviews common methods that exist within each category.

\subsection{Time Domain-Based Methods}
In time-domain-based methods, features are calculated from the signal's amplitude, representing specific aspects of the signal's dynamics over its time period. This allows for the quantification of amplitude variations from one period to another. Conventional methods \cite{td1, td2, td3, td4, td5, td6, td7, td8, td9, td10, td11, td12, td13, td14, td15, oc2_9} involve extricating shape characteristics--- such as maximum, peak-to-peak value, and crest factor--- and statistical properties of the input signal. Statistical features describe characteristics of the probability distribution of the input signal; common statistical features involve mean, root-mean-square (RMS) value, variance, and high-order statistics (HOS), such as skewness and kurtosis. Compared to first and second-order statistics (mean and variance), HOS features have more distinguishable characteristics as they can identify more complex aspects of the input signal, such as the shape of its distribution. Furthermore, HOS are less sensitive to noise. \
Besides shape-based and statistical methods, time-domain entropy methods \cite{td_en1, td_en2, td_en3, el20} are commonly utilized for feature extraction. Entropy methods rely on the observation that signals belonging to different classes or conditions typically exhibit different levels of irregularity or dynamic changes depending on the underlying state. Hence, different entropy measures can be utilized to quantify the irregularity of the input signal. In contrast to statistical features, entropy-based features are more sensitive to changes and complexities that arise within the signal. Thus, they can capture variations and irregularities in the signal that may be difficult to detect using traditional statistical measures. Additionally, entropy-based features are resilient to noise and outliers. \
In general, time-domain feature extraction approaches work directly on input signals, making them conceptually straightforward and relatively easy to implement. Further, they are computationally efficient since no further processing steps are required, making them advantageous in applications where real-time processing is crucial, especially with limited computational resources. However, time-domain analysis is highly susceptible to noise since noise in the signal would alter its amplitude and mask the dynamic characteristics of the signal. Furthermore, to achieve reliable performance, input segments of relatively long duration are usually required to capture the changes and complexities that evolve within the signal over time.

\subsection{Transform-Based Methods}
Transform-based approaches use advanced techniques to convert signals from the time domain to the frequency domain or time-frequency domain. This allows for the identification of frequency and frequency-temporal information, facilitating robust signal analysis and feature extraction.

\subsubsection{Frequency-Based Methods}
Frequency-based approaches represent signals in terms of their frequency content, revealing details that are not apparent in the time-domain waveform. Further, in contrast to time-domain analysis, frequency-based analysis allows the isolation and removal of noise or unwanted components by applying appropriate frequency filtering mechanisms to improve the signal-to-noise ratio and improve reliability of the extracted features. Frequency-based approaches involve the application of the Fourier transform (FT) to the input signal to compute its frequency content, which can be broadly categorized into two main categories: spectrum-based \cite{fd6, fd7, fd9, fd10, fd12, td2, fd1, fj14, fd4, fd5, fd8, fd11, fd13, fd14, fd15, fd16} and spectral density-based \cite{fd13, psd1, psd2, psd3, psd4, oc2_7}. Spectrum-based approaches use the frequency spectrum of the signal for feature extraction. Specifically, Some approaches directly use resultant Fourier coefficients of the input segment as input features \cite{fd6, fd7, fd9, fd10, fd12} since these coefficients describe the distribution of the signal's energy over the range of frequencies contained in the signal. Here, dimensionality reduction and feature selection techniques are often used to reduce the size of coefficients and select the most discriminative coefficients. 

Other approaches involve the extraction of various frequency-domain features from the frequency spectrum, such as statistical properties, energy,  entropy, and correlation coefficients\cite{td2, fd1, fj14, fd4, fd5, fd8, fd11, fd13, fd14, fd15, fd16}.  Spectral density-based approaches, on the other hand, utilize the power spectral density (PSD) of the input signal to extract the features \cite{fd13, psd1, psd2, psd3, psd4, oc2_7} since it provides a normalized measure of the power per unit frequency compared to the frequency spectrum. Entropy is frequently utilized in PSD analysis to calculate Spectral Entropy (SE) \cite{psd2, psd3}. SE is a very useful spectral feature that measures the irregularity or randomness of power distribution across signal frequencies and, hence, can be used to quantify the spectral complexity of the input signal. Other methods utilize Mel-frequency cepstral coefficients (MFCCs) as input features \cite{mfcc1, mfcc2, mfcc3, mfcc4, mfcc5, mfcc6, mfcc7, mfcc8, mfcc9, mfcc10} because they effectively capture the shape of the power spectrum. MFCCs are originally developed for audio signals \cite{mfcc} and are gaining attraction in other various applications as well. A thorough explanation of MFCCs, including computational steps and usage scenarios, can be found in \cite{mfcc}. Higher-order spectra \cite{spectra1, spectra2, spectra3, spectra4} are commonly used in frequency-domain analysis to conduct a more comprehensive spectral analysis compared to second-order spectral analysis. Common higher-order spectra include bispectrum and trispectrum \cite{hospectra1, hospectra2, hospectra3, hospectra4, hospectra5, hospectra6,  hospectra7, hospectra8, spectra2, hospectra10, hospectra11}, which correspond to the FT of the signal's third-order cumulant and the fourth-order cumulant, respectively. 

A key advantage of higher-order spectra over power spectrum is their ability to retain phase information \cite{hos1}. Specifically, the power spectrum, representing the second-order spectrum or the FT of the autocorrelation function, captures the magnitude of frequencies present in a signal but discards phase information. This limitation makes it challenging to identify phase-related properties of the signal, such as frequency coupling or interactions between different frequency components. In contrast, higher-order spectra provide insight into the phase relationships among frequency components by examining moments or cumulants of the signal beyond the second order, enabling the detection of nonlinearities, phase coupling, and other characteristics that are invisible in the power spectrum.  However, higher order spectra are more complex to calculate and interpret than traditional power spectral analysis because they require more data for reliable estimation and involve more sophisticated mathematical and computational methods. Furthermore, due to their multi-dimensional nature, visualisation and interpretation of higher order spectra can be challenging. Because of this complexity, specific cross sections or slices of the bispectrum or trispectrum are often analysed to extract meaningful information. This approach simplifies the analysis by reducing the dimensionality of the data, making it easier to visualise and interpret. \

While frequency analysis accurately distinguishes between different frequency components in the signal, the analysis spans the signal's entire duration, lacking the ability to provide temporal information about the timing of occurrence of these frequencies within the signal. Additionally, the FT is not ideal for analyzing signals that contain highly time-localized components, such as short bursts with high energy concentrations, because the components produce a wide range of frequencies in the frequency spectrum due to the inherent uncertainty principle associated with Fourier analysis \cite{ar14}.

\subsubsection{Time-Frequency-Based Methods}
In contrast to frequency-domain analysis, which lacks time resolution, time-frequency transforms produce representations that map the signal's energy across both time and frequency, allowing for a localized analysis in the time-frequency domain. This approach allows the identification of transient oscillatory components within the signal, providing a more complete understanding of its dynamics. Time-frequency transforms that are commonly used in the literature for feature extraction include short-time Fourier transform (STFT), \cite{tf1, tf2, tf5, tf6, tf7, tf8, tf9, tf10, tf11, tf12, tf13, tf19, tf20} Hilbert-Huang transform (HHT) \cite{tf3, tf14, tf21, tf22, tf23, tf31,  tf32,  tf33,  tf34,  tf35,  tf37}, Wigner-Ville distribution (WVD) \cite{tf17, tf18, tf28,tf29,tf30, tf38, tf39, tf40,tf41}, and continuous wavelet transform (CWT) \cite{tf4, tf7, tf15, tf16, tf24, tf25, tf26, tf27, tf39, tf42, tf43, wd1, tf44}. STFT is efficient in analyzing nonstationary signals whose spectral properties vary over time. However, it uses a fixed segment length in analyzing the signal, imposing a trade-off between time-domain and frequency-domain resolutions. In CWT, on the other hand, time-frequency analysis is based on a family of wavelets generated from a single mother wavelet through scaling and translation, offering a multi-resolution analysis of the signal at different frequencies (scales) and time intervals (translations). Moreover, in contrast to STFT, which uses complex exponentials as basis functions--- that extend infinitely in the time domain---, wavelet base functions are localized in time and frequency, making them more suitable for signals with transient or highly localized components. The HHT uses an adaptive approach for time-frequency analysis. Unlike STF and CWT, HHT does not impose a fixed basis function on the signal analysis, but instead uses an adaptive approach to analyze the signal.. This adaptivity makes it highly responsive to variations in the signal, facilitating very effective analysis of nonlinear and nonstationary signals. However, as an adaptive analysis method, its performance depends heavily on two factors: the stopping criterion for the sifting process \cite{it14} and the choice of the interpolation method for envelope estimation \cite{aa23}. 

A proper stopping criterion ensures that the decomposition process neither overfits nor underfits the signal, thereby preserving the signal's essential characteristics without introducing artifacts. On the other hand, the interpolation method for envelope estimation determines how well the upper and lower envelopes capture the actual oscillatory modes of the signal, heavily influencing the HHT's performance. The WVD  offers a time-frequency representation with high energy concentration, facilitating detailed analysis of a signal's energy distribution over both time and frequency domains. The high concentration of energy arises from the quadratic nonlinear structure of WVD compared to STFt and CWT. This energetic and correlative nature of the WVD \cite{lh92} makes it very useful in analyzing single-component signals. However, the quadratic nature of the WVD represents a major challenge when analyzing multi-component signals due to high cross-term interference \cite{ew17} caused by undesired cross-correlation between various signal components, leading to a false indication of non-existent signal components in the resultant time-frequency representations of the signal. Window-based WVD \cite{tf98, sf84} and kernel-based approaches \cite{ic89, tz90} are commonly used to reduce the cross-terms at the cost of some loss in resolution. \

From a feature extraction perspective, time-frequency-based methods can be grouped under three main approaches in which the generated energy-time-frequency mappings are treated differently: 
\begin{enumerate}
    \item Transformation coefficients as input features: In these approaches, the generated transform energy coefficients, representing signals' energy distribution across time-frequency instants, are used directly as input features for the ML model \cite{tf1, tf2,  tf3, tf4, tf31, tf34, tf35, wd1}. Feature selection techniques are commonly used here to select the most distinctive representations, which helps reduce redundancy and relaxes computational requirements for training and inference.
    \item Energy-time-frequency 2D heatmaps as imagery data: The generated energy-time-frequency representations are used to visualize 2D heatmaps that map the signal's energy across time-frequency instants. Accordingly, computer vision and DL techniques are leveraged where these heatmaps serve as input images \cite{tf5,tf6, tf7, tf8, tf9,  tf10, tf11, tf12, tf13, tf14, tf15, tf16, tf17, tf18, tf32, tf38, tf39, tf41, tf42, tf43, tf44}.
    \item Transformation coefficients as signal representations:  In these approaches, the generated mappings are treated as transformed representations of the signal. Subsequently, these representations are subjected to further signal-processing techniques to extract features of interest, such as entropy, energy, and statistical properties \cite{tf19,tf20, tf21, tf22, tf23, tf24, tf25, tf26, tf27,tf28, tf29, tf30, tf33, tf37, tf40}. Compared to the other two approaches, these methods generally result in a smaller number of features, making them more suitable for applications with low computational requirements.
\end{enumerate}\
Spectral kurtosis (SK) \cite{sk1, sk2} is a popular time-frequency analysis tool in signal processing that is commonly used to locate harmonics, transients, and repetitive impulses in the frequency domain. SK is based on computing the normalized fourth-order moment (kurtosis) of the STFT of the signal at each frequency bin, providing a statistical measure of the signal's characteristics at each time-frequency instant. Essentially, SK quantifies the "peakedness" or impulsiveness of the distribution of the power at each frequency bin. One of its primary applications is the detection of transient components in signals. Transients, due to their short and sharp nature, create distinct, non-Gaussian patterns in the signal's frequency content. When these features are present, the kurtosis values in the affected frequency bins will be significantly higher than those of normal behavior. Accordingly, high SK values can indicate the presence of impulses or bursts, as they cause a deviation from the Gaussian distribution typically observed in stable operating conditions. This is particularly useful in condition monitoring and fault diagnosis of rotating machinery and mechanical systems \cite{sk3, sk4, sk5}, where transient vibrations often indicate faults. Additionally, SK is an effective tool for distinguishing between harmonic components--- which often manifest as consistent, periodic components--- and noise in the signal. These harmonics will often show higher kurtosis values, indicating a non-Gaussian distribution, possibly due to the periodic and repetitive nature of the harmonics. In contrast, Gaussian noise tends to have a kurtosis value close to 0, indicating a Gaussian distribution.

\subsection{Signal Decomposition-Based Methods}
Signals generated by physical systems typically consist of multiple components or modes of time-varying nature in amplitude, phase, and frequency. These components contain meaningful information about the underlying structure of the signal. Decomposition techniques attempt to decompose the signal into elementary modes, enabling signal analysis at the level of its constituent components. This, in turn, allows for the extraction of highly distinctive features, that are closely related to the inherent structure of the signal. Common decomposition approaches include wavelet-based decomposition and adaptive mode decomposition (ADM) methods. In wavelet-base decomposition methods  \cite{aat22, wd1, wd2, wd3, wd&amd1, wd4, wd5, wd6, wd7, wd&amd2, wd8, wd10, wd11, wd12, wd13, wd14, wd15, wd16, wd17, wd18, wd19, wd20, wd21}, discrete wavelet transform (DWT), stationary wavelet transform (SWT), and wavelet packet transform (WPT) are commonly used to decompose the signal into elementary modes of high and low-frequency components using digital filter banks. The performance of wavelet decomposition is highly dependent on the base wavelet function and the level of decomposition. AMD methods \cite{amd1, wd&amd1, wd&amd2, amd2, amd3, amd4, amd5, amd6, amd7, amd8, amd9, amd10, amd11, amd12, amd13, amd14, amd15, amd16} utilize an empirical decomposition approach where the signal is adaptively decomposed into a set of simpler functions known as intrinsic mode functions (IMFs).  Hence, in contrast to wavelet-based methods, AMD  methods use an adaptive mechanism rather than a priori basis function to decompose the signal, leading to more 'natural" modes that are not influenced by a priori basis function \cite{ag17}. Moreover, the ability of AMD methods to adapt to varying signal characteristics makes them effective when dealing with nonlinear and nonstationary signals. However, this adaptability comes at the cost of higher computational complexity. Furthermore, AMD methods are generally sensitive to noise because they rely on the extrema and minima of the signal for signal decomposition. Additionally, they suffer from inherent limitations, such as end effect \cite{ez23} and mode mixing \cite{ag08}. Various methods have been introduced to address these limitations, such as ensemble empirical mode decomposition (EEMD) \cite{ew09} and variational mode decomposition (VMD)\cite{vd14}, local mean decomposition \cite{ts05}, and empirical wavelet transform (EWT) \cite{eg13}.\

In the existing literature, signal decomposition is utilized through two approaches for feature extraction; each approach serves a distinct objective:
\begin{enumerate}
    \item Extraction of highly discriminative features: In these methods, features are extracted from elementary modes resulting from the decomposition process. Accordingly, feature extraction methods can be generally grouped under four main categories: entropy-based \cite{wd16, wd17, wd21, amd6, amd7, amd16}, energy-based \cite{ wd5, wd11, wd18, wd19}, spectral-based \cite{aat22, wd6, wd20, amd5, amd8, amd13, amd15, amd16}, and statistical-based \cite {wd2, wd4, wd&amd1, wd5, wd7, wd&amd2, wd14, wd15, amd8, amd16}. Entropy as a measure of uncertainty, is used to quantify the irregularity within the decomposed modes. Energy-based approaches utilize the changes in energy content of the decomposed modes to extract the features. Spectral-based approaches use spectral characteristics of the elementary modes to extract the features. Statistical-based approaches rely on time-domain properties of the modes such as skewness, kurtosis, Root-Mean-Square (RMS), and crest factor for feature extraction.
    \item Obtaining a low-redundancy (highly informative) version of the composite signal \cite{wd1,amd1, wd3, wd8, wd10, wd12, wd13, amd2, amd3, amd4, amd9, amd10, amd11, amd12, amd14 }: This is accomplished by decomposing the signal into elementary modes and applying further processing to examine these modes based on predefined ranking criteria. The highest-ranked mode is thus selected and various feature extraction techniques can be applied to the selected mode accordingly. Here, the selected mode represents a low-redundancy and high-informativeness version of the original signal.
\end{enumerate}
An important aspect of the first approach is that it provides a convenient way to construct a feature vector of controllable size and highly discriminative nature \cite{aat22}, which is particularly helpful in applications with limited computational resources. This is facilitated by decomposing the signal into a finite number of modes and extracting a few features from each mode. For instance, decomposing the input segment of a signal into $8$ modes and taking the energy or entropy of each mode as a feature leads to a feature vector with a size of $1\times8$ distinctive features. It is also less computationally intensive than the second approach, making it more suitable for real-time processing and applications with limited processing capabilities. However, by focusing on the highest-ranked mode, the second method ensures that the extracted features are highly informative and less redundant. Moreover, the capability to rank and select modes based on predefined criteria allows for dynamic screening that can be automatically adapted according to the nature of the input signal, which becomes particularly useful in applications that involve multiple signals from various sources, such as advanced diagnostic systems and integrated condition monitoring applications. Additionally, selecting the most informative mode can result in more accurate and relevant feature extraction, potentially improving the performance.

\subsection{Envelope-Based Methods}
Envelope-based feature extraction methods \cite{env1, env2, env4, env5, env6, env7, env8, env9, env10, env11, env12, env13} primarily focus on analyzing the envelope of the signal to reveal essential characteristics of its amplitude, such as statistical properties, peak frequency, energy, and entropy. In these methods, The HT is commonly used to obtain the signal's envelope, which is then subjected to various feature extraction techniques. Envelope-based feature extraction is particularly useful in applications where amplitude modulation (AM) is a key characteristic of the signal; that's it, the amplitude of the signal is varied or altered in response to specific factors that are of interest, for instance, impacts generated by defects in rolling bearings \cite{env14}; in such situations, envelope-based analysis can effectively extract meaningful features. Generally speaking, as an amplitude-focused approach, envelope analysis offers limited frequency information about the signal. Furthermore, its effectiveness can vary significantly depending on the nature of the signal, as not all signal types are amenable to envelope-based analysis. Additionally, it is highly susceptible to noise that contaminates the signal's amplitude such as impulsive noise, making it difficult to extract effective features from the envelope \cite{env13}. \

The selection of an appropriate feature extraction method for a given signal depends on several factors, including the nature of the signal, the specific requirements of the application, and the available computational resources. For stationary signals, where the primary interest lies in frequency analysis, Fourier analysis provides a highly efficient approach to identifying dominant frequency components within the signal. In scenarios involving comparative spectral analysis of multiple signals that vary in length and/or bandwidth, PSD is particularly advantageous since it provides a normalized spectral density measure, facilitating effective comparison.  On the other hand, signals exhibiting time-varying spectral characteristics necessitate the use of proper time-frequency analysis for effective feature extraction. The STFT offers a computationally efficient tool for time-frequency analysis of the signal. However, its fixed resolution across all frequencies reduces its effectiveness for applications requiring variable-resolution analysis. In such contexts, wavelet analysis presents a practical solution due to its ability to perform multi-resolution analysis through scaled and shifted versions of a wavelet base function. For analyzing nonstationary and nonlinear signals characterized by rapidly changing components, the HHT provides an adaptive and more robust approach than Wavelet analysis but with more computational requirements. The WVD provides a practical approach for achieving a joint energy-time-frequency representation of the signal, although it is computationally intensive compared to the other transforms. It is particularly useful in high-resolution analysis scenarios or where energy concentration is a focal requirement. Signal decomposition techniques enable signal analysis at the level of its constituent components. This, in turn, allows for the extraction of highly distinctive features of a controllable size that are closely associated with the inherent structure of the signal. The envelope-based analysis is especially effective in applications where the envelope characteristics are of primary interest. 

\section{Challenges and Potential Solutions}
The practical deployment of signal-based ML systems faces significant challenges that can impede the efficiency and effectiveness of the system. Key among these challenges are the detrimental effects of noisy environments on signal integrity, the limited availability of abnormal samples, and computational complexity. These challenges require effective solutions to mitigate their effects and enhance signal processing techniques. This section delves into each of these challenges, exploring their implications and highlighting potential solutions.

\subsection{Effects of Noise } \label{nosie}
In real-world systems, signals are nonstationary, nonlinear in nature, and subject to noise \cite{ch1}. From a measurement perspective, noise can be defined as any unwanted signal present in the measurement process other than the desired signal \cite{ch2}. In signal processing, noise generally refers to any unwanted or random variation in the signal that can obscure or distort the intended information. It is an inevitable part of signal processing, particularly in machine learning applications where it can hinder feature extraction and significantly reduce its discriminative power. 
  
\subsubsection{Noise characteristics}
Noise, originating from various sources that are typically present in all measurement environments can be grouped into three main categories \cite{ch2}: \begin{enumerate}
    \item Intrinsic noise that arises from random fluctuations within measurement physical systems such as thermal and shot noise.
    \item Man-made noise sources such as motors, power lines, radio broadcasts, oscillators, laboratory equipment, and cellular phones.
    \item Noise caused by natural disturbances such as lightning and sunspots.
\end{enumerate}
Understanding the characteristics of noise and how it affects the desired signal is the first step toward an effective noise removal approach. Noise is typically modeled and characterized by various aspects such as spectrum, mode of occurrence, and statistical distribution.\newline 
Noise Spectrum (Noise Colors): Color of noise is a spectral-based description of noise where different “colors” of noise are classified according to their power spectral densities. The most important and common type in the context of signal processing is white noise. A defining characteristic of this type of noise is its flat power spectral density (PSD). Thus, it has a constant power at each frequency, affecting all frequencies in the desired signal equally. Pink noise, or $\frac{1}{f}$ noise, has a PSD  that decreases proportionally to the inverse of its frequency, making it more dominant at lower frequencies in the desired signal. In contrast, blue noise mainly impacts high-frequency comments in the signal since its PSD increases proportionally to the frequency. Purple noise has more energy at higher frequencies compared to blue noise since its PSD increases with frequency at a much faster rate than that of blue noise.\newline
Occurrence Nature: The presence of noise in the signal depends on the nature of the occurrence of noise. Continuous noise is present continuously during signal acquisition, persistently contaminating the signal over its entire duration. The severity of continuous noise is characterized by its amplitude (power); it is related to the signal power by the signal-to-noise ratio (SNR), which provides a measure to evaluate the severity of the noise and the performance of the signal denoising technique employed. Intermittent noise, in contrast, occurs at irregular intervals. Burst noise, on the other hand, is characterized by sudden occurrences, often of short durations. Impulse noise consists of random noise pulses. Compared to burst noise, impulse noise has extremely short durations. Burst and impulse noises are characterized by their large amplitudes compared to the desired signal.\newline
Statistical Distribution: Noise can be described in terms of amplitude distribution. Gaussian noise, commonly called additive white Gaussian noise (AWGN), is a common noise model in signal processing. Characterized by its Gaussian distribution, the amplitude of AWGN can be completely described by its mean and variance, which represent noise power. AWGN is commonly modeled as a zero-mean Gaussian process, maintaining the mean of the desired signal unchanged. As a white continuous noise, AWGN affects the spectral contents of the signal equally.

\subsubsection{Signal Denoising Techniques}
As mentioned earlier, the process of signal denoising involves eliminating noise while preserving the important features of the signal. The first step in signal processing-ML pipelines is to remove noise and outliers. Thus, the effectiveness of subsequent steps depends heavily  on the efficiency of the employed signal denoising technique. For impulse and burst noise, Order-Statistic Filtering (OSF) \cite{ch3}  is particularly effective; it is based on replacing each sample in the signal with a statistic (like the median, minimum, or maximum) calculated from the samples in its neighborhood, defined by a sliding window of a certain size that moves across the signal. The median filter \cite{ch4, ch5, ch6, ch7}, a specific type of OSF, is widely used to remove impulsive noise. OSF is highly robust to outliers such as impulse noise since order statistics are less sensitive to extreme values than linear operations such as averaging. Moreover, the median filter is highly effective in preserving signal edges (sudden changes in signal amplitude) while reducing noise, which is crucial for maintaining the integrity of the signal's features. The efficiency of the median filter is significantly influenced by the selection of the window size. A larger window size will include more samples, improving the ability to remove impulsive noise. However, a larger window could over-smooth the signal, potentially erasing important details. Further, it will increase the computational burden. Conversely, a smaller window size preserves more detail since fewer samples are involved in determining the median, but it may be less effective in removing noise as the median value will be influenced by the presence of noise in the window. Additionally, the performance of median filters is limited by the proportion of impulsive noise in the window. Filtering performance decreases as the proportion of impulsive noise increases \cite{ch8}. Specifically, with low levels of impulsive noise, the noise spikes are likely to be the minority within the window, and the median value computed within the window is likely to be a true signal value, effectively removing the noise. As the proportion of impulsive noise increases, there's a higher likelihood that the median value computed within the window will itself be a noise value, especially if the noise spikes constitute a considerable part of the samples in the window. In this case, replacing the original signal sample with the median does not remove the noise. Instead, it might even introduce additional distortion, especially if the median value is a noise spike. To overcome these problems, adaptive median filtering is commonly applied where the window size changes adaptively based on impulsive noise content \cite{ch8, ch9,  ch10}.\

For complex noise patterns such as intermittent and continuous noise, several techniques can be used to denoise the signal in the time domain, frequency domain, time-frequency domain, or deep learning (DL) latent space. Time-domain methods involve techniques that directly manipulate the signal in the time domain, aiming to identify and reduce noise components while preserving signal characteristics. Moving average (MA) filter \cite{ch12, ch13} is a simple yet effective method that smooths the signal by averaging neighboring data points. It is particularly useful for reducing simple random noise while preserving the signal's shape. Traditional MA filters operate by averaging a set of signal data points (referred to as the window or kernel) and sliding this window across the signal. This process smooths out short-term fluctuations, effectively reducing random noise. However, the primary limitation of a traditional moving average filter is its fixed window size, which may not be optimal for all signal conditions. Adaptive moving average  (AMA) filters  \cite{ ch14, ch15, ch16} address this limitation by dynamically adjusting the window size based on the characteristics of the input signal. Here, the window size changes in response to the variability or other statistical properties of the signal. This adaptivity is critical when dealing with nonstationary signals or when the noise characteristics are inconsistent throughout the signal. The design of the adaptive mechanism involves selecting proper criteria for window size adjustment. Common approaches include the use of local signal variance, rate of change, or other statistical measures that reflect the signal's nature.\
 
The Savitzky-Golay filter (SGF) \cite{ch17, ch18, ch19}  While it shares some conceptual similarities with MA filters, it provides a different approach to noise reduction and is particularly useful for preserving higher-order signal properties. Specifically, MA filters give equal weight to all points in the window, which can distort the features of the signal. On the other hand, SGF performs a local polynomial regression on the signal data points present in the moving window to provide a more nuanced smoothing that preserves the original shape and characteristics of the signal. This approach not only smooths the data but also preserves features of the data points, such as relative maxima, minima, and width, which are often obscured by noise. The design of an SGF involves the proper selection of the window size, and the order of the polynomial \cite{ch20, ch21, ch22, ch23}. The selection depends on the specific characteristics of the signal being processed and the nature of the noise. A higher polynomial order can better adapt to changes in the signal's trend, but it may also introduce noise if not chosen appropriately.\

Adaptive filters leverage adaptive noise cancellation (ANC) techniques to instantly isolate and remove unwanted noise from the signal, making them suitable for real-time applications such as autonomous driving, navigation, control systems, and robotics. ANC involves dynamically adjusting the filter parameters to minimize the difference between the desired signal (primary signal of interest) and the actual signal; the primary goal is to improve the SNR without distorting the primary signal. Common ANC techniques involve the least mean square (LMS) and recursive least square (RLS) adaptive filters \cite{ch24, ch25, ch26, ch27}, and Kalman filtering (KF) \cite{ch28,  ch29, ch30, ch31}. LMS adaptive filtering, a gradient-based method, attempts to minimize the mean square error between the desired and actual output of the filter. To facilitate the generation of an estimate of the unwanted noise, this process entails using a reference signal, highly correlated with the noise contaminating the primary signal and uncorrelated with the primary signal itself. The RLS algorithm provides a recursive solution to the least-squares problem, offering faster convergence than LMS. In contrast to LMS and RLM, which require a prior reference signal, KF estimates the state of the system, represented by the signal of interest, based on noisy measurements. The estimation is based on Bayesian estimation and state-space representation of linear systems \cite{ch32}, making KF particularly useful in real-time applications involving time-varying signals or systems with inherent uncertainties. KF consists of two main steps: prediction and update. In the prediction phase, KF predicts the next state of the system. In the update phase, the prediction is adjusted using the new measurement data, refining the system's state estimate. In the context of noise reduction, this process inherently filters out the noise from the measurements, isolating the primary signal. Standard KF assumes that noise is Gaussian and the system has linear dynamics. However, several extensions have been developed to handle more complex scenarios. Such as the extended Kalman filter (EKF) and the unscented Kalman filter (UKF) \cite{ch33}. However, the design of KF requires a good understanding of system dynamics and noise characteristics.\

Singular spectrum analysis (SSA), which is a well-known method to decompose time series into trends, seasonal components, and noise, has been explored within the signal processing field for signal denoising applications \cite{sg18, ch34, ch35, ch36, ch37,ch38, ch39} due to its effectiveness in separating useful oscillatory components from noisy signals. SSA works by embedding a signal in multidimensional space, constructing a trajectory matrix by sliding a window over the signal data points. The trajectory matrix is then decomposed using Singular Value Decomposition (SVD), which separates the signal into a series of matrices associated with singular values. These matrices represent different signal components, including deterministic structures and stochastic noise. Accordingly, the decomposed components are classified into groups that represent the signal and noise, respectively. This involves evaluating singular values and the corresponding eigenvectors to differentiate between signal information and noise components. The denoised signal is then reconstructed by summing the respective groups of matrices that contain signal information. Subsequently, the reconstructed components are transformed back into the time domain. The effective implementation of SSA in the context of signal denoising involves two main aspects: Selecting an appropriate window length and accurately grouping singular values and eigenvectors corresponding to signal information and noise components. These aspects are pivotal as they directly influence the quality and reliability of the signal-denoising process. The choice of window length determines the structure of the trajectory matrix and subsequently affects the decomposition and reconstruction phases. An appropriately chosen length ensures that the resulting trajectory matrix captures the underlying structure of the signal, effectively separating signal components and noise. A window length that is extremely short may not capture the signal's dynamics adequately, while a window length that is too long may reduce denoising efficiency and increase computational complexity. However, selecting an optimal length is not straightforward; it is highly dependent on the specific characteristics of the signal, including periodicity and stationarity \cite{ch34}. In practice, selecting an appropriate window length depends on prior knowledge of signals of interest and often ultimately relies on experimentation \cite{ch35, ch38} where empirical or heuristic methods are often employed to determine a suitable window length for a specific application. Additionally, the accurate grouping of decomposed components is fundamental for effective signal denoising as misclassification of these components can lead to either an inadequate removal of noise or the unintended discarding of useful signal information. There is no general rule for the grouping process; it is mainly determined by the specific requirements of the application, the statistical nature of the signal of interest, and associated noise \cite {ch38}.\

Frequency-domain noise reduction is based on the signal's frequency content, where noise is often more distinguishable due to its distinct spectral characteristics. Common frequency-domain techniques involve frequency-selective filtering and spectral subtraction (SS). Frequency-selective filters include low-pass \cite{ch44, ch45, ch46}, high-pass \cite{ch47, ch48, ch46}, band-pass \cite{ch50, ch54}, and notch filters (band-stop) \cite{ch52, ch53}. These filters attenuate noise components that are outside the primary frequency range. Specifically, low-pass filters allow components with frequencies lower than their defined cutoff frequency to pass through while attenuating components with frequencies higher than the cutoff. Conversely, high-pass filters perform the opposite operation, blocking frequencies below the cutoff. Band-pass filters pass frequencies within a certain range and attenuate frequencies outside that range. Notch filters attenuate specified narrow frequency bands. Frequency-selective filtering techniques are useful in applications where noisy frequency characteristics are known, such as noise caused by power-line and equipment interference. SS \cite{ch54, ch55,  ch56, ch57, ch58, ch59, ch60, ch61} relies on estimating the noise spectrum and subtracting it from the noisy signal's spectrum.  In contrast to frequency-selective filtering, which eliminates specific frequencies in the signal spectrum, SS provides global spectrum denoising mechanisms. The main steps in SS involve estimating the power spectrum of noise and subtracting it from the power spectrum of the original signal to obtain a new power spectrum. Negative values are eliminated by setting them to zero. The new power spectrum is combined with the phase of the original signal to reconstruct the denoised signal's time waveform. The noise spectrum is commonly estimated and updated during signal-inactivity periods where only noise is present. SS generally has moderate computational complexity  \cite{ch62}. Hence, it offers practical solutions in cases where noise is generally characterized as a stationary or a slowly varying process independent of the desired signal. However, in practice, random noise variations can lead to an overestimation of noise's power spectrum, leading to many negative values in the subtracted spectrum. While these values can be eliminated by replacing them with zeros, this nonlinear rectification process can cause significant distortion in the recovered signal \cite{ch62}, necessitating the adaptation of enhancement mechanisms \cite{ch54, ch59} that come at the cost of increased computational complexity.\

Time-frequency denoising methods are commonly applied for nonstationary signals since their varying spectral contents make traditional time-domain or frequency-domain methods less effective. The main concept of time-frequency denoising methods is centered around discriminating and eliminating noise-related coefficients within the time-frequency domain representations of the noisy signal. Time-frequency denoising involves transforming the signal into the time-frequency domain, representing it by a set of coefficients corresponding to each time-frequency instant. Since each coefficient reflects the signal's energy at that instant, signal denoising is achieved by screening these coefficients to distinguish between noise and meaningful signal components. Thresholding is a common technique employed in this stage; it involves setting a threshold value, below which coefficients are considered to represent noise and are, therefore, attenuated or set to zero. Coefficients above the threshold, which are assumed to represent the actual signal, are either preserved or adjusted. The key idea here is that small-magnitude coefficients. representing low energy,  are more likely to be associated with noise, while larger coefficients are likely to represent actual signal components. After thresholding, the time-domain signal is reconstructed by inversely transforming the modified coefficients to the time-domain. Wavelet-based thresholding \cite{ch65,ch66, ch67, ch68, ch69, ch70, ch71,  ch76, ch77} is commonly used for signal denoising, mainly attributed to its unique characteristics of multi-resolution analysis and the availability of a large set of basic functions. Multi-resolution analysis can capture global trends and fine details, making it useful in denoising. The diverse family of wavelet base functions allows the selection of a wavelet that matches signal characteristics. In thresholding-based denoising, the noise level is determined by the threshold value. Hence, excessively large thresholds could filter out some of the useful signal representations; on the other hand, too small thresholds could cause some noise to be retained in the signal.  There are several methods to guide the selection of an appropriate threshold, such as universal threshold, minimax threshold, and SURE threshold \cite{ch77}. Common thresholding functions include hard thresholding and soft thresholding \cite{ch68}. In hard thresholding, coefficients with absolute values less than the threshold are set to zero, while others remain unchanged. This can sometimes lead to discontinuities in the reconstructed signal. Soft thresholding, on the other hand, shrinks coefficients towards zero by the threshold value, providing a smoother result but potentially introducing bias. The choice of thresholding function and threshold values is influenced by the characteristics of the signal and the specific requirements of the application  \cite{ch77, ch79, ch67, ch81,  ch82, ch83}. In practice, it is an iterative process where thresholding and signal reconstruction are repeated, and the thresholding is adjusted at each iteration based on denoising performance.\

The use of DL for signal-denoising applications is an active research field \cite{ch84, ch85, ch86, ch87, ch88, ch89, ch90, ch91, ch92, ch93, ch94, ch95, ch96, ch97, ch98, ch99, ch100, papr_vib}; the existing approaches involve a wide range of models and methods, making them too numerous to mention. Nevertheless, the main theme is to use noise-free signals as training targets and corresponding noisy versions as training inputs. From a signal-processing perspective, such an approach can be centered around the concept of tunable denoising filters, Where a trained DL denoising model can be viewed as a tuned denoising filter whose coefficients, represented by model weights, are tuned through the training phase to remove noise from the signal. DL-based approaches are approved to be highly effective in noise removal due to their ability to learn complex, noisy patterns and related dependencies within the signal. Moreover, they can be designed to implement various operations on signal simultaneously, facilitating efficient end-to-end implementation of various signal-processing pipelines. This concept has been demonstrated in \cite{papr_vib}, where 1D convolutional autoencoders are utilized to implement a signal processing pipeline that smooths, compresses, denoises, and expands vibration signals.  The pipeline is intended to facilitate power-efficient remote condition monitoring. The smoothing and compression operations were implemented simultaneously using a smoothing autoencoder with a $\mu$-law compression-based activation function. The simultaneous denoising expansion of the compressed noisy signal is realised by a denoising reconstruction autoencoder.\ 

Despite these advantages, the effectiveness of DL-based denoising, in general, is limited to noise profiles that are present in the training set \cite{ch102}. Furthermore, as they lack inherent capabilities to adapt to varying noise patterns, conventional DL denoising approaches are impractical for real-time applications that involve varying noise characteristics.

\subsection{Availability of Labeled Abnormal Samples}
In supervised ML models, labeled samples that are large in size and diverse---in terms of class representation--- should be available to serve the training purpose. In real-world situations, the available labeled samples are usually limited in their size due to expensive and time-consuming labeling processes. Moreover, faulty or abnormal samples are not abundant since abnormal events, such as faults or anomalies, are infrequent, making it difficult to gather substantial abnormal signal data. As a result, real-world abnormal samples are usually insufficient to represent the corresponding classes during the training phase. There are several solutions in the literature to address this problem. DL-based solutions are becoming increasingly popular where transfer learning techniques are commonly adapted to avoid learning from limited labeled data \cite{ch106, ch107, ch108, ch109}. These approaches overcome the problem by adapting pre-trained models to extract deep features from the limited available signal samples, eliminating the need for model training. Recent trends utilize few-shot learning (FSL) \cite{ch110, ch111,  ch112, ch113}, enabling the pre-trained models to generalize over new categories of data using only a few labeled samples per class. While DL-based solutions show promising results, they are computationally intensive. Additionally, In contrast to computer vision, where mature pre-trained models can be developed using large-scale and well-labeled datasets \cite{oc2_5}, the development of pre-trained time-series models \cite{ch114, ch115, ch116} as generic feature extractors is still an evolving area of research.\

Non-ML solutions \cite{ch103, ch104, ch105, ch117, ch118, ch119, ch120, wd21} employ signal processing techniques for feature extraction and use similarity-based classification, providing a computationally efficient alternative to DL-based solutions. In these methods, features are extracted from available labeled normal samples, and a "normality" reference model is constructed from these features. Accordingly, signal classification is achieved by assessing the similarity between unknown samples and the reference model in the feature space. In tasks involving multiple classes, the process entails comparing unknown samples with reference samples representing different classes. The class of the unknown sample is then determined to be the class of the reference sample with the highest similarity score.\ 

Other solutions utilize data augmentation techniques \cite {ch121, ch122, ch123} to generate additional pseudo-samples. This is achieved by applying “label-invariant” transformations to available signal samples, which can considerably promote the final classification performance and generalization ability of the ML model \cite{ch123}. However, compared to image augmentation \cite{oc2_6}, which relies mainly on spatial transformations, signal augmentation involves unique challenges stemming from the complexity of the temporal and spectral characteristics of the signal. This complexity would reduce the augmentation reliability, consequently hindering the generalization of the employed model. To overcome this challenge, signals can be transformed into 2D time-frequency imagery representations, such as spectrogram and scalogram, where conventional image augmentation can be applied accordingly \cite{ch124, ch125, ch126}. However, although 2D time-frequency representations and classical images possess similar data structures, they fundamentally differ regarding what information is conveyed along the $x-y$-color space. Thus, employed data augmentation techniques should correlate with a real-world, physical transformation of the signal \cite{ch124}. While this is feasible for signals with well-defined characteristics, such as sound signals, it becomes impractical for complex signals.\

Advanced approaches leverage the powerful synthesizing capabilities of generative adversarial networks (GANs) to generate new synthetic samples with reliable characteristics \cite{ch128, ch129, ch130, ch131, ch132, ch133}. The advancement in computational capabilities enables the feasible implementation of these approaches despite their intensive computational requirements. Further, as an offline operation, synthesizing new samples does not impact online inferencing tasks.

\subsection{Computational Complexity}
Computational complexity is a major concern in the real-world deployment of signal-based ML systems. The complexity of these systems is twofold: the complexity associated with signal processing and the complexity inherent in the used ML model. From a signal processing perspective, the complexity is mainly attributed to three factors \cite{aat22}: the size of the signal segment $N_o$, algorithm-based computations involved in preprocessing and feature extraction, and the size of the resultant feature vector. While the complexity stemming from algorithmic computations is a fundamental aspect of the chosen approach, several techniques can be employed to reduce the influence of segment length and feature size on the complexity.

\subsubsection{Segment Length}
The segment length, represented by the number of samples in the segment, is a significant factor in signal processing since it directly impacts the computational complexity, system delay, and reliability of extracted features. Therefore, selecting a proper segment length is generally a balancing act that necessitates a thorough consideration of the trade-offs involved. It is often guided by theoretical analysis, empirical evaluation, and domain-specific knowledge. Considering oversampled signals, which is the common case in many applications, decimation can be used to reduce the signal's sampling rate, thereby decreasing the number of samples in the segment while maintaining the same time duration. The decimation process involves two steps: Applying low-pass filtering to the signal to prevent aliasing. The low-pass filter limits the signal's bandwidth to satisfy the Nyquist criterion, ensuring that the highest frequency in the signal is less than half of the new reduced sampling rate. Accordingly, the second step involves down-sampling the filter signal by retaining every $M$th sample and discarding the rest, where $M$ is the decimation factor. For instance, if  $M=4$, the process would keep every fourth sample and discard the three in between. Decimation is widely utilized in digital signal processing (DSP) applications  \cite{ch134, ch135}. However, the application of decimation requires a deep understanding of the frequencies of interest in the signal. To the best of our knowledge, the application of decimation in the context of feature extraction and, consequently, its effectiveness remain unexplored.

\subsubsection{Size of Extracted Features}
In contrast to the length of the signal segment, which influences the complexity of the signal processing pipeline, the dimensionality of the extracted features primarily affects the complexity of the employed ML model. There is a wealth of well-established methods in the ML literature that can effectively decrease feature size, thereby relaxing computational complexity and reducing feature redundancy. Common methods include dimensionality reduction \cite{ch136, ch137, ch138, ch139, ch140, ch141, ch142, ch143,ch136} and feature ranking and selection \cite{ch145, ch146, ch147, ch148, ch149, ch150, ch151, ch152, ch153, ch154, ch155,  ch156,  dl&sp6, ak23, ch159, ch136}. In DL-based feature extraction methods, the substantial size of the resultant deep-learned features necessitates dimensionality reduction since these features constitute a multi-dimensional latent space. Moreover, since these features lack interpretability, feature ranking and selection become essential to identify the most relevant features. In the context of signal processing-based features, these techniques are commonly employed when many features--- such as Fourier, wavelet, and energy-time-frequency coefficients--- are used. Despite their effectiveness in reducing the computational burden and improving the efficiency of the ML model, dimensionality reduction and feature selection come at the cost of extra processing, increasing the system's end-to-end complexity and time delay. An alternative solution is to utilize various signal processing tools to develop a small set of highly distinctive features. Although this typically demands domain knowledge and empirical evaluation, certain signal-processing tools can be particularly useful in this context. Specifically, Signal decomposition offers a reliable way to engineer a few highly-discriminative features from decomposed modes \cite{aat22} that are closely associated with the inherent structure of the signal. The size of the features extracted from these modes equals $1\times(K\times M)$ \cite{aat22}, where $K$ is the number of decomposed modes and $M$ is the number of extracted features for each mode. In \cite{aat22}, this approach was utilized for vibration-based condition monitoring of rolling bearings using wavelet-based decomposition. A comparative analysis shows that the approach would achieve excellent performance in fault detection with very few features compared to other approaches. Furthermore, the comparison shows that the decomposition-based method allows for shorter input segments, attributed to the high localization of the decomposed modes in both time and frequency.

\section{Applications}
This section delves into the practical application of signal-processing techniques for feature extraction in signal-based ML systems. By providing actual use cases, the section aims to bridge the theoretical foundations of signal processing with its tangible use in real-world scenarios. To foster a collaborative research environment and support reproducibility, the Python codes related to these use cases are made publicly accessible on the Github site of the Optimized Computing and Communications (OC2) Laboratory\footnote[4]{\url{https://github.com/Western-OC2-Lab/Signal-Processing-for-Machine-Learning}}. The first part of the section is a demonstration of condition monitoring of rolling bearings, where the power spectral density (PSD) of generated vibration is analyzed to extract features indicative of the operational health of the machine. The second part delves into epilepsy detection using EEG brain signals, where wavelet packet transform (WPT) is used to decompose the signals into their elementary modes. Accordingly, the energy of each mode is calculated, facilitating efficient EEG energy-based epilepsy detection.  

\subsection{Vibration-Based Condition Monitoring of Rolling Bearings}
\begin{figure*}[!htbp]
\centerline{\includegraphics[width=0.7\textwidth]{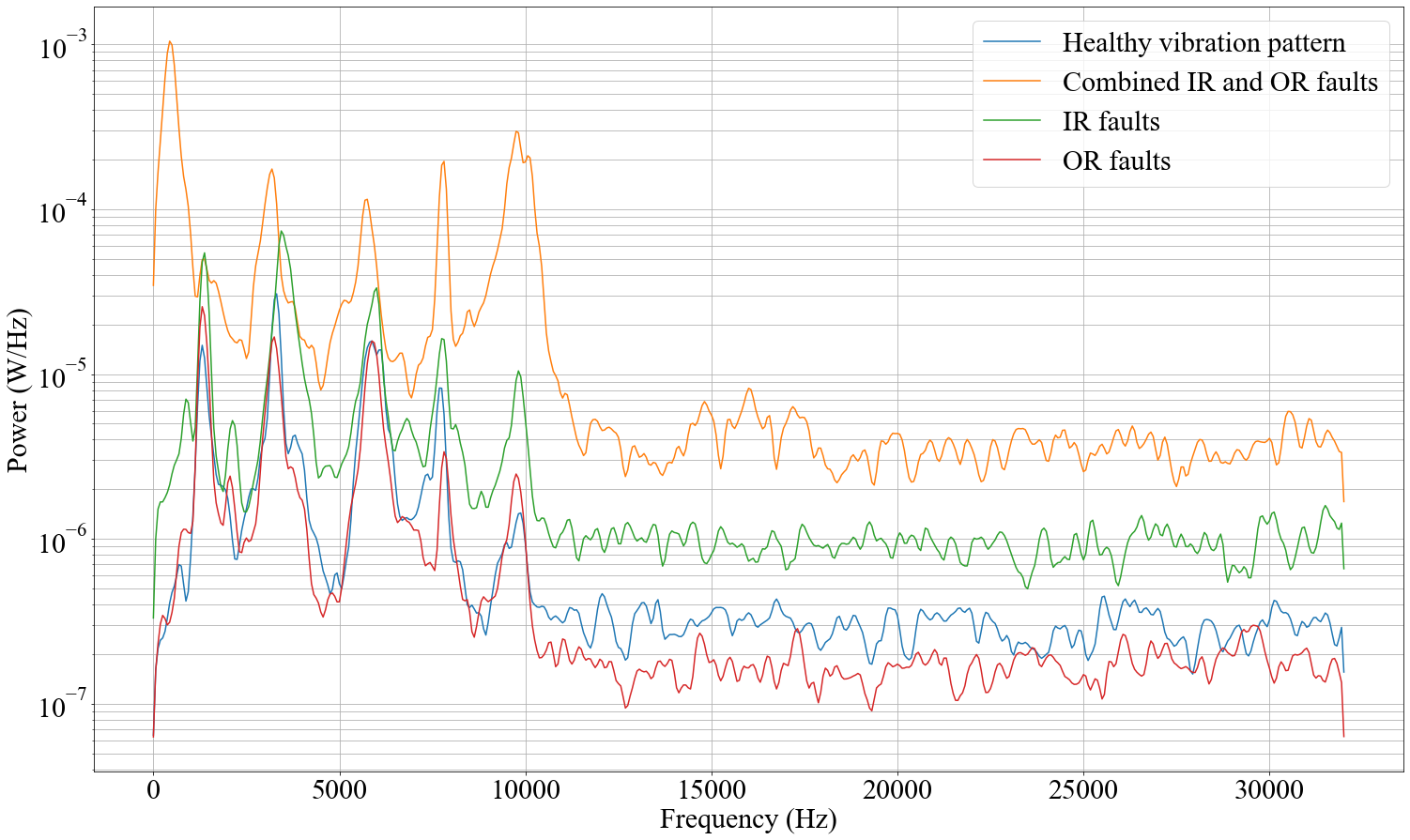}}
\caption{PSD estimates of four vibration samples representing healthy and abnormal operational conditions in the dataset.}
\label{psd_classes}
\end{figure*}

\begin{figure*}[!htbp]
\centerline{\includegraphics[width=0.7\textwidth]{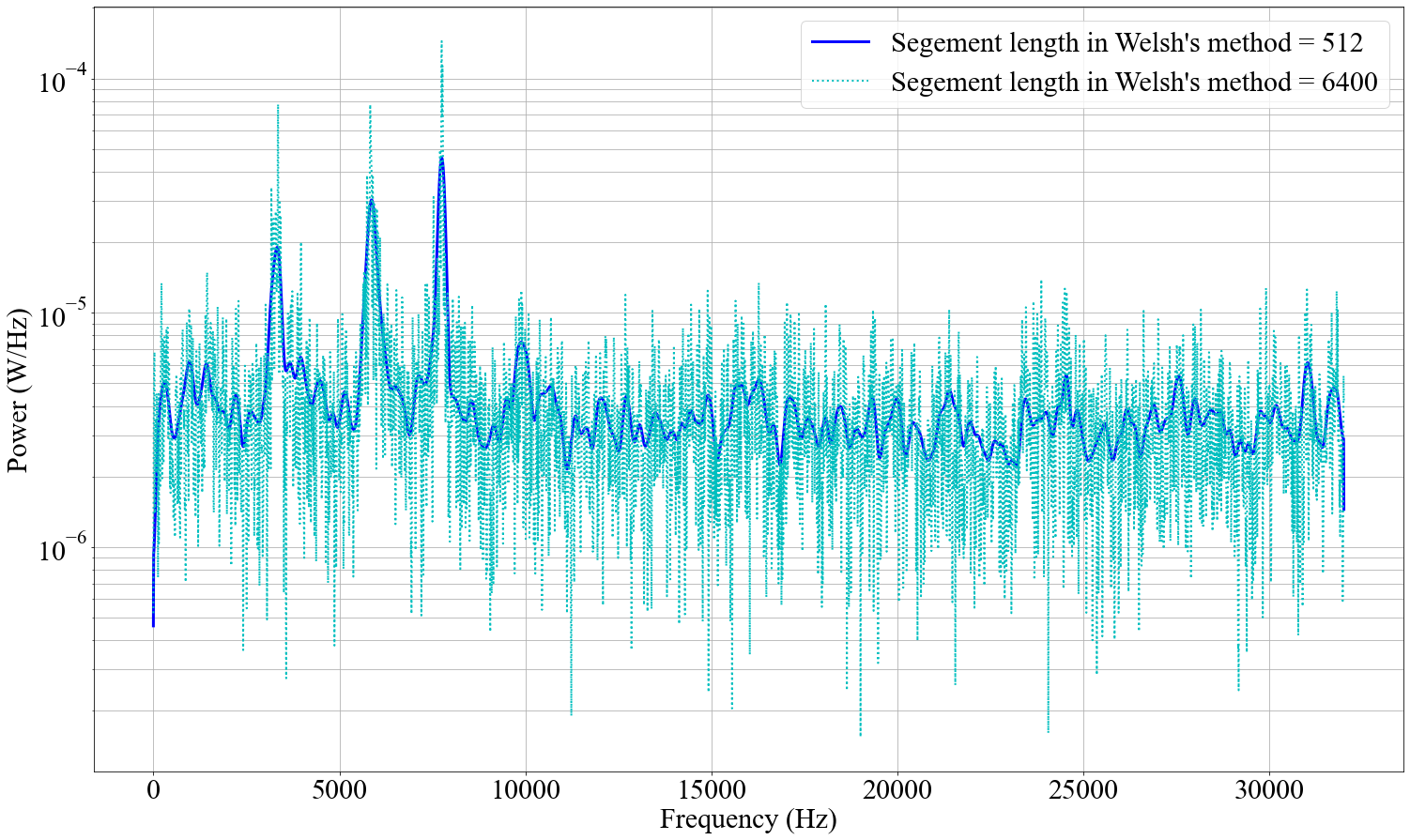}}
\caption{Two different estimations of the PSD using short and large segments in the Welsh method.}
\label{psd_seg_len}
\end{figure*}

\begin{figure}[!htbp]
\centerline{\includegraphics[width=0.5\textwidth]{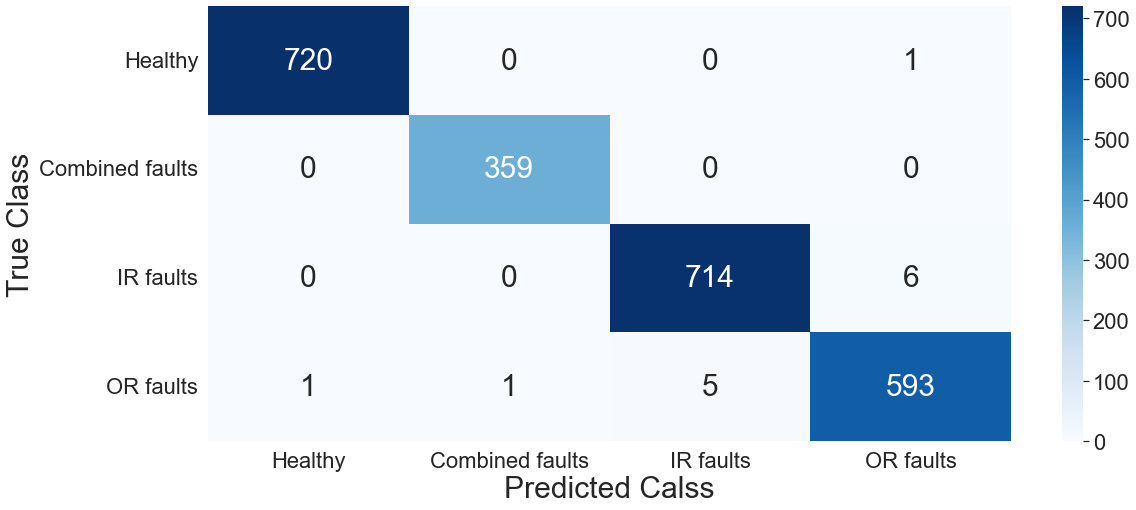}}
\caption{Confusion matrix (CM) of VBCM of rolling bearings.}
\label{cm}
\end{figure}

Vibration-based condition monitoring (VBCM)is commonly used in predictive maintenance (PdM) due to its inherent advantages over alternative forms of condition monitoring that include \cite{hos20, ran11}:
\begin{itemize}
    \item Vibration sensors are non-intrusive and can be contactless, facilitating non-destructive condition monitoring.
    \item Real-time acquisition of vibration signals can be conducted in situ, allowing for online local condition monitoring.
   \item Vibration sensors are cost-effective and widely available, offering various specifications to suit various requirements.
   \item Vibration waveform responds instantly to changes in the monitored condition and, therefore, is suitable for continuous and intermittent monitoring applications. 
   \item Of paramount significance, signal processing techniques can be applied to vibration signals to mitigate corrupting noise and extract weak condition indications from other masking signals.
\end{itemize}
The Paderborn University (PU) bearing dataset \cite{pu} is used to facilitate a real-time VBCM. The dataset includes actual vibration signals from a real system during healthy and faulty operations. The fault types include inner race (IR), outer race (OR), and combined IR and OR defects. Thus, the dataset comprises four classes of operational conditions: 1- healthy, 2- IR faults, 3- OR faults, and 4- combined IR and OR faults. 
\begin{figure*}[!htbp]
\centerline{\includegraphics[width=0.7\textwidth]{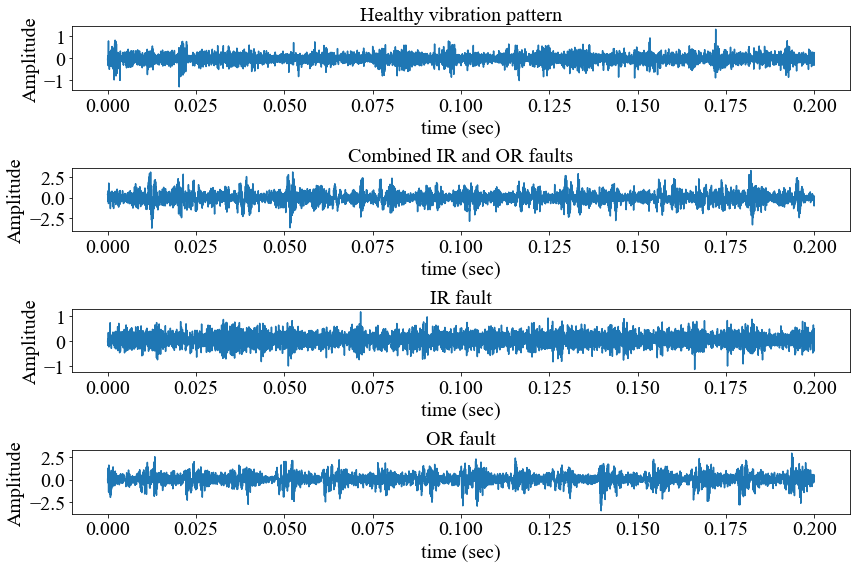}}
\caption{Sample vibration patterns representing operational conditions in the PU dataset.}
\label{pu_vib}
\end{figure*}
The vibration signals of the dataset were acquired at $64,000$ samples per second. Hence, in the signal processing pipeline, the vibration signals are segmented into non-overlapping segments of $N_o=12,800$ samples so that the resultant segment duration is $0.2$ seconds. This duration is selected to facilitate precise real-time monitoring with moderate computational requirements. This segmentation results in a dataset of  $7,997$ vibration samples in total. Fig. \ref{pu_vib} shows a healthy vibration pattern and faulty patterns for the PU dataset. The aim is to detect changes in the generated vibration (oscillation) patterns that may indicate a fault in the equipment. Here, power spectral analysis helps identify these oscillations since they appear as dominant frequencies in the PSD. Accordingly, PSDs of the vibration samples are estimated by Welch's method using the Hann window and a segment length of $512$  with $50\%$ overlap. The Number of FFT points (NFFT) is set equal to $1024$, which is calculated using the conventional method where NFFT is commonly set to be equal to $2^P$, where $P$ is the smallest power of $2$ that is greater than or equal to the Welsh segment length ($512$ samples). Thus, the value of $p$ is set to $10$.  Fig. \ref{psd_classes} shows PSD estimates of four vibration samples that represent healthy and abnormal operational conditions in the dataset. It is noticeable that a fault alters the PSD characteristics, and different faults have different effects on the PSD.
It is worth noting that the segment length of $512$ data points used in the Welsh methods represents a short time duration compared to the duration of the vibration segment. While such a short duration would reduce frequency resolution, it increases the number of estimates. This, in turn, helps to reduce noise \cite{ap2} and variance of the PSD estimate, leading to a smoother and more reliable estimate of the PSD, as demonstrated in Fig. \ref{psd_seg_len}, which compares two different estimations of the PSD of a vibration sample. 
The first estimation is based on a short segment length of $512$ data points, while the second estimation employs a larger segment of $6400$ data points. After PSD estimation, relevant spectral features are extracted from the PSD to characterize the varying vibration patterns. These features include:
\begin{itemize}
    \item Spectral Centroid (SC): The SC represents the center of gravity of the spectral contents of a signal, indicating the average frequency at which the spectral power is concentrated.  Since SC describes the spectral position of dominant oscillations in the generated vibration, it helps identify healthy and abnormal vibration patterns. SC is calculated from the PSD using the following formula \cite{ap3}:
    \begin{equation}
         CS = \frac{\sum_{k=0}^{K-1} f_k   P_k}{\sum_{k=0}^{K-1} P_k} \;\textit{Hz}
    \end{equation}
    \item Spectral Spread (SS): It describes the deviation of spectral contents with respect to the SC, providing a measure to assess how dispersed the spectral power is in relation to the SC. Mathematically, SS is obtained by calculating the spectral standard deviation with respect to SC \cite{ap4}:
    \begin{equation}
         SS = \sqrt{\frac{\sum_{k=0}^{K-1} (f_k-CS)^2   P_k}{\sum_{k=0}^{K-1} P_k}} \;\textit{Hz}
    \end{equation}
    The SC and SS are useful spectral features as they describe the position and the shape of spectral contents \cite{ap4} and, hence, characterize dominant oscillatory components within the signal.
    \item Spectral Entropy (SE): SC is obtained by calculating the Shannon entropy of the power spectral values, $P_k$, to asses the irregularity of vibration patterns. 
    \begin{equation}
        H\left [ x[n] \right ]=-\sum_{i=0}^{m-1}p\left ( x_i \right )\,log\left ( p\left ( x_i \right ) \right ),
    \end{equation}
    where, $p\left ( x_i \right )$ are probabilities associated with $x_i$ representing $m$ different values in $x[n]$.Typically, defects generate vibration with high regularity compared to healthy vibrations \cite{ap5}.
    \item Peak Power (PP): It represents the maximum value of spectral power in the PSD; it is expressed mathematically as $max(P_k), k=0,1,\dots, K$.
    \item Peak Frequency (PF): PF is the frequency pin that corresponds to the PP. The values of PP and PF characterize the dominant spike (maximum peak) in the PSD, which would be a discriminative indicator of healthy and faulty conditions as suggested by the comparison of healthy and abnormal PSDs in Fig. \ref{psd_classes}.
\end{itemize}
After the feature extraction stage, the resultant set of features is split into $5,597$ ($70\%$) samples for training and $2,400$ ($30\%$) samples for testing, and a random forest (RF) classifier is trained accordingly. The choice of the RF classifier is motivated by its higher performance compared to other classifiers, as reported in \cite{aat22}. The achieved performance results and the confusion matrix (CM) are shown in Table \ref{vbcm} and Fig. \ref{cm}, respectively. Further, a comparison, in terms of achieved accuracy (\%),  with recent deep learning (DL)-based methods on the PU dataset is presented in Table \ref{comp_pu}. The results show that the engineered features are highly effective in achieving excellent performance, similar to DL-based methods. Additionally, the signal processing approach introduced in this section is much more efficient than DL-based approaches. Specifically, unlike DL approaches, The introduced method employs a very short duration of the acquired vibration signal (only 0.2 seconds) and results in five features, sufficiently carrying out the tasks of fault detection and diagnosis. Moreover, in contrast to DL approaches, where extensive training on a large augmented dataset is required, the introduced method involves moderate training of an RF classifier.
\begin{table*}
\caption{Performance results of the proposed and other methods on the PU dataset.}
\begin{tabular}{c c c}
\hline
Accuracy (\%) & F1 score & AUROC \\
\hline
99.42\% & 0.995 & 1.00  \\
\end{tabular}
\centering  
\label{vbcm}
\end{table*}
\begin{table*}
\caption{Performance comparison between the proposed signal processing (SP)-based method and other DL-based methods on the PU dataset.}
\begin{tabular}{c c c c c c }
\hline
Method & Approach & Achieved accuracy (\%) \\
\hline
Proposed & SP-based & 99.42\%  \\

\cite{comp1} & DL-based & 99.44\% \\

\cite{comp2} & DL-based  & 97.05\% \\

\cite{comp3} & DL-based & 96.51\% \\

\cite{comp4} & DL-based & 99.50\% \\

\cite{comp5} &  DL-based & 96.67\% \\

\cite{comp6} & DL-based & 97.68\% \\
\end{tabular}
\centering  
\label{comp_pu}
\end{table*}

\subsection{Epilepsy Detection using EEG Brain Signals}
\begin{figure*}[!htbp]
\centerline{\includegraphics[width=0.7\textwidth]{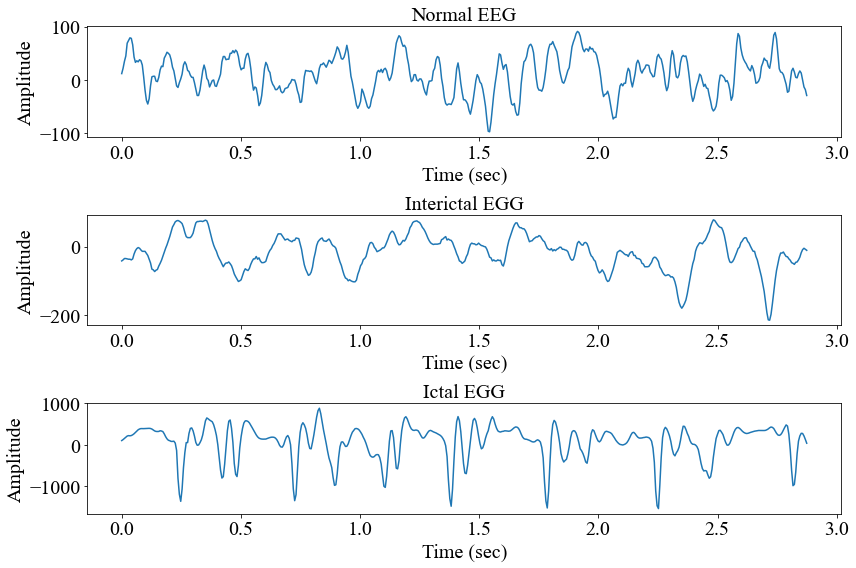}}
\caption{Sample EEG signals representing the three classes of the UoB dataset.}
\label{uob_eeg}
\end{figure*}

\begin{figure*}[!htbp]
\centerline{\includegraphics[width=0.7\textwidth]{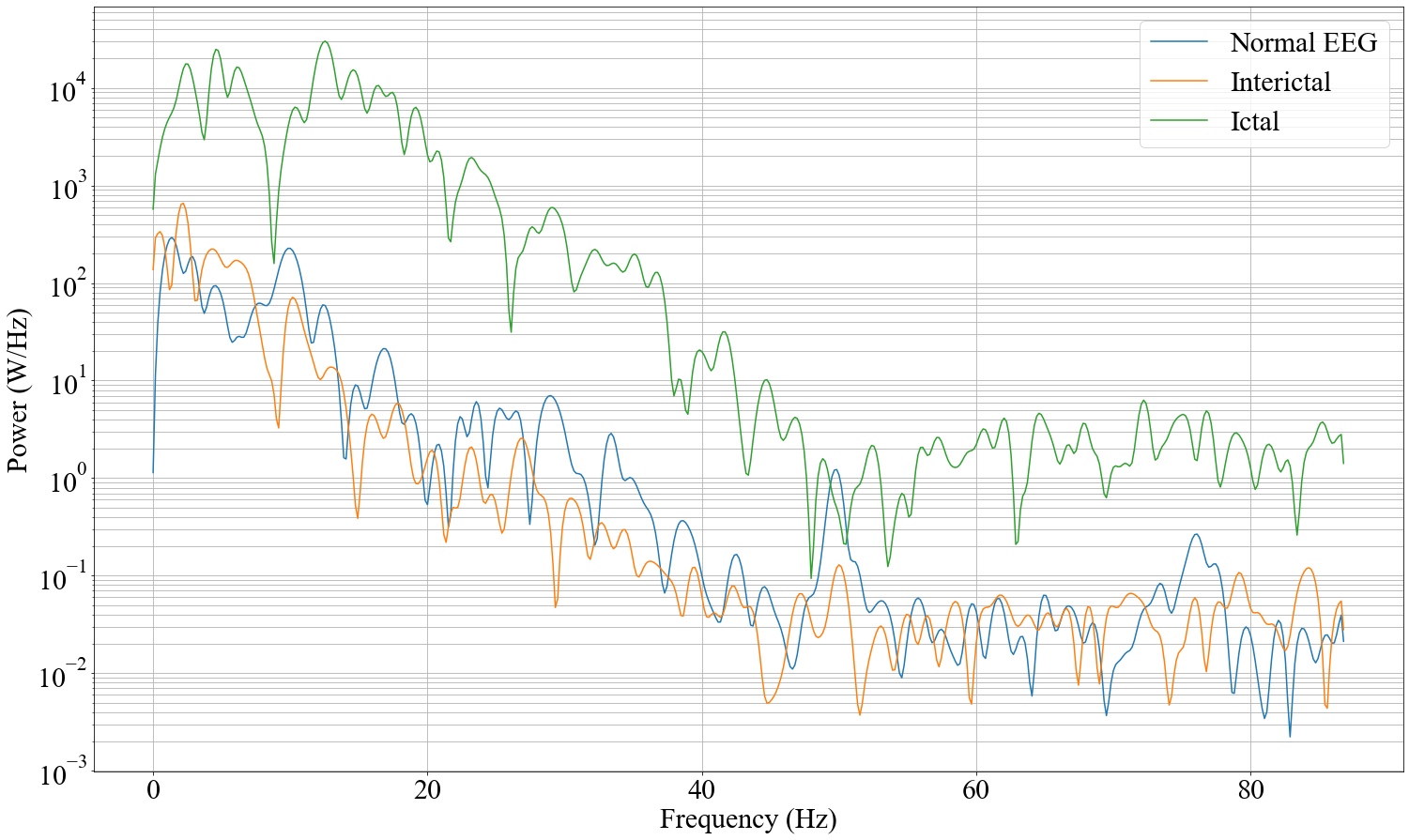}}
\caption{Samples of estimated PSDs representing the three classes of the UoB dataset.}
\label{psd_eeg_raw}
\end{figure*}

The electroencephalogram (EEG) is a technique for recording electrical activity arising from the human brain. EEG signals are typically obtained by placing electrodes on the scalp to measure the differential voltage fluctuations resulting from ionic current flows within the neurons of the brain. EEG signals can detect changes over milliseconds \cite{ap6}, thereby providing insights into the rapid dynamics of brain activity \cite{ap7}. Thus, they are commonly used as a medical diagnostic tool for conditions like seizures and epilepsy due to their high temporal sensitivity. This section introduces a signal processing-based method for epilepsy detection using EEG signals. The EEG signals are obtained from the University of Bonn (UoB) EEG dataset \cite{ap8}, which is a well-known dataset in the field of epilepsy research. The dataset consists of five subsets, labeled A through E, each containing 100 single-channel EEG segments of $23.6$ seconds. The signals were recorded with a sampling rate of $173.61$ Hz, leading to $4097$ data points per signal. sets C, D, and E were recorded from five patients; EEG signals in set D were taken from the epileptogenic zone, while set C was recorded from the hippocampal formation on the opposite hemisphere of the brain.  Sets C and D comprise EEG signals measured during seizure-free intervals (interictal), whereas the EEG signals in set E were recorded exclusively during seizure activity (ictal) \cite{ap12}. Thus, the EEG signal in the dataset can be categorized under three classes: normal (sets A and B), interictal (sets S and D), and ictal (set E). Fig. \ref{uob_eeg} shows an EEG sample from each class. 
The PSD-based method introduced in the previous section is adapted for EEG certification as a first attempt. The signals are segmented into non-overlapping segments, with a segment length $N_o$ of $500$ data point, which results in a segmented dataset of $4095$ total EEG samples. With a sampling rate of $173.61$ Hz, this gives a segment duration of $2.88$ seconds of duration. For PSD estimation, the Welsh's segment length is set to $\frac{N_o}{2}$ data points with segment overlap of $\frac{N_o}{3}$ data points; NFFT is set to $1024$. Samples from estimated PSDs are depicted in Fig. \ref{psd_eeg_raw}.
Accordingly, SC, SS, SE, PP, and PF features are extracted from the estimated PSDs. An RF classifier is trained on the extracted features using $70\%$ of the total samples ($2866$ samples) for training and the remaining $30\%$ samples ($1229$) for testing. The achieved classification results are $93.409(\%)$, $0.934$, $0.989$ representing, Accuracy (\%), F1 score, and AUROC, respectively.\

\begin{figure*}[!htbp]
\centerline{\includegraphics[width=0.7\textwidth]{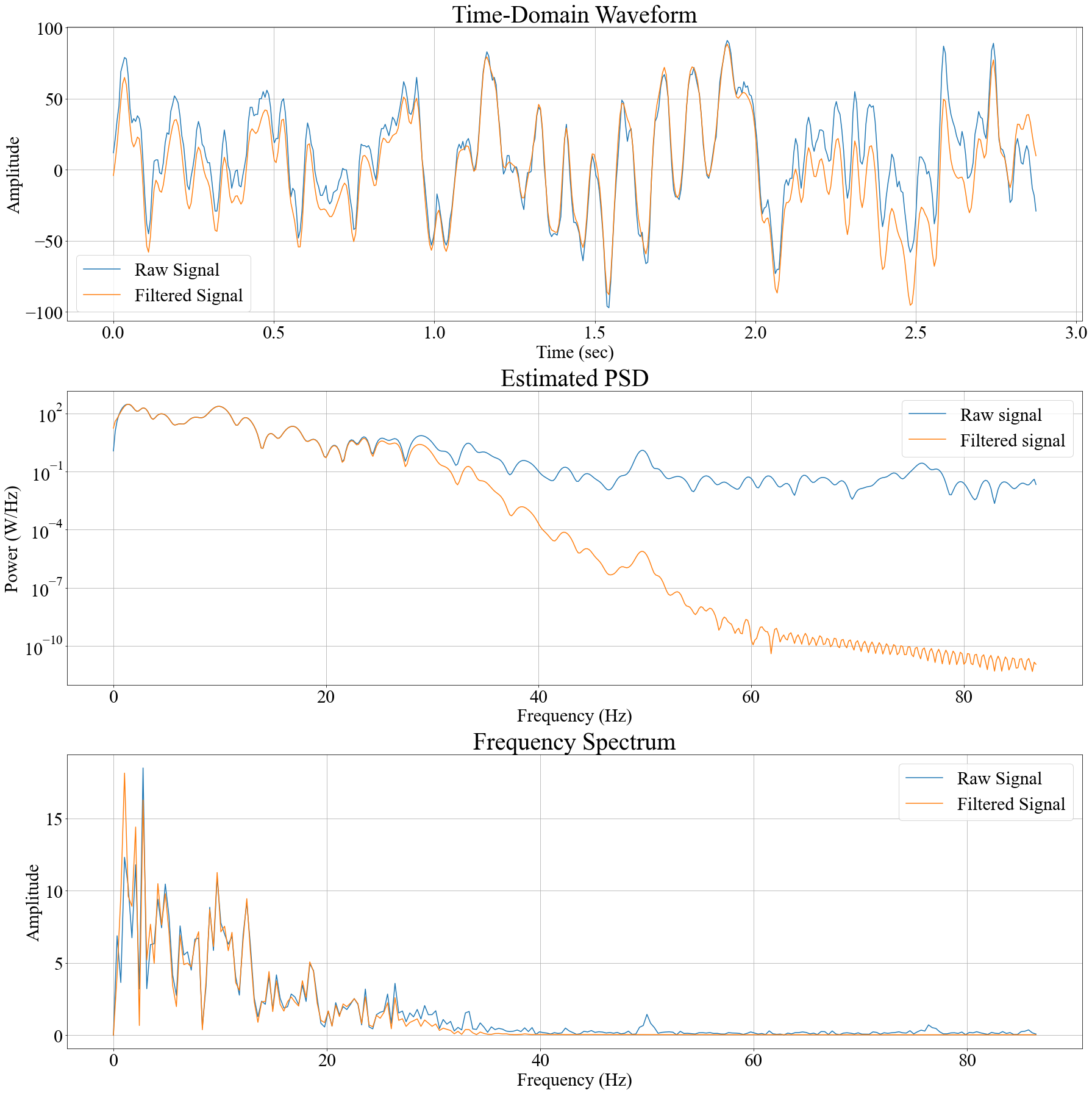}}
\caption{A comparison between unfiltered (raw) and filtered EEG signals.}
\label{eeg_comp}
\end{figure*}

\begin{figure}[!htbp]
\centerline{\includegraphics[width=0.5\textwidth]{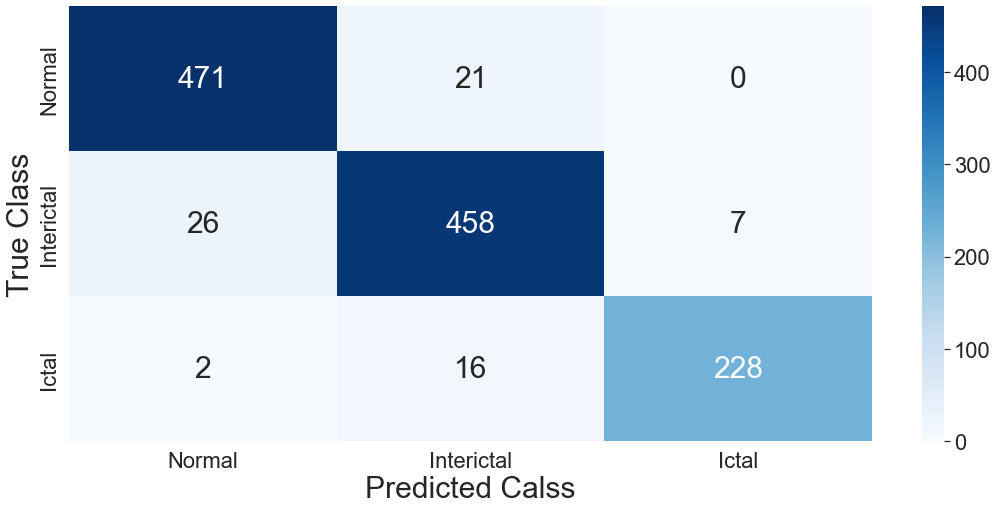}}
\caption{Confusion matrix (CM) of the PSD-based method using filtered EEG signals.}
\label{eeg_cm}
\end{figure}

In contrast to vibrations of rolling bearings, EEG signals are characterized as low-frequency signals with high temporal sensitivity, making them particularly susceptible to various forms of artifacts, such as eye and body movements. This places the need to employ a suitable signal-denoising method to filter out these artifacts and maintain the frequency range that is of clinical significance. Typically, EEG components such as infra-slow oscillations (ISO) (less than $0.5$ Hz) and high-frequency oscillations (HFOs) (greater than $30$ Hz) are outside the conventional bandwidth of clinical EEG \cite{ap9}. Considering this, a band-pass filter with pass and cut-off frequencies of $0.5$ HZ and $30$ Hz, respectively, is implemented to filter the EEG signals. Fig. \ref{eeg_comp} compares the time-domain waveforms, estimated PSDs, and frequency spectrum of unfiltered (raw) and filtered signals.
After denoising the signals using the implemented filter, the PSDs of the filtered signals are estimated, and the features are extracted accordingly. Table \ref{epilepsy1} shows the achieved results and compares them to those obtained using the unfiltered signals.
\begin{table*}
\caption{Performance results of PSD-based epilepsy detection.}
\begin{tabular}{c c c c}
\hline
Input signal & Accuracy (\%) & F1 score & AUROC \\
\hline
Raw signal & 93.41\% & 0.934 & 0.989  \\
Filtered signal & 94.55\% & 0.945 & 0.991  \\
\end{tabular}
\centering  
\label{epilepsy1}
\end{table*}
The comparison shows that filtered EEG signals brought a noticeable improvement in epilepsy detection, demonstrating the effectiveness of the employed signal-denoising in removing irrelevant frequency components and, consequently, improving detection accuracy. The CM of the filtered EEG signals case is presented in Figure \ref{eeg_cm}. 
It can be seen that there are high misclassifications between classes 1 and 2 (normal and interictal) in the CM. Examining PSD contents depicted in Fig. \ref{psd_eeg_raw} reveals a high degree of similarity between the PSDs of these two classes. Furthermore, EEG signals are generally nonstationary and nonlinear \cite{ap10}. These observations suggest that using time-frequency analyses for feature extraction would capture time-varying dynamics of EEG signals, thereby improving accuracy compared to pure frequency-domain approaches like PSD-based analysis. Thus, wavelet decomposition is employed to obtain elementary modes of EEG signals and the mode's energies are computed accordingly, This approach facilitates energy-based analysis with high time-frequency localization to capture energy-time-frequency varying dynamics within the EEG signal. The detailed steps of the method include:
\begin{enumerate}
    \item Selection of a proper mother wavelet function to decompose the segmented EEG signals using WPT to the maximum decomposition level, denoted as $k$. This step involves obtaining approximation and detail coefficients. 
    \item Use the obtained coefficients to construct elementary modes of the decomposed signal. This results in $M=2^k$ constructed elementary modes with high-time frequency localization.
    \item Compute energy $E_i$ for each mode:
    \begin{equation}
         E_i = \sum_{l=0}^{L-1} |M_i[l]|^2, i =1,2,\dots,M
    \end{equation}
    where $L$ is the length of the constructed mode $M_i[l]$. Accordingly, a feature vector of size $S=x\times2^K$ that contains energy values of each mode is created and fed into the RF classifier.
\end{enumerate}
The selection of the mother wavelet is a critical step in wavelet analysis, which is often based on empirical evaluation and prior knowledge. In \cite{ap11}, a comparison is conducted between various mother wavelets to determine the most suitable wavelet for analyzing EEG signals. Results indicate that $db4$ would be a proper selection for EEG energy analysis. Such existing results are particularly beneficial as they set guidelines for the proper selection of mother wavelets. Hence, using these findings as a guideline, Daubechies wavelets (db4 - db7) are used to decompose the EEG signals, and energy features are extracted accordingly. 
\begin{figure*}[!htbp]
\centerline{\includegraphics[width=0.7\textwidth]{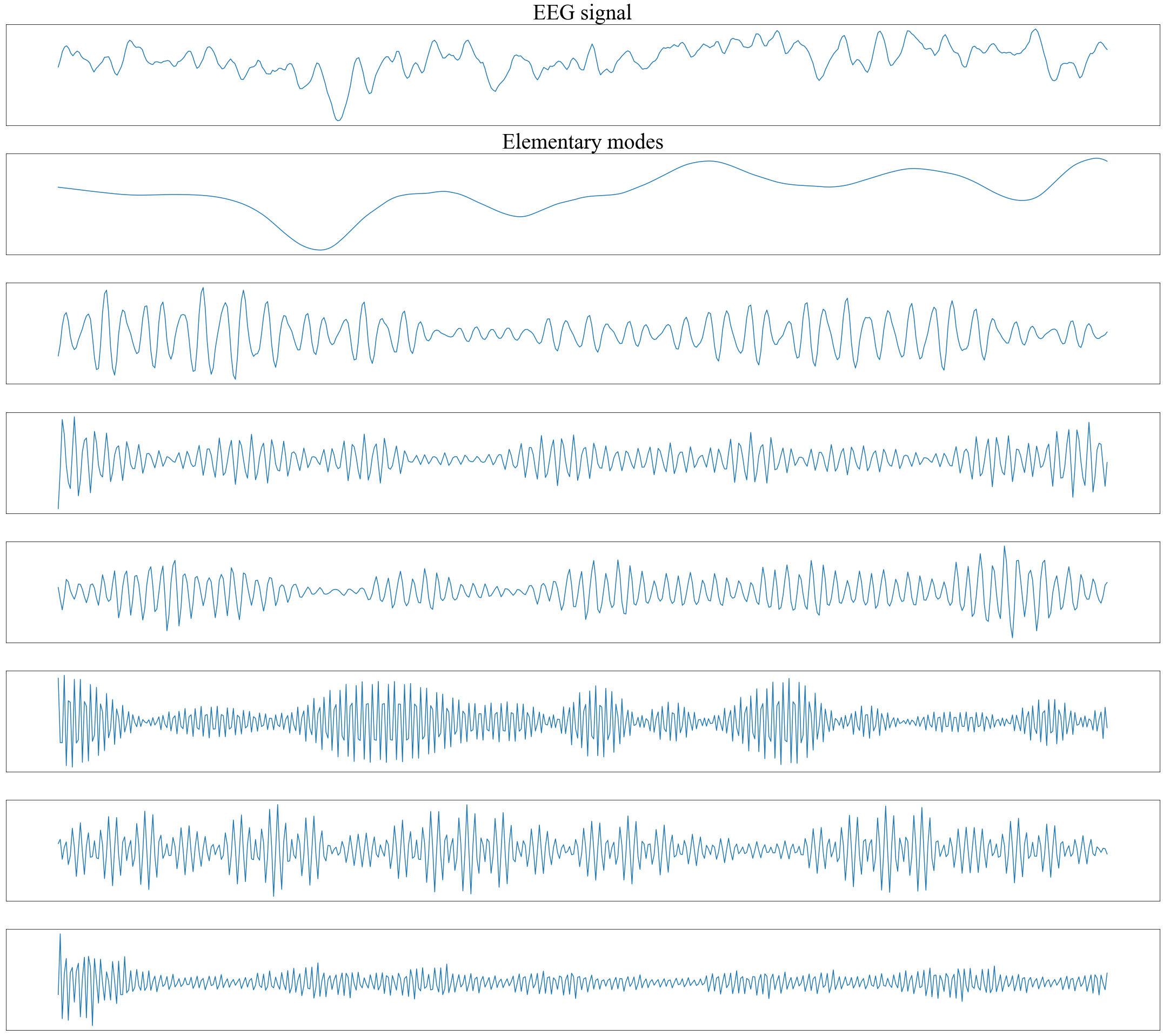}}
\caption{EEG signal along with some of its constructed elementary modes.}
\label{eeg_decompse}
\end{figure*}
The aim is to empirically evaluate these wavelets to determine the most suitable one for this application. The obtained results are presented in Table \ref{eeg_wpt_filetred}. 
\begin{table*}
\caption{Performance results of wavelet-based epilepsy detection.}
\begin{tabular}{c c c c}
\hline
Mother wavelet & Accuracy (\%) & F1 score & AUROC \\
\hline
$db4$ & 96.83\% & 0.968 &0.998   \\
$db5$ & 96.95\% & 0.969 & 0.997   \\
$db6$ & 96.59\% & 0.996 & 0.997  \\
$db7$ & 96.46\% & 0.965 & 0.996 \\
\end{tabular}
\centering  
\label{eeg_wpt_filetred}
\end{table*}
\begin{table*}
\caption{Performance Results of PSD-based and Wavelet-based Methods for Epilepsy Detection.}
\begin{tabular}{c c c c c}
\hline
Approach & Input Signal & Accuracy (\%) & F1 score & AUROC \\
\hline
PSD analysis & Raw signal & 93.41\% & 0.934 & 0.989  \\
PSD analysis & Filtered signal & 94.55\% & 0.945 & 0.991  \\
WPT-$db4$ & Filtered signal & 96.83\% & 0.968 & 0.998   \\
WPT-$db5$ & Filtered signal & 96.95\% & 0.969 & 0.997   \\
WPT-$db6$ & Filtered signal & 96.59\% & 0.996 & 0.997  \\
WPT-$db7$ & Filtered signal & 96.46\% & 0.965 & 0.996 \\
WPT-$db4$ & Raw signal & 97.56\% & 0.976 &  0.999  \\
WPT-$db5$ & Raw signal  & 98.17\% & 0.982 & 0.999   \\
WPT-$db6$ & Raw signal & 98.66\% & 0.987 &  0.999 \\
WPT-$db7$ & Raw signal & 97.92\% & 0.979 &  0.999   \\
\end{tabular}
\centering  
\label{eeg_all}
\end{table*}
A sample EEG signal and some of its constructed elementary modes are depicted in Fig\ref{eeg_decompse}. The performance results show that all wavelets achieved comparable performance with noticeable improvements compared to PSD-based analysis. These results are based on filtered EEG signals where irrelevant frequency components (ISO and HFO) are filtered out. However, in recent years, these bands gained clinical importance with the advancements in DSP \cite{ap9}. Motivated by this, the wavelet analysis is conducted on the unfiltered signals, and energy features are extracted accordingly. Table \ref{eeg_all} compares the obtained results with previous findings, including the results of PSD-based analysis. The comparison shows that in contrast to PSD-based analysis, where ISO and HFO components represent irrelevant frequency components, the presence of these components in wavelet analysis has considerably improved epilepsy detection. The results demonstrate that wavelet multi-resolution analysis reveals significant time-frequency dynamics within these bands, underscoring the significance of including these bands in wavelet analysis. 

\section{Conclusion}
Addressing the existing gaps related to the role of signal processing in ML, this paper has undertaken a comprehensive, integrated-article approach to present several contributions that aim to enrich the existing literature. First, the paper made a solid foundation on the topic through a comprehensive tutorial on signal processing fundamentals. Written for a diverse readership, the tutorial allows interested readers to grasp essential concepts and develop a proper background in signal processing.\

Furthermore, the paper provided a comprehensive overview of a typical signal-processing pipeline, introducing a structured workflow for signal-based ML applications by categorizing tasks into preprocessing, processing, and application phases. Additionally, the paper introduced an exhaustive review of feature extraction methods through a new taxonomy that clearly distinguishes between two main concepts in feature extraction: feature learning and feature engineering, thereby offering new insights into the topic of feature extraction. Focused on signal processing-based feature extraction, the paper reviewed various available techniques in terms of their main aspects, advantages, and limitations. The paper also addressed major challenges associated with the practical deployment of signal-based ML systems, explored their implications, and highlighted existing and potential solutions.\

In an effort to connect the introduced theoretical concepts with real-world applications, the paper demonstrated the practical application of signal processing-based ML systems through two use cases: vibration-based condition monitoring of rolling bearings and epilepsy detection from EEG signals. Moreover, this work contributes to the research community by introducing a public repository of relevant Python and MATLAB codes for various signal-processing techniques.

\bibliographystyle{IEEEtran} 

\bibliography{References}{}

\end{document}